\documentclass[a4paper,11pt]{article}
\pdfoutput=1 

\usepackage{jcappub}

\usepackage[T1]{fontenc} 

\usepackage{slashed}

\usepackage{bm}
\usepackage{xcolor}
\usepackage{graphicx}
\usepackage{amsmath}
\usepackage{amssymb}
\usepackage{mathtools}
\usepackage{soul}
\usepackage{braket}
\usepackage{xfrac}

\title{\boldmath 
The post-inflationary cosmology of the VISH$\nu$ axion-majoron model
}

\author[a,1]{Alexei H. Sopov,\note{Corresponding author, sopova@student.unimelb.edu.au.}}
\author[b,2]{Carlos Tamarit,\note{ctamarit@uni-mainz.de}}
\author[a,3]{Raymond R. Volkas\note{raymondv@unimelb.edu.au}}
\affiliation[a]{ARC Centre of Excellence for Dark Matter Particle Physics, School of Physics,\\ 
The University of Melbourne, Victoria 3010, Australia}
\affiliation[b]{PRISMA+ Cluster of Excellence and MITP, Johannes Gutenburg University,\\
D-55099 Mainz, Germany}

\abstract{It was previously shown how several explanatory deficiencies of the Standard Model (including the origin of dark matter, matter-antimatter asymmetry, small active neutrino mass, strong CP-conservation and the seeds for large-scale structure formation)  may be economically resolved when an experimentally-accessible QCD axion also plays the role of the majoron, and the scalar partner of the axion is dynamical during inflation. In this paper, we complete this general study of the cosmological history for a unit domain-wall number option of the DFSZ-type, dubbed VISH$\nu$, by performing a detailed lattice-informed analysis of the reheating era. In doing so, we make inflationary and leptogenesis predictions more precise through estimates of the reheating temperature and the expansion history. The viable reheating scenarios, which at the same time satisfy strict conditions for naturalness (radiative stability), are also shown to respect dark radiation bounds. We also characterise the high-frequency spectrum of gravitational waves, and mention other phenomenological implications that distinguish VISH$\nu$ from alternative proposals.
}

\makeatletter
\gdef\@fpheader{\\}
\makeatother
\begin{document}
\maketitle
\flushbottom

\section{Introduction}
\label{sec:intro}

The gravitational influence of dark matter, the baryon asymmetry of the universe (BAU) and small active neutrino masses together constitute strong empirical evidence for the existence of new physics which may be feebly coupled to the Standard Model (SM). In addition, it is thought that the SM has a fine-tuning issue, known as the strong-CP problem, due to the stringent experimental upper bound on the neutron electric dipole moment~\cite{Baluni:1978rf, Crewther:1979pi}.\footnote{This is disputed in Refs.~\cite{Ai:2020ptm,Ai:2024cnp,Ai:2024vfa}.} The Peccei-Quinn (PQ) solution~\cite{Peccei:1977hh} to this problem may be implemented in a phenomenologically successful way in invisible QCD axion models~\cite{Weinberg:1977ma,Wilczek:1977pj}, with the added benefit that axions may constitute the entire dark matter density~\cite{Preskill:1982cy,Abbott:1982af,Dine:1982ah}. As we review below, this can be configured to also generate tiny neutrino masses~\cite{Langacker1986,Shin1987} and the BAU, thus resolving an important theoretical problem and all three empirical problems in an economical framework~\cite{Volkas1988,nuDFSZ}. The purpose of this paper is to progress the cosmological study of one such realisation, the VISH$\nu$ model~\cite{Sopov:2022bog}, by demonstrating the consistency of the inflationary and post-inflationary epochs it may in principle describe.

In PQ models, the QCD axion is the pseudo-Nambu-Goldstone boson of a spontaneously broken colour-anomalous rephasing symmetry, \textit{viz.}\ $U(1)_{PQ}$. This is phenomenologically acceptable if the low-energy interactions of the axion are suppressed by a new UV mass-scale $\gtrsim 10^8$ times heavier than the electroweak scale, thus rendering the axion ``invisible''. There are essentially two options for how the colour anomaly is realised: either by chiral transformations of SM quarks, requiring at least one additional scalar electroweak doublet, or instead by exotic coloured fermions.  In the former case, the axion model is DFSZ-type~\cite{DFSZ,Zhitnitsky:1980tq} and the latter is KSVZ-type~\cite{Kim:1979if,Shifman:1979if}. In either, the axion necessarily emerges after symmetry-breaking alongside a heavy scalar partner, and the fact that both axion realisations feature new heavy states is consequential for the following two reasons. 

On the one hand, the new sector offers solutions to the other explanatory deficiencies of the SM. In particular,  masses for heavy Majorana neutrinos may be dynamically generated at the high $U(1)_{PQ}$-breaking scale,
realising not only small active neutrino masses by the familiar Type-I see-saw mechanism~\cite{Minkowski:1977sc,Yanagida:1979as,GellMann:1980vs,Mohapatra:1979ia}, but also facilitating BAU generation through leptogenesis~\cite{Fukugita:1986hr} in a certain region of parameter space. There are, in principle, as many possibilities for these \textit{axion-majoron} models as there are axions within which majorons can inhere: initial proposals were of the DFSZ-type, \textit{viz.}\ the $\nu$DFSZ model (or 2hd-SMASH)~\cite{Volkas1988,nuDFSZ,Sopov:2022bog}, later complemented by the SMASH (or $\nu$KSVZ-type) models~\cite{Salvio:2015cja,Ballesteros:2016euj,Ballesteros:2016xejSMASH,Ballesteros:2019tvf,Salvio:2018rv}, and followed by other variants~\cite{Berbig:2022pye}. Fortunately, these models may be distinguished not only by their particle content, but by their predictions and theoretical merits. 

On the other hand, both tree-level and radiative corrections involving the new high-mass states can threaten the naturalness of the much smaller electroweak scale. However, this can be avoided if the model decomposes into \textit{hidden sectors} whenever there are scale hierarchies, such that the inter-sector couplings take values that are sufficiently small to guarantee the radiative stability of the sub-Planckian scales. This situation is automatically technically natural, up to Planck-scale corrections, due to an enhanced spacetime symmetry arising in the limit in which the sectors decouple~\cite{Foot:2013hna,Sopov:2022bog} (dubbed \textit{Poincaré-protection} in flat-space).\footnote{Note that this does not do away with the SM electroweak hierarchy problem caused by the high Planck scale. The reader will nonetheless recognise that a naturally small electroweak scale 
must also of necessity be independently protected from the perturbative corrections of \textit{sub}-Planckian physics.} This is a non-trivial parameter space restriction. While it is compatible with the high-scale validity of $\nu$DFSZ-type models~\cite{Oda:2019njo}, even a heavy neutrino mass-scale suitable for hierarchical leptogenesis~\cite{nu2HDM}, this is not the case for the SMASH model~\cite{Ballesteros:2016xejSMASH}, which requires a sufficiently large inter-sector coupling.

Now, the explanatory success of the axion-majoron sector also depends on cosmological evolution through the expansion and thermal histories. It is therefore greatly enhanced if the model supports a viable inflaton. The latter leads to predictions for the energy scale of inflation, the reheating temperature, and the matter equation of state on the way to radiation domination. Recall that a sufficiently early phase of inflationary expansion~\cite{STAROBINSKY198099,Guth1981}, driven by the slow-roll of an inflaton field, can not only explain the homogeneity and flatness of the Hubble volume, but also generates a characteristic spectrum of primordial density fluctuations ultimately supporting the formation of large-scale structure. A compelling way to realise inflation is when at least one scalar field is non-minimally coupled to the Ricci scalar curvature $R$ with sufficient strength, a generic expectation in curved spacetime~\cite{Birrell:1982ix}. While the Higgs can assume the role of the inflaton~\cite{HiggsInflation1}, this scenario suffers a unitarity problem~\cite{Barbon:2009ya,Burgess:2009ea} which may be resolved if the modulus partner to the axion is also dynamical~\cite{Fairbairn:2014zta,Ballesteros:2016xejSMASH}.   

An important issue is whether or not the $U(1)_{PQ}$ symmetry is restored after inflation. If not, the axion abundance depends on an initial misalignment angle, thus compromising predictivity, and there can also be stringently-constrained isocurvature fluctuations in the axion dark matter. The restoration of $U(1)_{PQ}$ after inflation may thus be argued to be the favoured outcome. However, $U(1)_{PQ}$-breaking realises a network of topological defects which are cosmologically stable if the axion domain-wall number ($N_{DW}$) is greater than unity.  The presence of stable domain walls would render post-inflationary $U(1)_{PQ}$-restoration unviable~\cite{Sikivie:1982qv}, which is a well-known issue in the standard DFSZ realisation. Fortunately, there exist viable $N_{DW}=1$  implementations of both KSVZ-type and DFSZ-type in which the number of colourful fermions generating the anomaly is restricted or the Yukawa sector is restructured to cancel contributions~\cite{Peccei:1986pn,KRAUSS1986189,Davidson:1983tp,Davidson:1984ik,Geng1989,Geng:1990dv}. As axion-majoron models, these are the SMASH and VISH$\nu$ models, respectively. In these cases, the topological defects produced after inflation are unstable to decay into an additional non-thermal axion population~\cite{Davis}, which in addition to an averaged contribution from the misalignment mechanism~\cite{MisAlign1,MisAlign2,MisAlign3}, results in a more predictive axion dark matter mass window.

In Ref.~\cite{Sopov:2022bog}, two of the authors considered a general inflationary epoch entailed by the VISH$\nu$ model, and some basic consequences. In the most general version of the model, both of the electroweak Higgs doublets and the PQ scalar are non-minimally coupled to the Ricci scalar, yielding a three-field scalar sector. Candidate inflatons were identified by determining the parameter conditions, and the corresponding admixture of the neutral scalar parts of the multiplets, that give rise to an effectively single-field inflation that accords with cosmic microwave background  (CMB) observations, including successful fitting of the scalar spectral index ($n_s$) and adherence to the upper bound on tensor fluctuations ($r$) in the general window of 50 to 60 e-folds of inflation.\footnote{The precise value will be fixed in this work for certain parameter choices, see Figure~\ref{fig:rnsplot}.} 
In particular, the identification of the inflaton with a small-angle displacement from the modulus of the PQ scalar field was found to be compatible with the theoretical motivations of the model, allowing for non-minimal scalar gravitational couplings small enough to avoid unitarity issues.

In this paper we extend the analysis to consider the detailed dynamics of the post-inflationary epoch in VISH$\nu$, including preheating and reheating, as well as a study of possible scenarios for leptogenesis.\footnote{
During the development of this paper, Ref.~\cite{Barenboim:2024xxa} appeared which pursued a very similar line of investigation. There are nonetheless important differences which make our works function complementarily.   
We analyse a different area of parameter space as we do not introduce non-minimal gravitational couplings which break PQ symmetry, resulting in different leptogenesis and reheating scenarios, as well as a post-inflationary axion. 
Our analysis is also informed by explicit lattice simulations of the reheating epoch.} While our study complements that already performed for the SMASH model \cite{Ballesteros:2016xejSMASH,Ballesteros:2016euj,Ballesteros:2019tvf}, the post-inflationary epoch to be described has important differences owing to the suppressed portal couplings, which greatly delay the onset of the thermal era.
We find that, in order to transfer energy from the dark sector to the SM radiation bath in time for successful leptogenesis, while suppressing dark radiation, the hidden-sector structure is typically constrained to realise long-lived heavy Majorana neutrinos which ultimately reheat the thermal bath through their decays, scenarios for which we obtain viable parameter space.

The remainder of the paper is then structured as follows. The VISH$\nu$ model is reviewed in Sec.~\ref{sec:vishnu}. In Section \ref{sec:part1}, we present our lattice-informed study of the reheating epoch in VISH$\nu$ models: after reviewing standard analytical results, we characterise the relevant non-perturbative and non-linear dynamics, which then inform our study of the end of reheating, where we identify viable scenarios compatible with the hidden-sector structure. In Section \ref{sec:part2}, we tie inflationary predictions discussed in Ref.~\cite{Sopov:2022bog} to explicit reheating scenarios,  demonstrate the consistency of these with the implementation of leptogenesis, dark radiation radiation bounds and other constraints on post-inflationary axion cosmology.\footnote{The last point will be further elaborated in Ref.~\cite{Companion}.}
In Section \ref{sec:part3}, we estimate the distinctive stochastic background of gravitational waves originating from inflation and reheating. We also devote two appendices to the finer details of our analysis: Appendix \ref{sec:appA} further concerns details relevant to the lattice simulations, while Appendix \ref{sec:appB} summarises further details of our analysis of the end of reheating.
In Section~\ref{sec:con}, we conclude with a summary of our key findings and review how the new phenomenological targets for VISH$\nu$ models we have established can be probed by future experiments.  

\section{The VISH$\nu$ model}
\label{sec:vishnu}

A VISH$\nu$ extension~\cite{Sopov:2022bog} to the Standard Model (SM) is an $N_{\text{DW}} = 1$ variant of the $\nu$DFSZ model~\cite{Volkas1988,nuDFSZ}, with non-minimal gravitational coupling of at least one scalar field.

The SM matter content is extended by one complex scalar singlet ($S$), a second electroweak doublet (with the two doublets denoted $\Phi_i$) and three right-handed sterile neutrinos ($\nu^i_R$), while renormalisable interactions are restricted by a global $U(1)$ symmetry, with unit colour anomaly, which assigns the charges $+1$ and $-\frac{1}{2}$ to $S$ and the $\nu^i_R$, respectively. This achieves an alloy of lepton number with a Peccei-Quinn symmetry, which is spontaneously broken to realise Majorana mass terms for the sterile neutrinos and a dynamical DFSZ-type axion. The unit colour anomaly condition, which obviates a cosmological domain wall problem, requires a choice of quark-flavour-dependent charge assignments (``avatars'') which have been enumerated in Ref.~\cite{Cox:2023squ}.

As many of our cosmological results will generalise, we will continue to focus on the well-motivated ``top-specific''~\cite{Chiang:2015cba, Chiang:2017fjr} VISH$\nu$ avatar for definiteness, wherein only the top quark generates the colour anomaly.\footnote{Due to a coincidence discussed in our Ref.~\cite{Sopov:2022bog}, this choice allows us to partly address the flavour puzzle.} This particular $U(1)_{\text{PQ/L}}$ fixes the Yukawa sector as:
\begin{equation}
\begin{split}
    - \frac{\mathcal{L}_Y}{\sqrt{-g}} &= \overline{q_L}^{j} y_{u1}^{j3}  \widetilde{\Phi}_1 u_R^3 + \overline{q_L}^j y_{u2}^{ja}  \widetilde{\Phi}_2 u_R^a   + \overline{q_L}^{j} y_{d}^{jk} \Phi_2 d_R^k\\  &\quad + \overline{l_L}^j y_{e}^{jk}  \Phi_2 e_R^k  + \overline{l_L}^{j} y_{\nu}^{jk}  \widetilde{\Phi}_2 \nu_R^k  + \frac{1}{2} \overline{(\nu_R)^c}^j y_{N}^{jk} S \nu_R^k   +
    \text{h.c.},
\end{split}
\end{equation}
where flavour indices $j, k =1,2,3$ and $a = 1,2$ have been used for clarity with $u_R^3$ being the right-handed top quark. All avatars share the same gravi-scalar sector (with $i=1,2$),
\begin{equation}\label{graviscalar}   
       -  \frac{\mathcal{L}_{\text{gravi-scalar}}}{\sqrt{-g}} = \left(\frac{M^2}{2} + \xi_1\Phi^\dagger_1 \Phi_1 + \xi_2\Phi^\dagger_2 \Phi_2 + \xi_S|S|^2 \right)R + V(\Phi_1, \Phi_2, S),
\end{equation}
where
\begin{equation}\label{potential}
    \begin{split}
       V(\Phi_1, \Phi_2, S) &= M_{ii}^2\Phi^\dagger_i \Phi_i + M_{SS}^2 |S|^2 + \frac{\lambda_i}{2}(\Phi^\dagger_i \Phi_i)^2 + \frac{\lambda_S}{2}|S|^4 + \lambda_{iS}(\Phi^\dagger_i \Phi_i)|S|^2 \\ 
        &\quad + \lambda_3 (\Phi^\dagger_1 \Phi_1)(\Phi^\dagger_2 \Phi_2) + \lambda_4 (\Phi^\dagger_1 \Phi_2)(\Phi^\dagger_2 \Phi_1)  - \lambda_{12S} \Phi_1^\dagger\Phi_2 S\ + \text{h.c.},
    \end{split}
\end{equation}
and the theory has been written in the Jordan frame, with $m_P \simeq M $ the reduced Planck mass. The $U(1)_{\text{PQ/L}}$ symmetry is broken when $S$ develops a real vacuum expectation value $\langle S \rangle \equiv v_S/\sqrt{2} \sim 10^{11}$ GeV, realising an axion decay constant $f_a = v_S$. Electroweak symmetry breaking is implemented by $\langle \Phi_1 \rangle = (0,v_1/\sqrt{2})^{\text{T}}$ and $\langle \Phi_2 \rangle = (0,v_2/\sqrt{2})^{\text{T}}$, where $v \equiv \sqrt{v_1^2 + v_2^2} \simeq 246$ GeV, and we define $\tan \beta \equiv \frac{v_1}{v_2}$. The large fundamental-scale hierarchy in the model ($v \ll v_S$) is preserved by making the inter-sector couplings $\lambda_{iS}$, $\lambda_{12S}$ and $y_\nu$ sufficiently small. This parameter-space is technically natural due to a larger group of isometries realised in the sector-decoupling $\lambda_{iS},\lambda_{12S},y_{\nu} \rightarrow 0$ limit~\cite{Foot:2013hna, Sopov:2022bog}. 

The dimensionless non-minimal couplings $\xi_1$, $\xi_2$ and $\xi_S$ of the scalar fields to the Ricci scalar, often simply set to zero, are nonetheless required to furnish counterterms for the renormalisation of the interacting scalar field theory in a curved spacetime~\cite{Chernikov:1968zm,CALLAN197042,Birrell1980,Bunch_1980,Bunch_1980_2,Birrell:1982ix,Nelson1982,Ford1982,Parker1984,Parker1985,Buchbinder2017,faraoni2004cosmology}, making them compulsory in an inflationary universe. In a modulus-phase parameterisation for the complex scalar fields, it is the real scalar modulus fields which are non-minimally coupled, while the compact phase field directions (including the axion) are minimally coupled as we do not include possible $U(1)$-breaking dimension-four terms.\footnote{The alternative has been considered in Ref.~\cite{Barenboim:2024xxa}} In the \textit{large-field} limit of the background fields,
\begin{equation}\label{largefieldlimit}
    \xi_1 \rho^2_1 (t) + \xi_2 \rho^2_2(t) + \xi_S \sigma^2(t) \gtrsim \frac{m_P^2}{2}.
\end{equation}
where $\sigma(t) \equiv \sqrt{2}|S(t)|$ and $\rho_i(t) \equiv \sqrt{2}|\Phi^0_i(t)|$ (we neglect the charged Higgs components), the \textit{Einstein frame} potential is effectively flattened into a plateau in the modulus field-space. 

For sufficiently large non-minimal coupling, $> \mathcal{O}(10^{-2})$, this results in a viable high-scale inflation model, meeting slow-roll criteria consistent with measurements and exclusion limits on the primordial power spectra (see Figure~\ref{fig:rnsplot} and the discussion in Section \ref{sec:efolds} below for details). In this multi-field embedding of  Higgs Inflation~\cite{HiggsInflation1}, the quartic field self-interactions orient radial valleys etched in the large-field plateau, which can confine classical field trajectories with large-field initial displacement, inducing effectively single-field inflation along their floors~\cite{Kaiser:2013sna}.\footnote{The situation is described as ``effective'' in the following sense: the orthogonal modulus directions get super-Hubble masses, the trajectory does not turn in field space (suppressing non-Gaussianity) and initial isocurvature perturbations developed in the pseudoscalar directions, uncorrelated with the curvature perturbation, are sufficiently suppressed. } Parameter conditions ensuring effectively single-field inflationary valleys in a given modulus-field direction (inflation \textit{scenarios}) were derived in Ref.~\cite{Sopov:2022bog} for general non-minimal coupling in VISH$\nu$-like models. In this work, however, we exclusively focus on the \textit{small} non-minimal coupling parameter regime for the model, namely $\xi_i, \xi_S \lesssim 1$.
As explained in Ref.~\cite{Sopov:2022bog}, this results in three viable scenarios:
\begin{align}
\label{eq:infscenarios}
    \Phi_1 \Phi_2 S\text{-Inflation:}&\quad  
    (\lambda_3 + \lambda_4) \xi_1 - \lambda_1 \xi_2 < 0\ \text{and}\ (\lambda_3 + \lambda_4) \xi_2 - \lambda_2 \xi_1 < 0,\nonumber\\
    \Phi_1 S\text{-Inflation:}&\quad 
    (\lambda_3 + \lambda_4) \xi_1 - \lambda_1 \xi_2 > 0,\nonumber\\
    \Phi_2 S\text{-Inflation:}&\quad 
    (\lambda_3 + \lambda_4) \xi_2 - \lambda_2 \xi_1 > 0.
\end{align}
named for the scalar fields which have components displaced from zero as non-trivial components of the classical inflaton field admixture, later given in Eq. (\ref{eq:infmisalign}) for $\Phi_1 S$-Inflation.
The corresponding conditions are valid in the limit of very small $\lambda_{iS}$ mentioned above.

\begin{figure*}[t]
\begin{center}
\includegraphics[width = 0.6\textwidth]{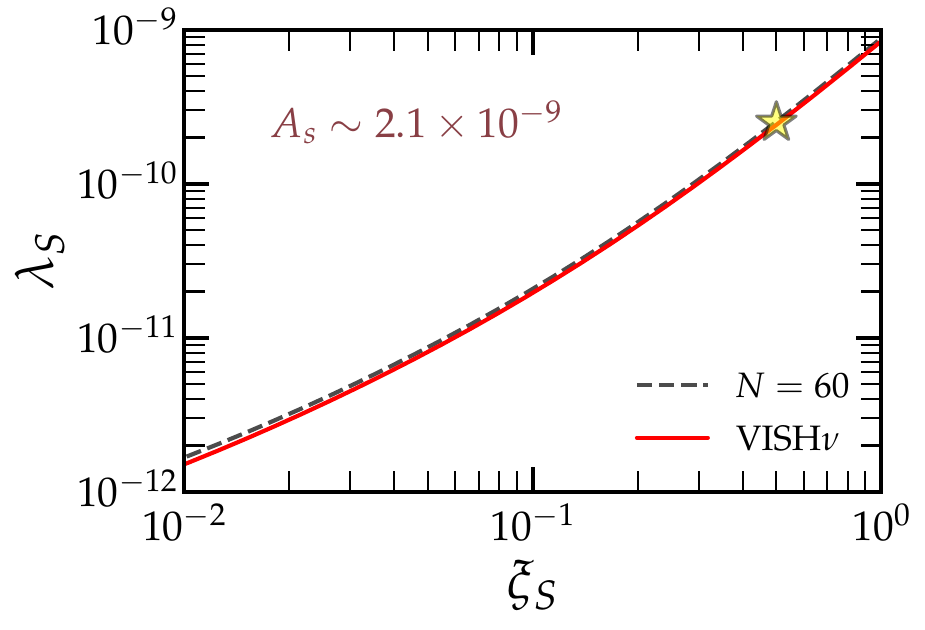}
\end{center}
\caption{\label{fig:lxiplot} We plot the numerical relationship between the $S^\dagger S$ non-minimal gravitational coupling ($\xi_S$) and the $S$ self-coupling ($\lambda_S$). The latter must be highly suppressed for the consistency of the inflation model (the fitting of the scalar spectral amplitude $A_s$~\cite{Planck:2018jri}). The allowed domain of $\xi_S$ values is further explained in Section \ref{sec:efolds}, alongside the dependence on e-folds of inflation ($N$). The starred point represents our benchmark value in what follows.}
\end{figure*}

The complementary non-minimal coupling regime, $\xi_i, \xi_S > O(1)$, which is beyond the scope of this paper, is the subject of an ongoing debate over unitarity violation during inflationary dynamics and preheating.\footnote{The argument is made by, e.g. Refs.~\cite{Barbon:2009ya,Burgess:2009ea}, while the result has been disputed in e.g. Refs.~\cite{Bezrukov:2010jz,Bezrukov_2013}.} While this makes the status of Higgs inflation~\cite{HiggsInflation1} uncertain, the issue can be entirely avoided in models such as VISH$\nu$ or SMASH, where the non-minimally coupled inflaton $\sim \sigma(t)$ can have the very weak, but nonetheless radiatively stable, self-coupling $\lambda_S \lesssim 10^{-9}$ required to suppress the scalar spectral amplitude for small $\xi$. (The exact numerical dependence is depicted in Figure~\ref{fig:lxiplot}.) This obviously makes a dedicated study of the \textit{small} $\xi$ regime the more interesting one in the VISH$\nu$ context and also means that important phenomenological departures from the related SMASH model can be identified.

\section{Reheating epoch}
\label{sec:part1}

\subsection{Preliminaries}
Having briefly reviewed the model and some parameter regimes, we now turn to the post-inflationary dynamics that precede the thermal radiation era, beginning at the moment when the accelerated expansion ends ($\epsilon_H = 1$) and the homogeneous background inflaton field, which still dominates the energy budget, starts to oscillate around the potential minimum. As a helpful preliminary to our numerical study, we establish (and largely review) results at zeroth and linear-order in field perturbations which elucidate the subsequent dynamics. We first give some relevant details of the inflationary dynamics established in Ref.~\cite{Sopov:2022bog}.

\subsubsection{The end of inflation}

As we will need to choose an inflationary scenario to study, we focus for definiteness and simplicity on $\Phi_1 S$-Inflation, though this choice incurs very little loss of generality for small $\xi$. During the accelerated expansion, the inflaton trajectory 
is initially stabilised in a valley of the Einstein frame potential where $\langle \Phi_2\rangle \simeq 0$ satisfying
\begin{equation}
    \phi(t) = \cos( \alpha )\cdot \sigma(t) + \sin( \alpha ) \cdot \rho_1 (t)
\end{equation}
where, in the degenerate limit for the non-minimal couplings $\xi_S \simeq \xi_i$,
\begin{equation}
    \alpha \simeq \arctan 
    \sqrt{\frac{2\lambda_S}{\lambda_1} \left(1 + \frac{\lambda_S}{\lambda_1} \right) },
\end{equation}
Working equivalently with Cartesian components, $\sigma e^{i\theta} \equiv  \sigma_1 + i \sigma_2$ and $\rho_1e^{i\theta'} \equiv h + iG^0$, we choose initial misalignment angles ($\theta$ and $\theta'$) so that, without loss of generality, the inflaton direction is in the $\sigma_1 h$-plane, while $\sigma_2$ and the imaginary part of $\Phi^0_1$ can be identified with the axion and the longitudinal polarisation of the $Z$ gauge-boson (in unitary gauge), respectively. Due to the radiatively-stable coupling regime $\frac{\lambda_S}{\lambda_1}\lesssim 10^{-9}$ required to saturate the scalar spectral amplitude with $\xi_S < 1$~\cite{Sopov:2022bog}, it follows initially that
\begin{equation}\label{eq:infmisalign}
    \phi \simeq \sigma_1 + \sqrt{\frac{2\lambda_S}{\lambda_1}} h \sim \sigma _1, \quad 
    \phi_\perp \simeq h - \sqrt{\frac{2\lambda_S}{\lambda_1}} \sigma_1 \sim h,
\end{equation}
where $\phi_\perp$ is the scalar direction orthogonal to the inflaton in the $\sigma_1 h$-plane. Hence, we may initially treat, to a very good approximation, the classical part of $\sigma_1$ as the inflaton, and $h$ as an initially displaced spectator field. 

Under these assumptions, it can be shown that the universe exits the inflationary phase ($\epsilon_H = 1$) when
\begin{equation}\label{finalvalue}
    \phi^2_{\text{end}} = \sigma^2_{\text{end}} \simeq \frac{m_P^2}{\xi_S} \left[\frac{\sqrt{1 + 32\xi_S + 192\xi_S^2}-1}{2(1+6\xi_S)}\right],
\end{equation}
so that, for $\xi_S < 1$, the large-field limit (\ref{largefieldlimit}) is no longer satisfied, and this will be true at all subsequent times after inflation due to spatial expansion. This suggests that the non-minimal couplings may be neglected in the reheating dynamics~\cite{Ballesteros:2016xejSMASH}, which we assume in our lattice simulations.\footnote{The effects of the non-minimal coupling will be felt during the first few oscillations of the background field, when the fluctuations are in a linear regime. So, considering a linear approximation for the modes, we confirmed that the corrections to the Hubble rate and effective masses after this point are indeed negligible for our benchmark value of $\xi_S$. In particular, there is no tachyonic instability induced by the Ricci scalar. } Additionally, the valley in the $\sigma_1 h$ plane, which had stabilised the perturbed inflaton trajectory with the help of the non-minimal couplings, no longer does so. 
The sub-dominant $h$ component, being essentially decoupled from the $\sigma_1$ component, now evolves independently in the oscillatory regime and is rapidly depleted through thermalisation into a subdominant bath of SM radiation. In fact, this manifestation of the hidden-sector structure is ultimately responsible for the substantial difference between the reheating epochs in the VISH$\nu$ and SMASH models, where the reheating is ultimately by the Higgs component.

Analogous conclusions clearly follow for $\Phi_2S$-Inflation and $\Phi_1\Phi_2S$-Inflation and make the reheating analysis scenario-independent. 
The inflation scenario will be ultimately tied to low-energy inputs for electroweak-sector parameters through the non-trivial parameter conditions, Eq. (\ref{eq:infscenarios}) supporting a particular valley orientation. 

\subsubsection{Background dynamics}

To assist in the subsequent presentation of our numerical results, we initially review standard analytical results applicable to the very first stages of the reheating when field fluctuations are small. We will first derive the expected background dynamics to zeroth-order in perturbation theory and then review, using a linear approximation, how the vacuum fluctuations developed during the inflationary expansion are subsequently amplified in a process of preheating.

Let us denote the classical inflaton by $\phi(t)$. Neglecting the radiation bath and its interaction rate with the condensate, the energy density and pressure of the scalar fluid,
\begin{equation}
    \rho_\phi(t) = \frac{\Dot{\phi}^2}{2}  + V(\phi)\qquad \text{and}\qquad   p_\phi(t) = \frac{\Dot{\phi}^2}{2} - V(\phi),
\end{equation}
satisfy a simple continuity equation,
\begin{equation}
    \Dot{\rho}_\phi + 3H(1+w_\phi) \rho_\phi = 0,
\end{equation} 
with $w_\phi(t) = \frac{p_\phi}{\rho_\phi}$ being the inflaton equation-of-state. The Hubble parameter $H(t) = \frac{\Dot{a}}{a}$ satisfies the Friedmann equation $3m_P^2H^2 \simeq \rho_\phi$. Hence, to zeroth-order, the equation of motion is
\begin{equation}\label{inflatoneom}
    \Ddot{\phi} + 3H\Dot{\phi} + \lambda \phi^3 = 0,
\end{equation}
where $\lambda \simeq \lambda_S$ and we neglect dimensionful parameters.
It is standard to study this system using a rescaled conformal time $\tau$ and rescaled field value $\tilde{\varphi}$ \cite{Greene:1997fu},
\begin{equation}
    \tau \equiv \sqrt{\lambda}\phi_{\text{end}} \eta \equiv \sqrt{\lambda} \phi_{\text{end}} \int_0^t \frac{\mathrm{d}t'}{a(t')}\qquad \text{and}\qquad  \tilde{\varphi}(\tau) \equiv \frac{\varphi(\tau)}{\phi_{\text{end}}} \equiv \frac{a(\tau)\phi(\tau)}{\phi_{\text{end}}},
\end{equation}
choosing $t = 0$ such that $\Tilde{\varphi}'(0) = 0$, where $'$ denotes a rescaled conformal time derivative, so that $\phi_{\text{end}} \equiv \phi(0)$ and $a(0) = \Tilde{\varphi}(0) = 1$ (with $\phi_{\text{end}} \sim \phi|_{\epsilon_H = 1}$). Hence, (\ref{inflatoneom}) becomes
\begin{equation}\label{conformalinflatoneom}
    \Tilde{\varphi}'' + \Tilde{\varphi}^3 \simeq 0.
\end{equation}
This has a Jacobi elliptic function solution with elliptic modulus $k= \frac{1}{\sqrt{2}}$,
\begin{equation}\label{jacobi}
    \Tilde{\varphi}(\tau) \simeq \text{cn}\left(\tau + \tau_0, \frac{1}{\sqrt{2}}\right),
\end{equation}
 and $\tau_0$ is determined by the initial conditions. The solution has period $T =4K\left( \frac{1}{\sqrt{2}} \right) \simeq 7.42$, where $K(k)$ is a complete elliptic integral of the first kind. In the small-limit with $\xi \lesssim 1$ and $V(\phi) = \lambda\phi^4$, the coherent field oscillation-averaged energy-momentum tensor is traceless \cite{Turner:1983he} and the approximate conformal symmetry is reflected in the (oscillation-averaged) solution to the Friedmann equation,
\begin{equation}
    a(t) \simeq  1 + \frac{\phi_{\text{end}}}{\sqrt{m_P}}  \left(\frac{\lambda}{3}\right)^{1/4} \sqrt{t}.
\end{equation}
This shows that the dominant inflaton zero-mode drives a radiation-like expansion.

Of course, this makes the canonical definition of the end of reheating as the onset of radiation domination after inflation rather uninformative. To be more precise, we define the end of reheating as the moment where a thermal radiation bath in equilibrium with photons first becomes the dominant energy component after inflation. \textit{Reheating} is then the intervening epoch spanning the end of inflation to this moment in the post-inflationary evolution. In a toy model, the growth and backreaction of the radiation bath can be modelled at the level of the background dynamics using the coupled system of equations,
\begin{equation}\label{adiabaticsystem}
    \Dot{\rho}_\phi + 3(1+w_\phi)H\rho_\phi = - \Gamma_{\phi}\rho_\phi,\quad \Dot{\rho}_R + 4H\rho_R = \Gamma_{\phi}\rho_\phi \quad \text{and} \quad H^2 = \frac{\rho_\phi + \rho_R}{3m_P^2},
\end{equation}
and leads to an additional effective friction term in the classical inflaton equation of motion,
\begin{equation}
    \Ddot{\phi} + 3(H + \Gamma_{\phi})\Dot{\phi} + \lambda \phi^3 = 0,
\end{equation}
where $\Gamma_{\phi}$ is the inflaton total interaction rate with species in the bath. In practice, the end of reheating will also be when the total average interaction rate of the inflaton with this bath becomes comparable to the expansion rate, \textit{i.e.} $\Gamma_{\phi} \sim H$. 

However, a completely perturbative study of the reheating along these lines is flawed when the energy of the inflaton zero-mode is far more efficiently redistributed into the fluctuations in the matter fields. In this case, the non-trivial backreaction of these fluctuations on the background dynamics should be taken into account. 

\subsubsection{Fluctuations in the linear regime}
\label{linearanalysis}

We should now consider the effect of the fluctuations. Initially, the background inflaton oscillations are frictionised by the rapid creation of particles with a highly non-thermal distribution in a process called preheating. The effects are non-perturbative, and the corresponding growth in field fluctuations often results in the breakdown of the linearised approximation, calling for numerical studies using lattice simulations of the classical matter wave scattering (see Section \ref{sec:lattice}). 

\subsubsection*{Scalar fields}

To exhibit the preheating analytically, let us first consider the Heisenberg-picture theory of a free quantised scalar perturbation field 
\begin{equation}\label{Fourierexpansion}
    \delta\Hat{X}(t,\mathbf{x}) = \int \frac{d^3\mathbf{k}}{(2\pi)^3} \left[\delta X_k(t) e^{i\mathbf{k}\cdot\mathbf{x}}\Hat{a}_\mathbf{k} +  \delta X^*_k(t) e^{-i\mathbf{k}\cdot\mathbf{x}}\Hat{a}^\dagger_\mathbf{k} \right],
\end{equation}
in the presence of the time-dependent homogeneous inflaton background $\phi(t)$, which we treat as classical. We may initially ignore particle interactions (mode-mode couplings) to linearise the equations of motion in momentum-space. Explicitly, the Fourier modes of the conformally-rescaled $\delta\Hat{\chi} = a(t)\delta\Hat{X}$ with some comoving wavenumber $\mathbf{k}$ satisfy
\begin{equation}\label{modeeqn}
    \delta\chi''_{\mathbf{k}} + \omega^2(|\mathbf{k}|,\eta) \delta\chi_\mathbf{k} = 0,
\end{equation}
where $'$ now denotes a conformal time ($\eta$) derivative. The frequency is a function of the background field and, after the onset of reheating, is therefore periodic in conformal time:
\begin{equation}
    \omega^2(|\mathbf{k}|, \eta) \equiv |\mathbf{k}|^2 + m^2_{\text{eff},0} [\varphi(\eta)],\quad \text{with}\quad  m^2_{\text{eff},0} \equiv \frac{\partial^2V(\varphi, \delta\chi)}{\partial[\delta\chi]^2}\Big|_{\delta \chi = 0}.
\end{equation}
(We have again justifiably ignored a term due to spatial expansion~\cite{Greene:1997fu}.) Note that
the frequency, and hence the mode function, is only dependent on the wavevector magnitude $k = |\mathbf{k}|$. A vacuum state is then chosen in the sub-horizon limit of all relevant modes, where curvature is negligible. When $\omega(k, \eta)$ evolves adiabatically, that is $|\omega'| \ll \omega^2$, the mode functions are given in the WKB approximation, and the positive frequency solution for $k\eta \rightarrow - \infty$,
\begin{equation}\label{WKB}
    \delta\chi_{\mathbf{k}} \rightarrow \frac{e^{- i \int^\eta \omega(|\mathbf{k}|, \eta') \mathrm{d}\eta'}}{\sqrt{2\omega(|\mathbf{k}|, \eta)}},
\end{equation} 
defines the \textit{Bunch-Davies} vacuum state consistent with negligible particle creation. At the end of inflation, the perturbed field remains linear (and free) meaning that vacuum expectation values are well-approximated by ensemble averages over classical random field realisations sampling an initially Gaussian distribution \cite{Gorbunov:2011zzc}.  While the mean value (1-point function) for the field perturbation $\langle \delta \chi \rangle $ may be taken to vanish, the variance (coincident 2-point function) of the homogeneous and isotropic random field is then characterised by the dimensionless \textit{power spectrum} $\Delta_{\delta\chi} (\tau, k)$ for the  fluctuations in momentum-space:
\begin{equation}
    \langle \delta\chi^2 \rangle = \lim_{\mathbf{x}' \rightarrow \mathbf{x}} \bra{0} \delta\Hat{\chi}(\tau,\mathbf{x'})
 \delta\Hat{\chi}(\tau,\mathbf{x}) \ket{0} = \frac{1}{(2\pi)^2a^2}\int \mathrm{d}k k^2 |\delta\chi_k(\tau)|^2 \equiv \int \frac{\mathrm{d}k}{k} \Delta_{\delta\chi} (\tau, k).
\end{equation}
We may assume ergodicity, in which case the ensemble averages may be treated as \textit{volume} averages (which is almost always the sense in which we use $\langle \dots \rangle$ hereafter). It is also possible to define an  effective comoving \textit{occupation number}~\cite{Lozanov:2020zmy},
\begin{equation}
    n_{\mathbf{k}} (\eta) \equiv \frac{\omega(|\mathbf{k}|,\eta)}{2} \left( |\delta\chi_\mathbf{k}|^2 + 
 \frac{|\delta\chi'_\mathbf{k}|^2}{\omega^2(|\mathbf{k}|,\eta)}\right) - \frac{1}{2},
\end{equation}
by computing the vacuum expectation value of the Hamiltonian operator. As modes become densely-occupied due to preheating dynamics, mode commutators become negligible with respect to their arguments~\cite{Lozanov:2020zmy} and the generically \textit{non-linear} evolution of the quantum field, including backreaction on the oscillating inflaton, is effectively described by the classical equations of motion for the matter waves (in which case the $- \frac{1}{2}$ is dropped).

With initial vacuum fluctuations seeded by the inflationary expansion, solutions to (\ref{modeeqn}) are dynamical. For a periodic frequency $\omega(k,\eta)$, there are two linearly-independent solutions given by Floquet's theorem (see e.g. Refs.~\cite{magnus_hills_2004,Amin:2014eta}): 
\begin{equation}
    \delta\chi^\pm_{k}(\eta) = e^{\pm \mu_k \eta} p_\pm (k,\eta) 
\end{equation}
where $p_\pm(\eta)$ are periodic in time and $\mu_k$ is the $k$-dependent \textit{Floquet exponent} associated to $\omega$. \textit{Parametric resonance} occurs when there is a subset of modes satisfying $\text{Re}\,\mu_k \neq 0$, called \textit{instability bands}, where the linear solution inevitably becomes exponentially growing as the oscillating inflaton exerts a driving effect through the periodic frequency, resulting in exponentially growing occupation numbers $n_k \sim e^{2|\text{Re}\, \mu_k|\eta}$. This evolution is consistent with a violation of the adiabatic condition when the classical inflaton crosses zero and quickly results in a highly non-thermal, non-Gaussian particle distribution. Intuitively, the otherwise unbounded growth comes to be physically limited by the background energy density being transferred to the fluctuations, and is ultimately hampered by the onset of non-linear effects. 

In preheating, the Floquet exponents ultimately depend on the couplings of the field perturbations to the inflaton (through their effective masses) and the periodic function representing the classical inflaton solution \textit{e.g.} (\ref{jacobi}). The case of reheating with a quartic inflaton potential with a four-point interaction \textit{viz.} $V \supset \frac{g^2}{2}X^2\phi^2$, is well-known and originally studied in Ref.~\cite{Greene:1997fu}. Let us introduce a rescaled wavenumber:
\begin{equation}
    \kappa \equiv \frac{k}{\sqrt{\lambda_S}\phi_{\text{end}}}.
\end{equation}
The relevant findings of Ref.~\cite{Greene:1997fu} which one expects to apply to the VISH$\nu$ context at the onset of the preheating are as follows:
\begin{itemize}
    \item The $\sigma_1$ ($\sim$ inflaton) perturbation satisfies $g^2 = 3\lambda_S$, which has an associated unstable band of wavenumbers (\textit{self-resonance}) $3/2 < \kappa^2 < \sqrt{3}$ with $\mu_{\max} \simeq 0.0359$ at $\kappa \simeq 1.27$.
    \item The $\sigma_2$ ($\sim$ axion) perturbation satisfies $g^2 = \lambda_S$, which has an associated unstable band of wavenumbers $0 < \kappa^2 < 1/2 $ with $\mu_{\max} \simeq 0.1470$ at $\kappa \simeq 0.47$ (the growth is \textit{stronger} than self-resonance).
    \item The Higgs components are all very weakly coupled to the background ($g^2 \ll \lambda_S$). In this regime the unstable band is extremely narrow, with $\mu_k \ll 0.01$, making resonant production of Higgs particles from the inflaton negligible.
\end{itemize}

The periodicity of the oscillating inflaton is not the only driver of non-perturbative particle production. Another relevant effect is called \textit{tachyonic instability}, and occurs when the inflaton probes a region of the potential where $V_{,\chi\chi} < 0 $. In principle, this situation is realised in VISH$\nu$ by the hitherto neglected dimensionful couplings, provided the effective potential is not minimised at zero. To see this consider
\begin{equation}
    V(\sigma_R,\sigma_I) = \frac{\lambda_S}{4}(\sigma^2_R + \sigma^2_I - v_S^2)^2
\end{equation}
with the inflaton being the classical part of $\sigma_R$ so that in the linear regime,
\begin{equation}\label{frequencieslinearregime}
    \omega^2_{\sigma_R} (|\mathbf{k}|,\eta) 
    \simeq |\mathbf{k}|^2 + 3\lambda_S\varphi^2(\eta) - a^2 \lambda_S v_S^2,\quad \omega^2_{\sigma_I} (|\mathbf{k}|,\eta) \simeq |\mathbf{k}|^2 + \lambda_S\varphi^2(\eta) - a^2 \lambda_S v_S^2.
\end{equation} 
Let us take $\eta = 0$ to be the moment where the inflaton first crosses the origin (where e.g. $V_{,\sigma_{R}\sigma_{R}} <0$ and adiabaticity is first violated). Using the positive frequency solution in (\ref{WKB}), it follows that $\Sigma_{R,I;\mathbf{k}} \propto e^{- i(|\mathbf{k}|^2 - a^2\lambda_S v_S^2)^{1/2}\eta}$, which are exponentially growing solutions for all modes in the instability band $|\mathbf{k}| < a\sqrt{\lambda_S}v_S = |m_{\text{eff},0}(0)|$  (where $\Sigma_{R,I} = a\sigma_{R,I}$). Correspondingly, one initially has that $n^{\Sigma_{R,I}}_{\mathbf{k}}(\eta) \sim e^{2|\omega_{R,I}|\eta}n^{\Sigma_{R,I}}_{\mathbf{k}}(0) $.  As the modes eventually get a real-valued mass in the course of the background oscillation, this becomes \textit{bona fide} exponential particle production. The situation is dangerous in axion models for $v_S \gtrsim 10^{16}$-$10^{17}$~GeV~\cite{Ballesteros:2021bee}, however we find that the effect is completely negligible if the dimensionful coupling is much smaller in magnitude than the inflaton oscillation amplitude (which is the case for the post-inflationary $v_S$ regime under investigation~\cite{Companion}).

Whether amplified by parameteric resonance or tachyonic instability, if the fluctuations (particles) are not dissipated by annihilation and decay, the initial perturbation grows to become comparable in size to the background inflaton field and the linearised approximation breaks down. For example, this build-up occurs in VISH$\nu$ as the parametrically amplified fields are the components of $S$, which are feebly coupled to other species. Two important back-reaction effects are to be mentioned which eventually terminate the preheating. The first non-trivial alteration of the background dynamics (which is often simply called \textit{back-reaction}) can be understood in a quasi-linear regime using the Hartree approximation, where modes and fields evolve independently (admitting only self-correlations in time~\cite{Lozanov:2020zmy}). In this approximation, the back-reaction can be estimated by mass corrections due to one-loop Hartree diagrams proportional to the field variances~\cite{Greene:1997fu}. For example, following Ref.~\cite{Greene:1997fu}, we find:
\begin{equation}
    \omega^2_{\sigma_R} (|\mathbf{k}|,\eta) 
    \simeq |\mathbf{k}|^2 + 3\lambda_S\varphi^2(\eta) - a^2 \lambda v_S^2 + 3\lambda_S\langle \delta \sigma_R^2 \rangle + \lambda_S \langle \delta \sigma_I^2 \rangle.
\end{equation}
Accordingly, the instability bands are shifted (affecting the parametric resonance) and the fluctuations can terminate the tachyonic instability by inducing a global minimum of the effective potential at the field-space origin, in a process of non-thermal symmetry restoration.  

Eventually, this approximation will also break down as mode-mode couplings become important and the fully non-linear regime is reached. The ensuing \textit{rescattering} redistributes the energy accumulated by a given unstable mode into the other modes and fields to which it is coupled, including those outside the instability bands. At this point, the evolution of the system is only tractable numerically, motivating the lattice study we perform in Section \ref{sec:lattice}.

\subsubsection*{Fermion fields}

The non-perturbative production of fermions from the inflaton condensate, though greatly limited in its efficiency by Pauli blocking, may still differ appreciably from the perturbative approximation of decaying inflaton particles during preheating~\cite{Fermions1, Giudice:1999fb_Fermions2, Peloso:2000hy_Fermions3}, occuring, for example, even for heavy fermion species otherwise kinematically inaccessible as final states. In particular, the parametric excitation of fermion modes coupled to the inflaton background, due fundamentally to their background-dependent periodic frequency, has been demonstrated in a linearised analysis of the toy reheating model $\frac{\lambda}{4} \phi^4 + y\overline{\psi}\phi\psi$ in Ref.~\cite{Fermions1}.  (There is an obvious parallel to VISH$\nu$ model, rendered explicitly as $\phi \rightarrow \sigma$, $\psi \rightarrow N_i$ and $y \rightarrow \frac{y_{N_i}}{2}$.) This effect is modulated, as in the bosonic case, by a resonance parameter $q\equiv \frac{y^2}{\lambda}$.  

\begin{figure*}[t]
\begin{center}
\includegraphics[width = 0.49\textwidth]{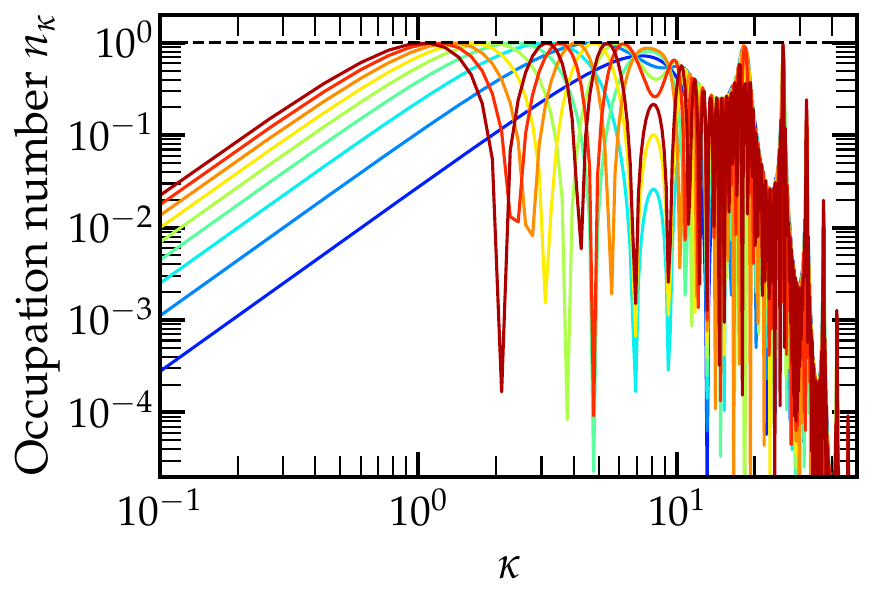}
\includegraphics[width = 0.49\textwidth]{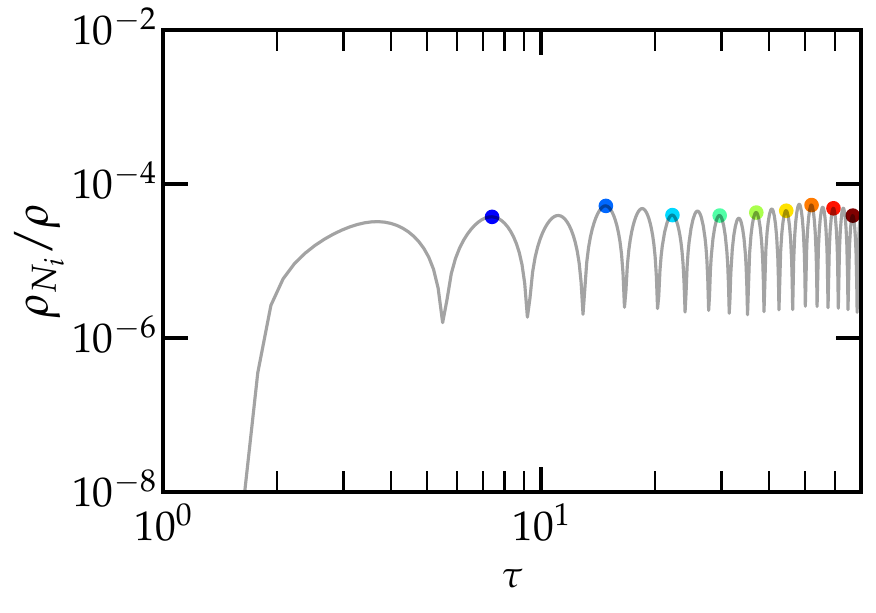}
\caption{\label{fig:fermionicpreheating1} In the left panel we plot the fermionic occupation number spectrum after 16 oscillation periods of the background field ($\tau < 100$), choosing $y_{N_i} = 10^{-2}$ and $\lambda_S = 2.5\times10^{-10}$ (see Section~\ref{sec:neutrinos} for details). The apparent Pauli blocking effect, restricting $n_k \leq 1$, results in the almost immediate saturation of the neutrino energy density fraction, as seen in the right panel. Colours are used in each plot to correlate the time point (energy fraction) with the spectral contours. Note that, in this regime, perturbative production of neutrinos from the inflaton is otherwise kinematically blocked, and the neutrinos are produced with low momentum ($ \kappa < y_N/\sqrt{\lambda_S}$).}
\end{center}
\end{figure*}

For the purposes of this paper, we are principally interested in the energy accumulated by the neutrino baths (which later thermalise with the SM bath), and the perturbative limit has been shown to provide a good guide to the dissipation of inflaton energy density over many e-folds for $q^2 \ll 1$~\cite{Garcia:2021iag_Fermions4} (our benchmark regime for the first generation $N_1$). Nevertheless, for $q^2 \gg 1$, a non-perturbative study is essential for kinematic reasons, and interesting, in that it predicts a small primordial non-thermal energy fraction (see Figures~\ref{fig:fermionicpreheating1} and \ref{fig:fermpreheating}) which becomes matter-like after the PQ phase transition. What is more, essentially due to the bound on fermion occupation numbers, the effect may be studied numerically without the use of lattice simulations and does not constitute an important backreaction effect for feasible simulation times. In essence, the parametrically excited fermions will simply spectate much of the non-linear evolution of the bosonic fields discussed in Section \ref{sec:lattice}, before they can ultimately facilitate the end of reheating for a subset of the model parameter space, as explained in Section \ref{sec:reh}.

\subsection{Lattice study}
\label{sec:lattice}

We used a modified version of the CLUSTEREASY code~\cite{Felder:2000hq,Felder:2007kat} to numerically solve lattice-discretised and canonically-rescaled versions of the following system of equations
\begin{equation}\label{latticeeqns}
\begin{split}
    \Ddot{X}_i + 3 \frac{\Dot{a}}{a} \Dot{X}_i - \frac{1}{a^2}\nabla^2 X_i + \frac{\partial V(X_j)}{\partial X_i} &= - \Gamma_i \Dot{X}_i, \\
    \Dot{\rho}_{\text{R}} + 4 \frac{\Dot{a}}{a}\rho_{\text{R}} &= \sum_j \Gamma_j \Dot{X}_j,
    \\
    3m_P^2\left(\frac{\Dot{a}}{a}\right)^2 &= \rho_{\text{R}} + \frac{1}{2}\sum_j \Dot{X}_j^2 + \frac{1}{2a^2} \sum_j (\nabla X_j)^2 + V(X_j),
\end{split}
\end{equation}
where the $X_i \in \{ \sigma_R, \sigma_I, h, H, A, H^+_R, H^+_I \}$ are the seven real scalar Cartesian components
\begin{equation}\label{CartesianBasis}
    S =\ \frac{1}{\sqrt{2}}(\sigma_R + i\sigma_I),\quad \Phi_1 =\ \frac{1}{\sqrt{2}} \begin{pmatrix} 0 \\ h \end{pmatrix},\quad \Phi_2 =\ \frac{1}{\sqrt{2}} \begin{pmatrix} H^+_R + iH^+_I \\ H + iA \end{pmatrix},
\end{equation}
of the classical random fields in unitary gauge, $\Gamma_i$ are their decay rates to fermion final states (given in Appendix \ref{sec:appA}) and $V$ is given by (\ref{potential}). This is possible since, during $\Phi_1 S$-Inflation, the large-field background inflaton configuration is stabilised in a valley of the Einstein frame potential where $\langle \Phi_2\rangle \simeq 0$. As we have seen, the inflaton can then be taken to be a linear combination of predominantly the real part ($\sigma_1$) of the PQ scalar ($S$), and sub-dominantly the real part ($h$) of the neutral component of the active Higgs doublet ($\Phi_1$), while no other Higgs components acquire an effective vev. Hence, the inflationary regime of $\Phi_1 S$-Inflation automatically aligns the Higgs basis for the two electroweak doublets with the mass basis, which greatly simplifies our numerical study. By therefore consistently working in unitary gauge, we may omit Goldstone components that instantiate the longitudinal polarisations of the massive $W^\pm$ and $Z$ gauge bosons.\footnote{The standard picture of resonant electroweak gauge boson production from the oscillating Higgs spectator was confirmed using auxiliary $\mathcal{C}$osmo$\mathcal{L}$attice~\cite{Figueroa:2020rrl,Figueroa:2021yhd} simulations. However, we saw no important effect on the dynamics of the PQ fields, and the bosonic electroweak-sector fields remained a highly subdominant energy component. Hence, the gauge bosons (including the longitudinal polarisations) could be neglected in our general lattice study.} 

\begin{figure*}[t]
\begin{center}
\includegraphics[width = 0.5\textwidth]{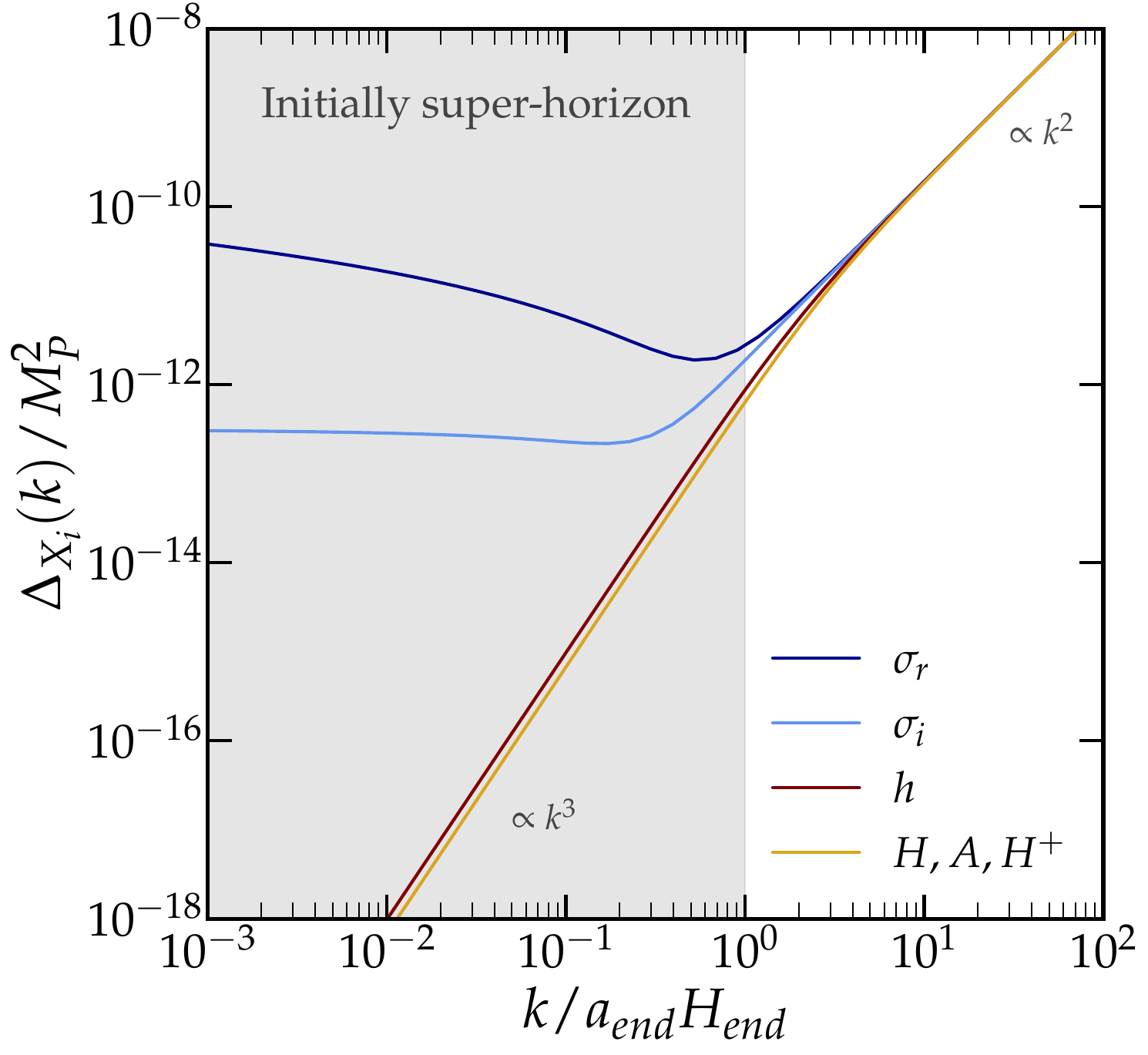}
\end{center}
\caption{\label{fig:initconds} We plot the power spectra for the field fluctuations at the end of inflation used in the initialisation scheme for the lattice simulation. It may be seen that all scalar fluctuations remain in vacuum on sub-horizon scales; the Higgs fluctuations are suppressed on super-horizon scales; the axion modes are excited into an initial, subdominant, isocurvature spectrum; while the radial modes may be identified with the inflaton. (This recapitulates the main points of the ``effectively single-field'' discussion in Section \ref{sec:vishnu}). }
\end{figure*}

Our modifications to CLUSTEREASY essentially follow Ref.~\cite{Ballesteros:2021bee} and correspond to the following additions to the program.
\begin{itemize}
    \item Implementation of interaction rates $\Gamma_i$ in the evolver kernels. These are given in Appendix~\ref{sec:appA}.
    \item Implementation of dynamical $\rho_R$ with backreaction in the Friedmann equation.
    \item Calculation of gravitational wave spectra (in a linearised approximation).
    \item Estimation of axion spectra and axion/modulus energy densities (see Section \ref{sec:latticept}). 
    \item Modification to initial conditions for the fields.
\end{itemize}
Note that the fields are initialised in momentum-space as Gaussian random fields. As we began our program at the end of inflation, we replaced the default initialisation scheme, which assumes a (minimally-coupled) scalar power spectrum valid only in the sub-horizon limit, with one that is also valid for super-horizon modes (of non-minimally coupled fields) because these will be dynamically relevant during preheating.\footnote{While another option is to set the initial time for the simulation several e-folds before the end of inflation, we found it expedient to do a separate linear analysis that included the non-minimal couplings (where our small-field limit approximation breaks down) so that they could be consistently ignored in the actual simulation (where the small-field limit applies). This was possible as the effects of the non-minimal couplings are negligible in the small-field regime, as discussed above.} The initial power-spectra we obtained numerically using the benchmark parameters are given in Figure \ref{fig:initconds}.

\subsubsection{Preheating}
\label{sec:latticeprefragturb}

\begin{figure*}[t]
\begin{center}
\includegraphics[width = 0.8\textwidth]{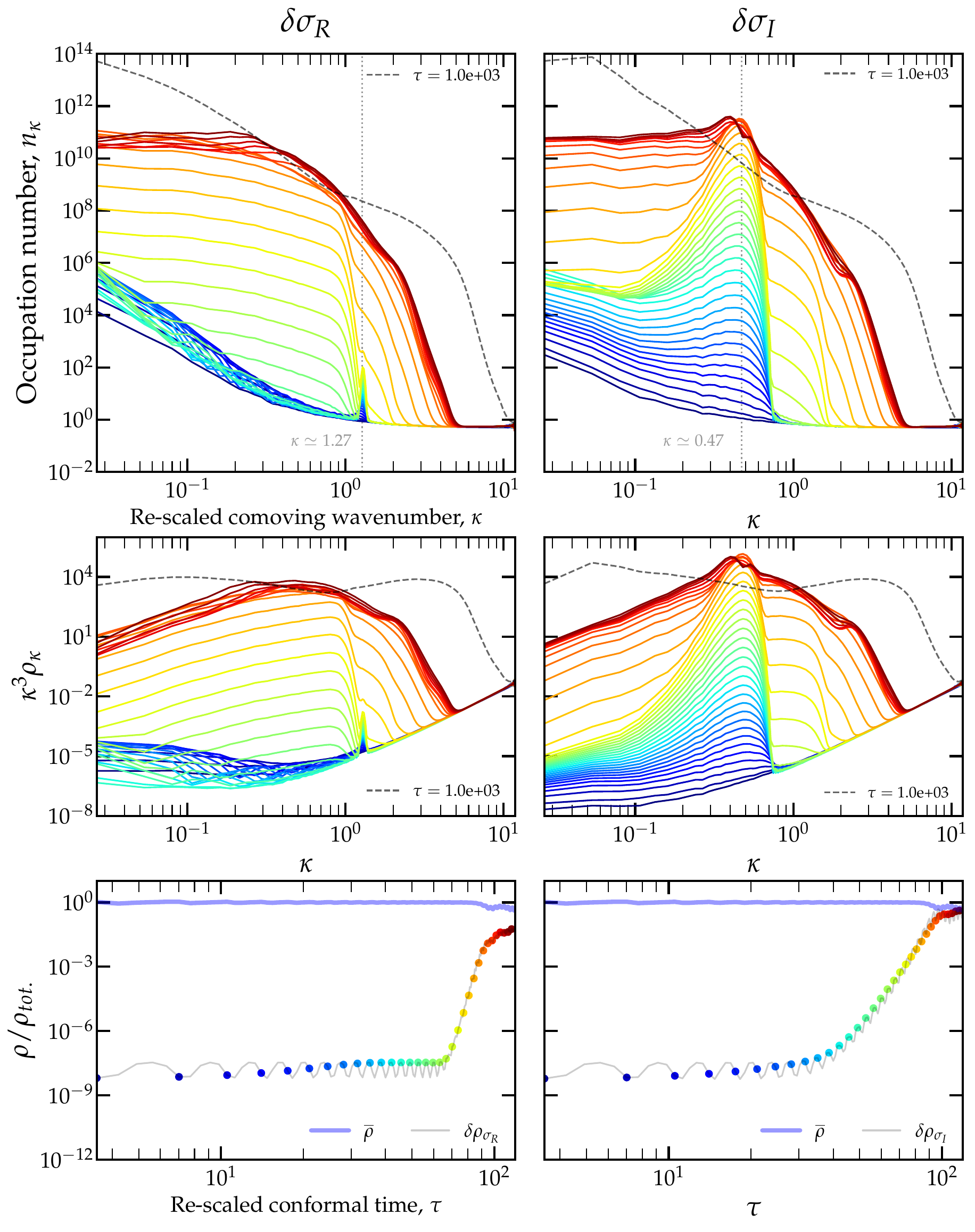}
\caption{\label{fig:spectrapreheating} 
We compare, for (\ref{eq:sim1}), the preheating of the inflaton ($\delta\sigma_R$, \textit{left}) and axion ($\delta\sigma_I$, \textit{right}) fluctuations. In the first row, we plot the occupation number spectrum, excited through parametric resonance from the initial inflationary spectrum, with a peak efficiency at the wavenumbers indicated by the dotted grey line (see text for details). In the second, we plot the corresponding energy density per wavenumber decade. In the third row, the growth in fluctuation energy is depicted, which becomes comparable to the background energy source at the end of preheating. As in Figure~\ref{fig:fermionicpreheating1}, colours are used to correlate spectral contours at the indicated times, while the dashed line in first two rows indicates a much later time point after preheating.   }
\end{center}
\end{figure*}

Our simulations (with benchmark values in Table \ref{tab:table1} of Appendix~\ref{sec:appA}) begin in the linear regime and initially confirm the picture we have outlined in Section \ref{linearanalysis}. There is an initial phase of \textit{preheating}, where the inflationary fluctuations in both $\sigma_R$ and $\sigma_I$ are amplified by parametric resonance due to an approximate periodicity in the inflaton background field; a process studied most informatively in Fourier-space (see Figure \ref{fig:spectrapreheating}). As expected, we find that the small dimensionful couplings and the feebly-coupled Higgs component furnish negligible corrections to a canonical preheating of the PQ scalar components. For the case of $\sigma_I$, the exponential growth of mode occupation numbers in the instability band $\kappa < \frac{1}{\sqrt{2}}$ is initially much faster than that of $\sigma_R$ due to the larger maximal Floquet index~\cite{Greene:1997fu}.\footnote{Note that in the long wavelength limit, $\mu_\kappa \rightarrow 0$, ensuring that in principle cosmologically-relevant axion isocurvature perturbations are not enhanced by parametric resonance~\cite{Bond:2009xx}.} However, mode-mode couplings soon become important, and we see a redistribution of the spectral power both within a more general phase-space and between the field components, swamping the weaker self-resonance in $\sigma_R$ and resulting in dominantly infrared spectra. As the fluctuations become highly non-linear, the back-reaction effects on the resonant production become increasingly severe, terminating the preheating at $\tau \sim 90$. Note that, as can be seen in Figure~\ref{fig:fermionicpreheating1}, preheating ends well-after the production of neutrinos through fermionic preheating, which constitute a miniscule fraction of the energy density at the time, so that the eventual disruption of the periodic background does not conflict with the analysis of Section~\ref{sec:neutrinos}.

\begin{figure*}
\begin{center}
\includegraphics[width = 0.5\textwidth]{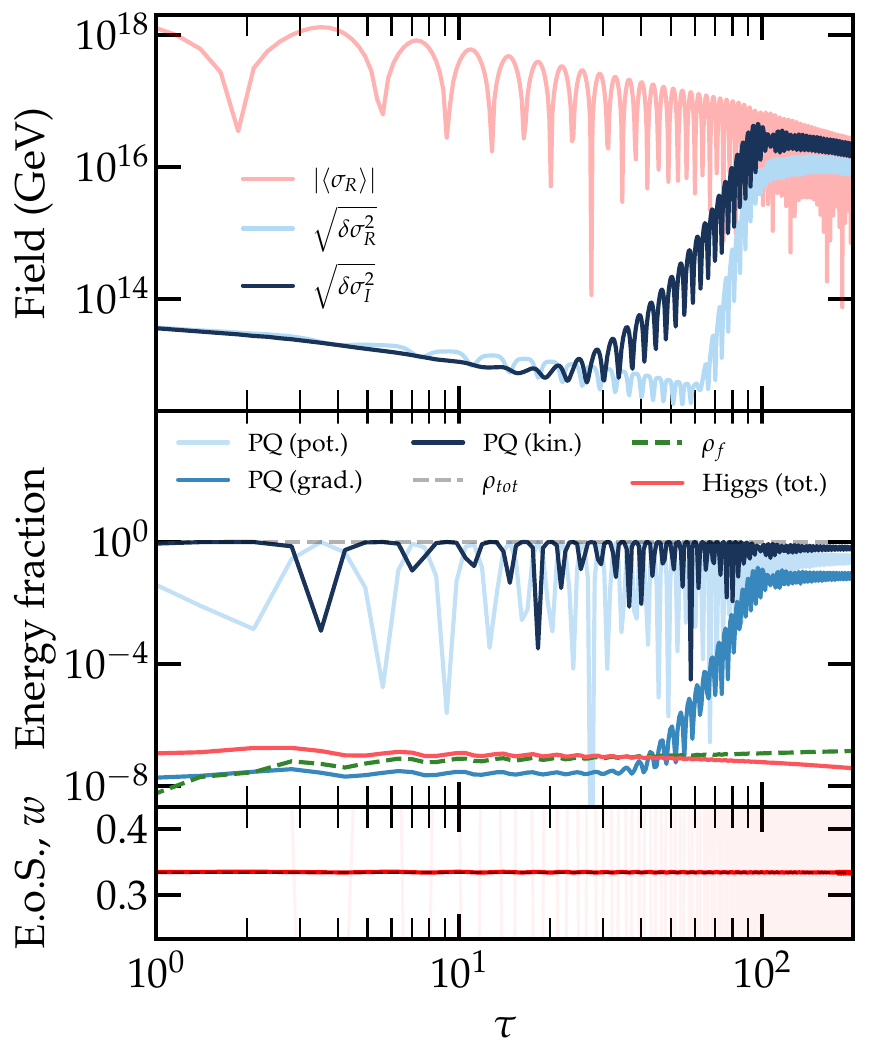}
\caption{\label{fig:varenpreheating} We depict the time evolution of the following quantities during preheating using (\ref{eq:sim1}): (\textit{top}) the inflaton zero mode, $|\langle \sigma_R\rangle|$, along side root-mean squared fluctuations of inflatons, $\sqrt{\delta\sigma^2_R}$, and axions, $\sqrt{\delta\sigma^2_I}$; (\textit{middle}) the energy fractions in potential, gradient and kinetic energy for the PQ sector, as well as the Higgs and radiation baths (into which the Higgs are quickly depleted); and (\textit{bottom}) the time-smoothed equation of state (solid), alongside the oscillating lattice output (semi-transparent), compared to radiation $w=\frac{1}{3}$ (dashed black). 
}
\end{center}
\end{figure*}

Considering now spatially-averaged coordinate-space quantities (see Figure \ref{fig:varenpreheating}), there is a corresponding growth in the variances of the PQ components, ultimately reducing the amplitude of the background field from which the energy in fluctuations has been drained. Consequently, there is a corresponding growth in gradient energy, with the radiation-like equation-of-state maintained by a slight reduction in potential energy. During this time, there is negligible growth in Higgs fluctuations, which are essentially decoupled from the inflaton, and the energy density stored in the post-inflationary Higgs components is rapidly transferred into a sub-dominant bath of SM radiation (with $T_{\text{max}} \sim H_{\text{inf}}$),
which spectates the remaining evolution until it is populated by late-time decays of $N_i$ at the end of reheating. 

In Figure \ref{fig:varenpreheating}, we also show that, during the preheating, the variance of the inflationary fluctuations may be seen to non-thermally restore a global minimum of the effective PQ-sector potential at the field-space origin for $f_a <  10^{13}$~GeV, 
extending into the post-inflationary era the well-known result that the de Sitter temperature during inflation restores $U(1)_{\text{PQ}}$ during inflation if $T_{GH} > f_a$, with $T_{GH} = \frac{H_{\text{inf}}}{2\pi}$~\cite{Gibbons1977}. Indeed, the growth in variance by parametric resonance further stimulates PQ restoration~\cite{Tkachev:1998dc}, as the non-thermal fluctuations, despite being excited with typical energy below a corresponding thermal distribution (if one assumes instantaneous reheating), induce larger than $T^2$ corrections to the effective potential~\cite{Kofman:1995fi,Kolb:1996jr,Rajantie:2000fd}.

Nonetheless, whenever the background PQ modulus field is displaced from the symmetric minimum, the PQ symmetry is instantaneously broken with an effective axion decay-constant parametrised by the background value (with important consequences for axion and neutrino thermalisation discussed in Sections \ref{sec:reh}). 

In comparison to parametric resonance, tachyonic instability, which can enhance the very long wavelength inflationary modes in the regime probed by CMB data 
was less efficient over much of the general range of $f_a$ values; being completely absent during preheating for $f_a \sim 10^{11}$ GeV. In fact, while for $f_a > 10^{13}$~GeV 
the oscillating inflaton can probe a tachyonic region during preheating, we find no appreciable tachyonic instability for $f_a \lesssim 3 \times 10^{16}$~GeV in our simulations of this initial epoch, 
before the field-variances are grown by parametric resonance to the larger values ultimately responsible for restoration of a PQ-symmetric minimum~\cite{Tkachev:1998dc}. 
This seems to indicate that, in order for the effect to become efficient, the background field amplitude must have red-shifted to become comparable to the low-energy vev, because then the time interval that the modes are tachyonic becomes a non-negligible fraction of the oscillation period. In other words, the tachyonic instability is expected to be most efficient during spontaneous symmetry breaking (if it is not instead driven by some other dynamics like Ricci scalar oscillations~\cite{Opferkuch:2019zbd}), which is the regime in which it is has been previously studied, see e.g. Ref.~\cite{Felder:2001kt}.
For $f_a \sim 10^{11}$ GeV, the symmetry-breaking regime is only reached $\sim 16$ scale factor expansion e-folds after inflation (well after preheating). On the other hand, as has been noted elsewhere~\cite{Kasuya:1998td,Tkachev:1998dc,Ballesteros:2016xejSMASH,Ballesteros:2021bee}, the non-thermal symmetry restoration of the effective potential due to parametric resonance fails for $f_a \gtrsim 10^{17}$ GeV because the displaced minimum successfully attracts the background field before  $\tau \sim 90$.
As a consequence, preheating following Higgs-like inflation with a PQ scalar induces a dangerous tachyonic instability only for the \textit{pre-inflationary} axion scenario (see Ref.~\cite{Ballesteros:2021bee}). This is not a problem for the post-inflationary axion regime~\cite{Companion}.

\subsubsection{Early non-linear evolution ($\tau < \tau_{PQ}$)}
\label{sec:latticefragturb}

\begin{figure*}
\begin{center}
\includegraphics[width = 0.52\textwidth]{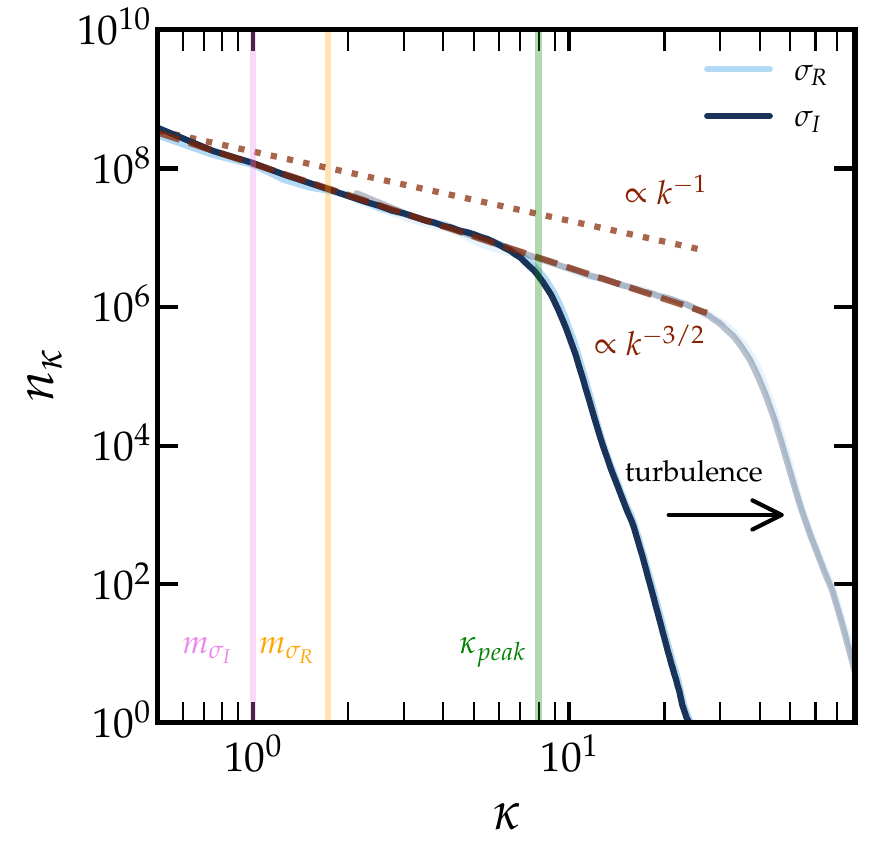}
\includegraphics[width = 0.45
\textwidth]{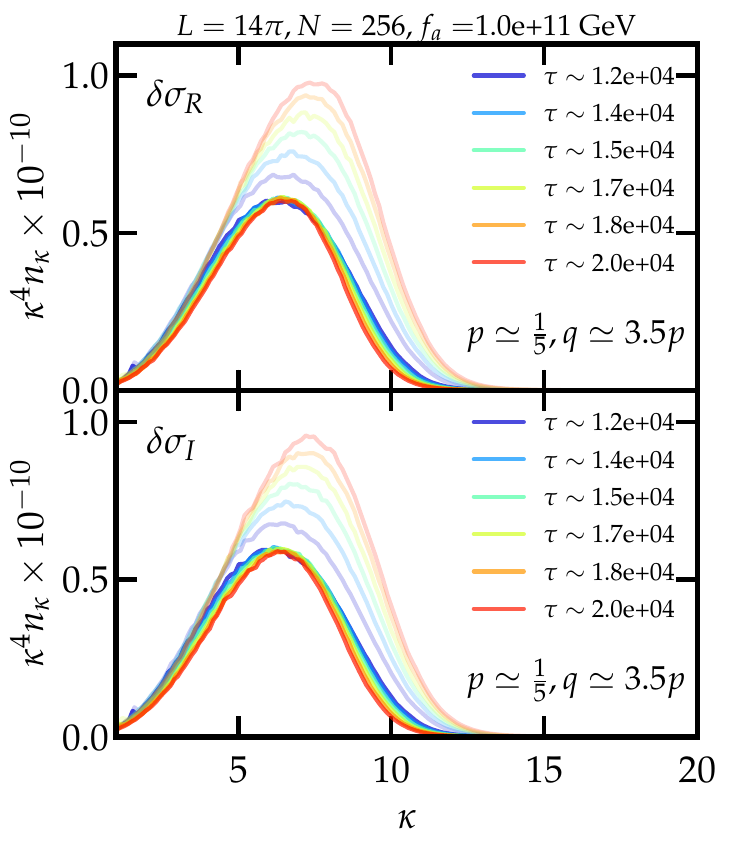}
\caption{\label{fig:nkevol1} In the \textit{left} panel, we plot the particle distribution function for the real and imaginary parts of the PQ field well into the turbulent era, $\tau \sim 5\times10^4$ with (\ref{eq:sim2}), along with several properties mentioned in the text: in red we demonstrate the $s = \frac{3}{2}$ tilt in comparison to the thermal value of $-1$; in green, yellow and pink we indicate landmark values for the wavenumbers including the peak location ($\kappa_{\text{peak}}$) of the energy distribution near the elbow; and we also depict an example self-similar evolution under turbulence with $p=\frac{1}{7}$. In the \textit{right} panel, we exemplify the turbulent scaling  in the domain of an example lattice simulation (transparent), using the energy per wavenumber decade to demonstrate the peak evolution. The opaque contours of corresponding colour correspond to scaling with the inverse of Eq.(\ref{eq:selfsimilar}).}
\end{center}
\end{figure*}

Once back-reaction has terminated the preheating in the non-linear regime, and for the prolonged period where the thermal friction on the inflaton oscillations from the SM radiation bath can be neglected, the evolution of particle distributions, energy densities, field averages and variances for the PQ scalar components must all be obtained empirically from lattice simulations. However, our integration time and spatial resolution were inevitably limited by our computational resources and we were not able to simulate the entire reheating period for our benchmark parameter choices, precisely due to the feeble coupling of the inflaton to the SM. Fortunately, the non-linear regime is also characterised by \textit{turbulence}, as explained in Refs.~\cite{Micha:2002ey,Micha:2004bv} whose analysis we follow and extend. The associated self-similar evolution of particle distributions, Eq. (\ref{eq:selfsimilar}), and consequent appearance of scaling laws for variances, Eq. (\ref{eq:turbvar}), as well as typical energies, then motivate extrapolations which we use in Section \ref{sec:reh} to analyse the final thermalisation process that ends reheating. 

During the initial stage of the non-linear era, \textit{fragmentation} of the relativistic inflaton condensate occurs~\cite{Felder:2006cc}, associated with a regime of ``driven turbulence'' in Ref.~\cite{Micha:2004bv}. 
When parametric resonance becomes inefficient, the oscillating inflaton zero-mode still comprises a sizeable fraction of the energy budget but is not yet in a dynamical equilibrium with the $k \neq 0$ modes. Indeed, the condensate energy continues to be siphoned into the particle fluctuations by, for example, effective 3-particle scatterings, \textit{i.e.} 2-2 mode scattering with a zero-momentum initial or final state~\cite{Kasuya:1998td}. 
As a result, the zero-mode oscillation amplitude is continually damped, along with the potential energy, which is transferred to the kinetic and gradient energy of the relativistic fluctuations, preserving $w\simeq\frac{1}{3}$. The individual contributions of the axion and inflaton quanta to the total energy then approach 50\%. 

During this time, we find that the particle distributions of the PQ components for the modes of interest, which were strongly non-thermal during the preheating, are organised by rescatterings into almost-thermal distributions. That is, there is an excess of power in the infrared modes, manifesting as a tilt in the spectrum away from the thermal expectation (e.g. $n_\kappa \propto \kappa^{-s}$ with $s > 1$), 
and a cutoff slightly larger than the peak of the energy distribution ($\kappa_{\text{peak}}$) representing insufficiency of power in the UV tail of the spectrum to truly be thermal (see Figure~\ref{fig:nkevol1}); the large-k modes remaining in the vacuum state until they are populated by collisions. As a consequence, the non-thermally averaged annihilation rates can be quite different to their thermal value.

\begin{figure*}[t]
\begin{center}
\includegraphics[width = 0.38\textwidth]{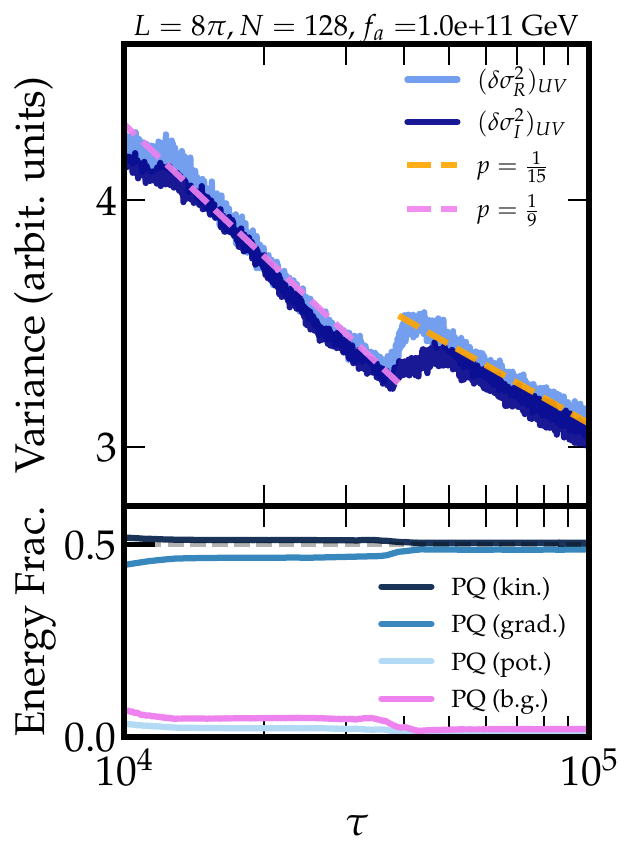}
\includegraphics[width = 0.45\textwidth]{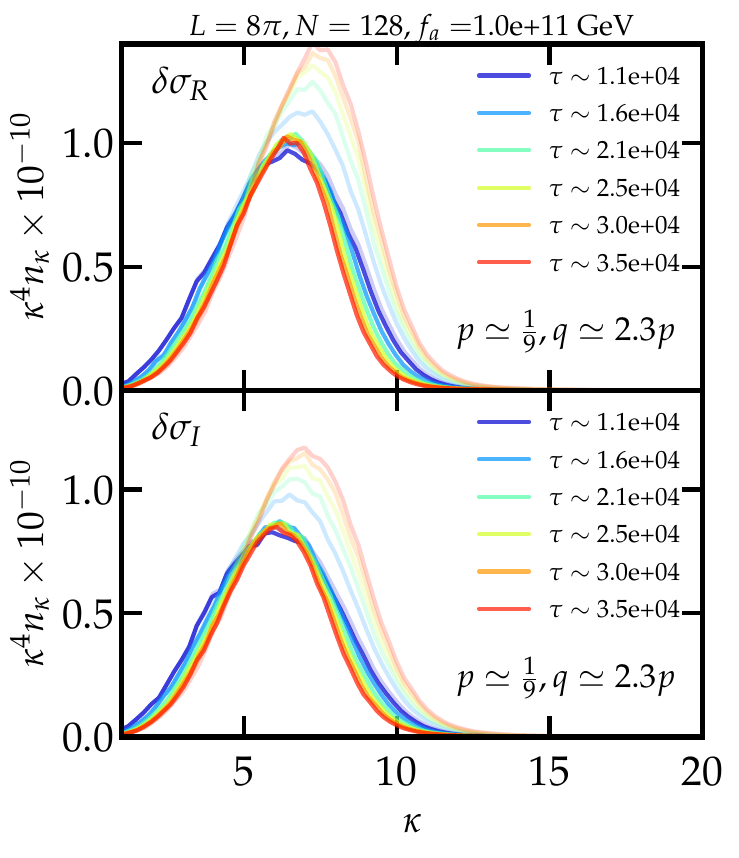}
\caption{\label{fig:nkevol2} In the top \textit{left} panel, we show the evolution of the variances at very late times, $\tau >10^4$ with (\ref{eq:sim3}), only integrating over the UV contributions ($\kappa > m_{R,I}$) to exemplify the relativistic scaling behaviour discussed in the text. In the bottom left panel, we show that the zero-mode energy (and with it the potential) energy has almost completely decayed away by $\tau\sim 10^4$. In the right panel, we repeat the procedure of Figure~\ref{fig:nkevol1}. }
\end{center}
\end{figure*}

Following Refs.~\cite{Micha:2002ey,Micha:2004bv},
the subsequent evolution after inflaton fragmentation may be quantified as follows. The particle distributions (after some reference time $\tau_*$) are assumed to follow a ``self-similar'' ansatz
\begin{equation}\label{eq:selfsimilar}
    n_{\kappa}(\tau) = \left(\frac{\tau}{\tau_*}\right)^{-q} n_{\tilde{\kappa}}(\tau_*),\quad \text{where}\quad \tilde{\kappa} = \left(\frac{\tau}{\tau_*}\right)^{-p}\kappa,
\end{equation}
where $q$ and $p$ are indices which may be theoretically predicted for certain approximations, or else inferred numerically (see Figures~\ref{fig:nkevol1} and~\ref{fig:nkevol2}). Hence, the typical energies $\langle \omega \rangle \sim \kappa_{\text{peak}}$ and, under some reasonable assumptions, the comoving field variances will also follow a power laws with index $\sim p$,
\begin{equation}\label{eq:turbvar}
    \langle \omega \rangle \propto \left(\frac{\tau}{\tau_*}\right)^{p}\qquad \text{and} \qquad \langle a^2 \delta\sigma^2_{R,I} \rangle \propto \left(\frac{\tau}{\tau_*}\right)^{-2p}.
\end{equation}
For our purposes, the exponents $p$, $q$ and $s$ are of primary importance to this epoch of the cosmological history for the following two reasons. First, we should interpret the red-shifting value of the inhomogeneous modulus field, which at later times is $f^{\text{eff}}_a \equiv \sqrt{\langle \sigma^2 \rangle}\propto \tau^{-(1+p)}$, as an effective vev responsible for time-dependent masses for the dark sector states and, hence, as an effective axion decay constant delaying thermalisation (see the discussion in Section \ref{sec:reh}); while the field variances, in so far as they are much larger than $v_S^2$, are responsible for the non-thermal symmetry restoration.\footnote{We consider it inappropriate to naively identify the time-dependent axion decay constant with the depleted zero-mode in the non-linear regime, and instead subsume the non-thermal corrections into the time-dependent effective value. } For example, in this way we can roughly predict the time at which the non-thermal restoration of the PQ symmetry fails (if thermalisation is sufficiently slow), resulting in an out-of-equilibrium phase transition, discussed below. Second, the values of $p$ and $q$ lead to extrapolations of the non-thermal particle distribution functions used in our annihilation rate estimates in Section \ref{sec:reh}. 

Eventually, the background energy density is only a small fraction of that stored in the fluctuations and plays a subdominant role in much of the subsequent evolution, which is assumed to approach a regime of ``free turbulence''. The dynamics are nonetheless initially dominated by 3-particle interactions for which it is expected that $p \simeq \frac{1}{5}$, $q\simeq4p$ and $s\simeq 3/2$~\cite{Micha:2004bv}. This is supported by our findings drawn from  simulations with parameters (\ref{eq:sim2}), as we summarise in the right panel of Figure~\ref{fig:nkevol1}.

As this initial regime is expected to completely decay the zero-mode, 4-particle collisions eventually dominate, for which it is expected that $p \simeq \frac{1}{7}$ in the relativistic regime~\cite{Micha:2004bv}. As a result, there is a very gradual approach to a thermal distribution, \textit{viz.} a transfer of power from the IR excess into the UV tail, which, for our purposes, does not terminate before interactions with fermions in the bath ultimately thermalise the hidden scalar components. To model this next stage, we halved the number of lattice sites per dimension, $N$, to feasibly increase the integration time by an order of magnitude for simulations with parameters (\ref{eq:sim3}). However, in doing so, we found that the depletion of the inflaton zero-mode becomes disrupted, resulting in fluctuating power laws for the variances. As summarised in Figure~\ref{fig:nkevol2}, while there is a brief period where $p$ is slightly below the expected value, the late time stationary behaviour was consistent with $p\simeq \frac{1}{15}$ (identified in Ref.~\cite{Micha:2002ey} as possibly spurious). 

There are two possible explanations for this discrepancy. On the one hand, it is possible that it is a finite-size effect stemming from poor IR capture, hence blocking the decay of the zero-mode. (Increasing the box size would compromise the UV capture necessary to keep the simulation accurate). On the other hand, this may be due to the excitation of a more infrared regime of axion fluctuations during preheating than self-resonance of the inflaton alone (which was considered in Ref.~\cite{Micha:2002ey}). Following rescattering, both components of the PQ scalar field then obtain a spectrum which is not wholly relativistic, so that the regime does not meet the criteria for $p=\frac{1}{7}$ within a feasible integration time. 

In either case, then, these findings do not conflict with the assumption $p\simeq\frac{1}{7}$ at sufficiently late times. For the purposes of extrapolation, see Appendix~\ref{sec:appB} and Figure~\ref{fig:nkextrap}, we assume that relativistic free turbulence with $p\simeq \frac{1}{7}$ is achieved for $\tau > 10^{5}$, the largest time we simulated, until the phase transition or thermalisation intervenes.\footnote{This value was also originally assumed in the analysis of the analogous S-Inflation scenario for the SMASH model to predict the onset of non-thermal phase transition ~\cite{Ballesteros:2016xejSMASH}, enabling a more standardised comparison.} The modelling of the IR contribution simply introduces a small theoretical error in the predicted thermalisation rates of Section \ref{sec:reh}. 

\subsubsection{Late non-linear evolution ($\tau > \tau_{\text{PQ}}$)}
\label{sec:latticept}

\begin{figure*}[t]
\begin{center}
\includegraphics[width = 0.75\textwidth]{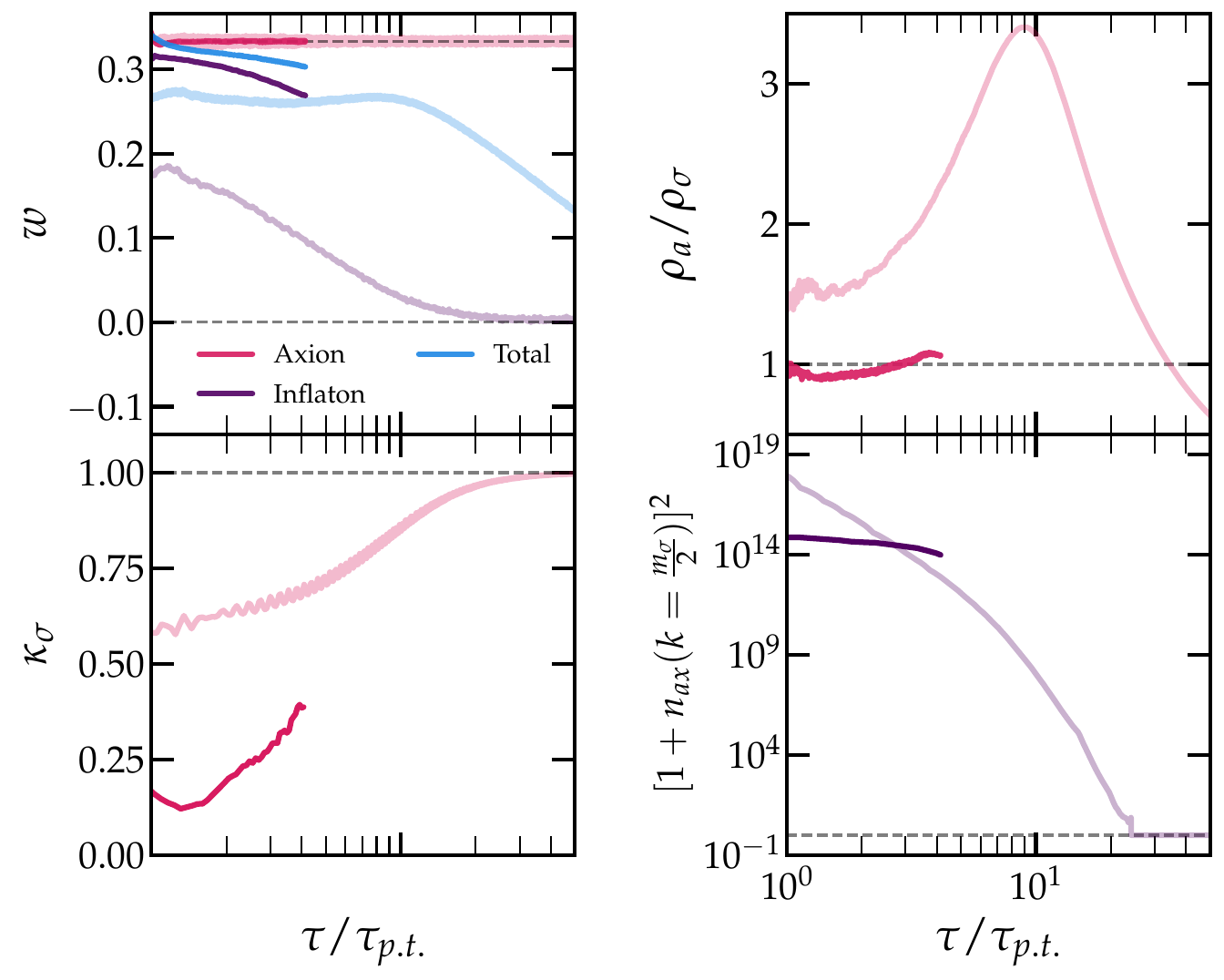}
\caption{\label{fig:afterpt} Plots from lattice simulations (\ref{eq:sim4}) and (\ref{eq:sim5}) of the non-thermal PQ phase transition (PT), rescaled with a reference time for the phase transition. For the transparent shading we use $f_a = 2\times10^{16}$~GeV yielding a longer integration time post-PT;
for comparison, in opaque, we use a smaller value $f_a = 5\times10^{14}$~GeV corresponding to a later PT (more appropriate for our regime).
In the \textit{top-left} panel, we plot the equation of state along with that of the axion and radial fractions; in the \textit{bottom-left} panel we plot the time dilation factor for $\sigma$ decays ($\kappa_\sigma$) defined in Eq. (\ref{eq:timedilinf}); in the \textit{top-right} panel we plot the ratio of the axion and modulus energy densities; while in the \textit{bottom-right} we estimate the enhancement factor of the modulus and axion interactions. }
\end{center}
\end{figure*}

If thermalisation has not yet occurred, a generic outcome for the VISH$\nu$ parameter space, then the field variances of the PQ scalar components are eventually reduced, by red-shifting and collisions in the non-linear turbulent regime discussed above, below a critical value $\sim f_a^2$.  For $f_a \ll 10^{14}$ GeV, so that the relativistic $p\simeq \frac{1}{7}$ scaling is expected to be reached after $\tau \sim 10^5$, this occurs at roughly:
\begin{equation}
    \tau_{PQ} \sim a_{PQ} \simeq   6.0 \times 10^6 \left( \frac{1 \times 10^{11}\ \text{GeV}} {f_a}\right)^{8/7}
\end{equation}
corresponding to $\sim 16$ e-folds after inflation has ended. Around this time, the non-thermal restoration of the symmetric minimum of the effective potential fails. The inflaton field, which, since the end of inflation, has been oscillating (in)homogeneously around an approximately quartic minimum at the field-space origin, instead becomes attracted and confined to the quadratic valley around the displaced $U(1)$ vacuum manifold, leading to field configurations of non-zero winding, and a massless axion.

\begin{figure*}[t]
\begin{center}
\includegraphics[width = 0.3\textwidth]{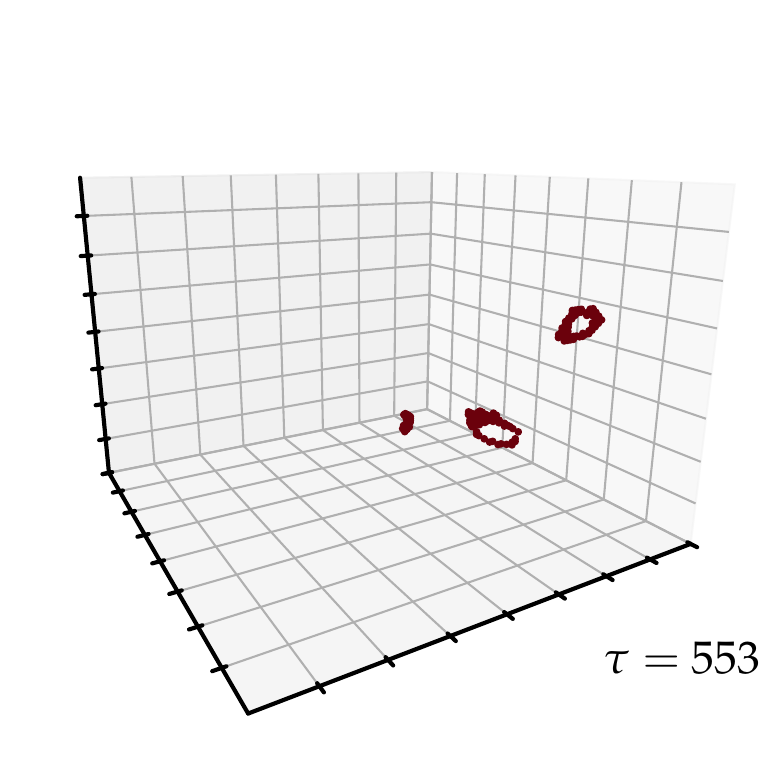}
\includegraphics[width = 0.3\textwidth]{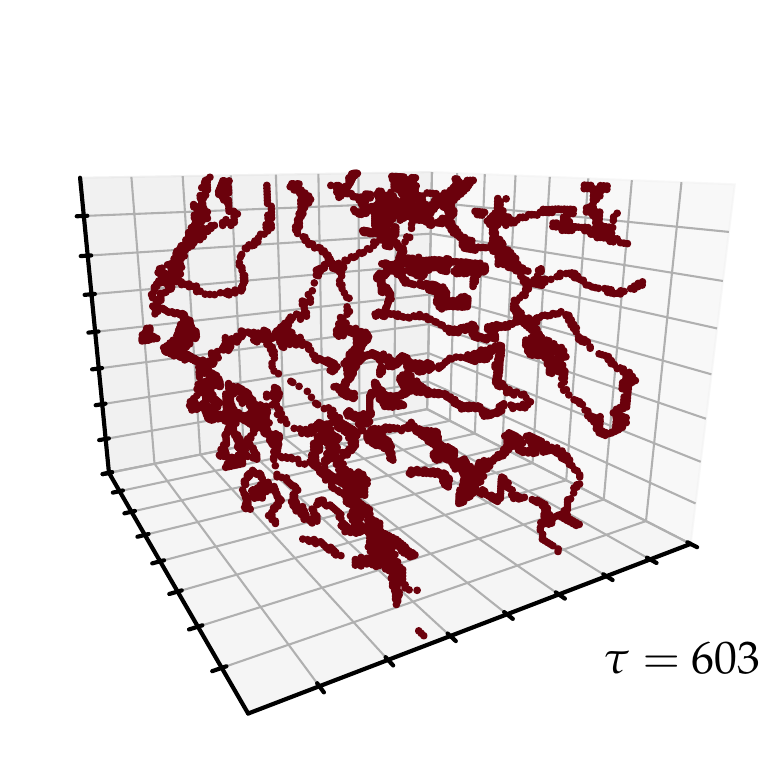}
\includegraphics[width = 0.3\textwidth]{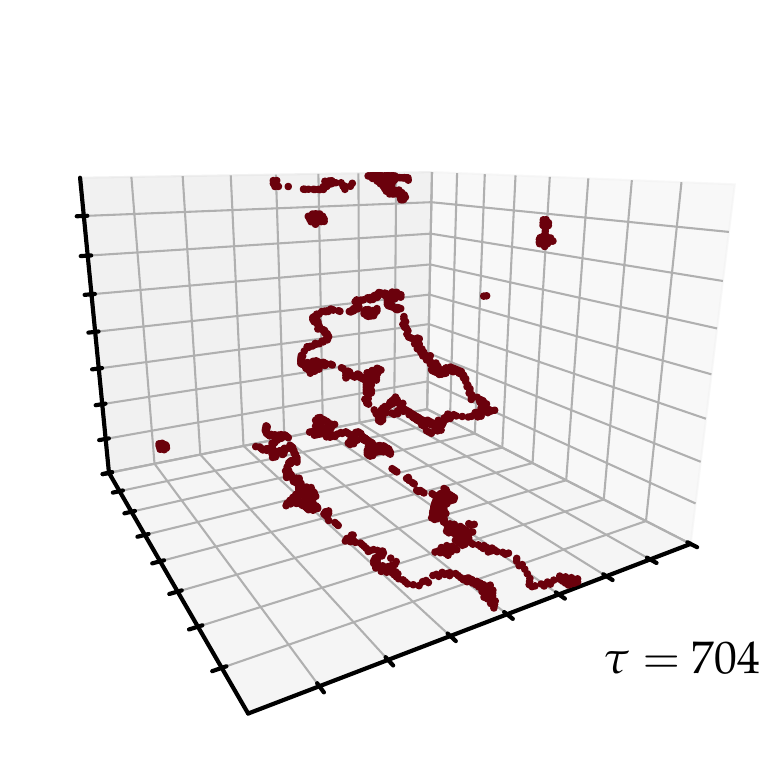}
\caption{\label{fig:strings1} We provide example visualisations of three-dimensional lattice slices from a simulation with (\ref{eq:sim6}), shaded regions correspond to lattice sites with $\sigma <0.1f_a $. In the left plot, we show a slice in the lead up to the phase transition with small-length string-loops, in the middle we show the dense string-network produced just after the phase transition, in the right plot we show a later configuration as strings begin to dissolve, along with a candidate ``infinite'' string loop.}
\end{center}
\end{figure*}

\begin{figure*}[t]
\begin{center}\includegraphics[width = 0.44\textwidth]{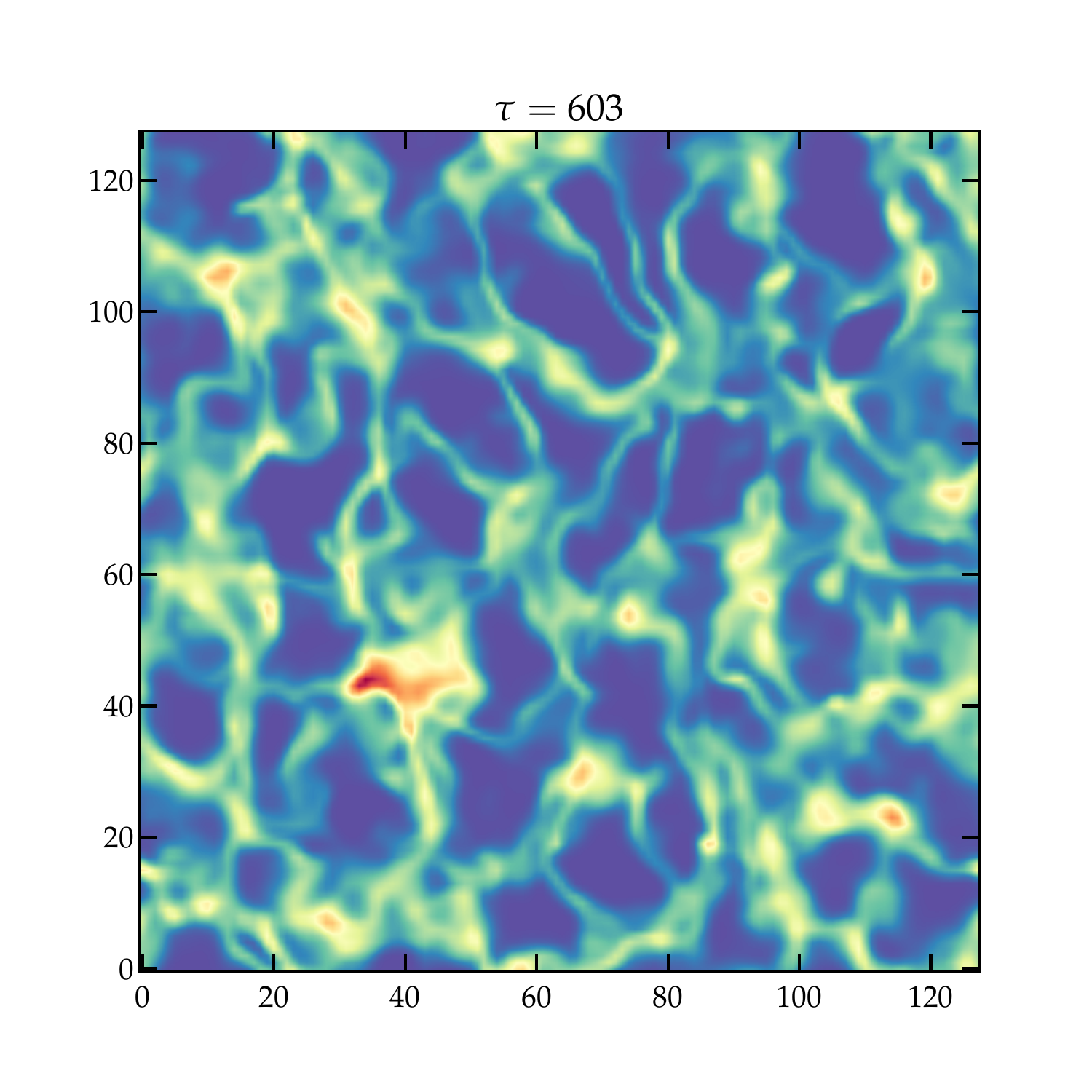}
\includegraphics[width = 0.46\textwidth]{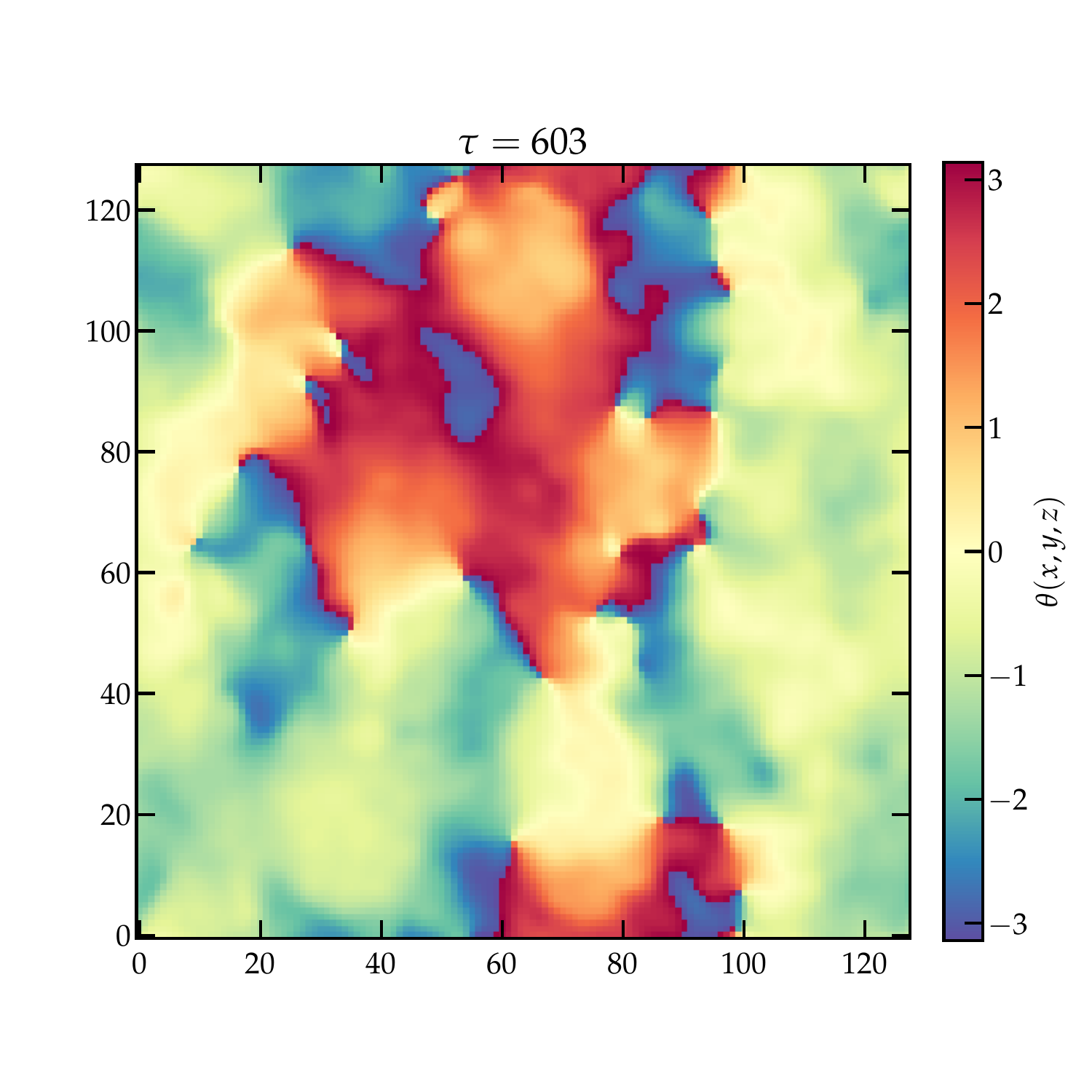}
\caption{\label{fig:strings2} We provide: on the left, a projection of the lattice box along the $z$-axis; on the right, a plot of the the phase field along an example cross-section of the lattice, demonstrating the winding configurations in $\theta$; both corresponding to the middle slice of Figure~\ref{fig:strings1} for simulation (\ref{eq:sim6}). For the left projection plot, a representative colour map has been chosen so that the  
blue shaded area is well-outside the string core, while green-red represents the cosmic strings, with red being closer to the core.}
\end{center}
\end{figure*}

As a consequence, as is shown in Figures~\ref{fig:strings1} and \ref{fig:strings2}, topological axion-majoron strings are then formed within the sub-Hubble lattice volume and organise into a dense network. (Small loops are periodically formed and dissolve in the oscillations leading up to $\tau_{PQ}$~\cite{Tkachev:1998dc,Fedderke:2025sic}, accelerating the energy transport from the residual zero-mode.) Subsequently, the string loops intercommute and are seen to dissolve into inflaton and axion fluctuations, representing additional non-thermal contributions to their energy spectra.    

We note that the correlation length has been reduced by the fragmentation process to a sub-Hubble length. This, in addition to the periodic formation of the strings, represents an interesting difference from the standard thermal Kibble-Zurek mechanism~\cite{Kibble:1976sj,Zurek:1985qw,Zurek:1993ek}, where the correlation length is limited only causally (for a second-order phase transition) resulting in infinite super-Hubble strings. Given the importance of the string contribution to the relic density of post-inflationary axion dark matter, it is then interesting to consider if this results in an important departure from the standard scenario, where the infinite string network, irrespective of initial configuration, is expected to follow an attractor solution, at late times, where the string length per Hubble patch is constant up to logarithmic corrections~\cite{Gorghetto:2018myk}. While we could not robustly confirm this in our simulations, and a more detailed study is beyond the scope of this paper (see also comments in Ref.~\cite{Tkachev:1998dc,Kasuya:1999hy,Lozanov:2019jff}); in our opinion, the standard results remain applicable to the present context, with percolation due to the large initial density of loops~\cite{Copeland:1998na,Conlon:2024uob,Srivastava:2024yxf} being an additional mechanism for super-Hubble string formation. (A candidate pair of long strings is depicted in the right panel of Figure~\ref{fig:strings1}.)

After $\tau \sim \tau_{PQ}$, the non-thermal effective mass corrections from the inflaton field $\propto \langle \delta\sigma^2\rangle$, which had been approximately red-shifting with the physical momenta of the field fluctuations and larger than the zero-temperature mass $\propto f_a^2$, are now reduced to a subdominant level, resulting in time-independent zero-temperature masses. As a consequence, so long as they remain non-thermal, the inflatons and $N_i$ become increasingly non-relativistic, lifting the decay rate suppression from time dilation, i.e. $\kappa_\sigma \rightarrow 1$ with Eq. (\ref{eq:timedilinf}), while the axions now constitute massless radiation until after the QCD phase transition. Eventually, if thermalisation is delayed, one expects the matter-like inflatons to dominate the energy density, so that $ w\rightarrow 0$. 

We were able to establish this picture at a quantitative level using our simulations of the phase transition (PT), summarised in Figure~\ref{fig:afterpt}. As our integration time was limited to $\tau \sim 10^4$, it was necessary to choose $f_a \gg 10^{11}$ GeV. To establish the late-time behaviour well-after the PT, we chose $f_a = 2 \times 10^{16}$ GeV (\ref{eq:sim4}), while to model the PT as close as possible to the late-time turbulent regime we used $f_a = 5 \times 10^{14}$ GeV (\ref{eq:sim5}). 

We extracted the (radial) inflaton and axion energy densities using a polar parametrisation of the PQ field $S=\frac{\delta\sigma+f_a}{\sqrt{2}}e^{ia/f_a}$, noting that the interaction energy is suppressed for $\langle\sigma\rangle \rightarrow f_a$. (Note that while the strings are present in the lattice volume, there is some small-scale contamination of these quantities, e.g. gradient energies and spectra, from string cores, which we did not mask.)  For the later PT, the radial inflatons were seen to have a more relativistic energy spectrum initially, having undergone more collisions in the turbulent regime, while the energy contributions remained $\sim 50\%$ each even in the presumed foreground of axion and inflaton production from the string network.\footnote{Hence we do not consider a separate source term in Section\ref{sec:reh} corresponding to strings.} This was not the case for the earlier PT, where the initially predominant axion contribution (more strongly excited by parametric resonance) was enhanced by a factor $\sim 3$ as the string network dissolved, resulting in a mildly relativistic equation of state for a few e-folds longer than might be expected. The late-time behaviour, however, was consistent with a transition to non-relativistic inflaton-domination. 

Note that the occupation numbers are initially $\sim 10^9$ in the phase-space Bose sphere populated after preheating, but after the phase transition, the mass-scale of the inflaton eventually exits this region of phase space due to red-shifting of the physical momenta. Hence, bosonic interactions may be considered to eventually re-enter a roughly perturbative regime, as scatterings and decays no longer experience Bose enhancement above kinematic thresholds. We can estimate this time by considering the usual stimulation factor for mono-energetic final states, with typical momenta $\sim m_\sigma$ (e.g. from  $\sigma\leftrightarrow aa$ or $\sigma\sigma \leftrightarrow aa$), \textit{viz.} $\sim [1+ n_{\text{ax}}(k=m_\sigma/2)]^2$, and the non-thermal distribution functions inferred from the lattice. As can be seen in Figure~\ref{fig:afterpt}, this enhancement is dramatically suppressed a few e-folds after the phase transition. Hence, we expect that the inflaton and axion fluctuations no longer efficiently re-scatter their momenta unless their perturbative interaction rates compete with the expansion rate. This helps to explain why the inflatons come to dominate the energy density for the simulation (\ref{eq:sim4}) instead of prematurely decaying to axions. 

Nonetheless, if the SM and $N_i$ energy densities remain suppressed, we expect that the short inflaton matter-domination ends once the inflatons perturbatively decay to axions. As $\lambda_S \ll 1$, this occurs well after the phase transition for the parameter space of interest, and was not accessible to our simulations. The end of reheating must then be studied perturbatively. However, as we have illustrated, non-perturbative effects must be taken into consideration, and we ensure that any downstream implications for the thermalisation process from the non-linear era are incorporated into our analysis in Section~\ref{sec:reh}, using extrapolations from the lattice simulations we have described above.

\subsection{Lattice-informed perturbative study}
\label{sec:reh}

It is clear from our non-perturbative analysis of the bosonic reheating that, without the inclusion of the right-handed neutrinos (RHNs) that couple the hidden scalar to the SM with sufficient strength, reheating can only be successful in the DFSZ model with the costs of either a radiatively \textit{un}stable electroweak scale, requiring $\lambda_{iS}, \kappa > \lambda_{S}$ for efficient annihilations and decays to thermal Higgs states; or possible unitarity violation, requiring $\xi \gg 1$ so that the inflaton can instead dominantly be a Higgs scalar).\footnote{Of course, there remains parameter space in this limit where the inflaton is again the scalar partner to the axion. We only require $\lambda_S \ll 1$ if $\xi \lesssim 1$.}

In fact, the decays of the $N_i$ to the SM realise at least three possible reheating scenarios for the model, which correspond to different regimes for the neutrino parameters $y_{N_i}$ (the diagonal and real entries of $y_N$) and $(y_\nu^\dagger y_\nu)_{ii}$ discussed below. In each case, the axions and inflatons generally do not thermalise, but can produce a freeze-in population of $N_i$, respectively through: inflaton decay to $N_1$, fermionic preheating and annihilations of non-thermal axions to $N_i$.  As the inflatons inevitably decay to hot axions which challenge stringent dark radiation constraints, the reheating is typically only observationally viable in the case that the $N_i$ are long-lived. This is because the $N_i$ become non-relativistic after production and hence suppress the energy fraction in axions at later times. The late $N_i$ decays then simultaneously produce a lepton asymmetry as they create the thermal bath to end reheating, which is sufficient for leptogenesis in a certain subset of the parameter region discussed below. We also find parameter space where standard thermal leptogenesis is a viable outcome. 

\subsubsection{The hidden neutrino portal}
\label{sec:neutrinos}

Let us first establish a relevant parameter region for the neutrino portal couplings: $(y_\nu)_{ij}$, between the RHNs and the SM, and $y_{N_i}$, between the inflatons, axions and RHNs. (We will work in a basis where $y_{N}$ is diagonal and real.) Explicitly, we consider naturalness, the lightest $\nu_L$ mass and CP asymmetry in $N$ decays. 

\paragraph{Naturalness} It is desirable that the neutrino Yukawa couplings are sufficiently small to ensure a radiatively stable electroweak scale, which is a well-known problem for many leptogenesis models~\cite{Vissani:1997ys}. In our context, the corresponding hidden-sector naturalness constraint is inferred from the 1-loop beta function for $\lambda_{2S}$ (see Refs.~\cite{nu2HDM,nuDFSZ,Sopov:2022bog}),
\begin{equation}\label{eq:naturalyukn}
     \sum_i y^2_{N_i} (y^\dagger_{\nu} y_\nu)_{ii} \lesssim \frac{ 4 \pi^2m^2_{22}}{f_a^2}\simeq 8.9 \times 10^{-15}\ \left(\frac{m_{22}}{\ 1.5\ \text{TeV} } \right)^2 \left(\frac{1 \times 10^{11}\ \text{GeV}}{f_a}\right)^{2},
\end{equation}
where $m_{22}^2 = M_{22}^2 + \lambda_{2S}\langle S \rangle^2$. 
Using the Casas-Ibarra parametrisation~\cite{Casas:2001sr} for $y_\nu$, namely
\begin{equation}
    y_\nu = \frac{\sqrt{2}}{v_2} U^\dagger \mathcal{D}_{\sqrt{m_\nu}} R \mathcal{D}_{\sqrt{m_N}}, 
\end{equation}
where $\mathcal{D}_{\lambda} \equiv \text{diag}\{\lambda_1, \lambda_2, \lambda_3\}$ is a diagonal matrix of eigenvalues, $U$ is a unitary matrix which diagonalises the mostly-active neutrino mass matrix $m_\nu \equiv v_2^2 y_\nu \mathcal{D}_{m_{N_i}^{-1}}y_\nu^T$ (so that $\mathcal{D}_{m_{\nu}} \equiv Um_{\nu} U^T$) and $R$ is an complex orthogonal matrix, we have that
\begin{equation}\label{eq:yvyvexp}
    (y_\nu^\dagger y_\nu)_{ii} = \frac{2m_{N_i}}{v_2^2}\sum_j  m_{\nu_j} |R_{ji}|^2,
\end{equation}
where the dependence on $U$ has dropped out. Consequently, Eq. (\ref{eq:naturalyukn}) may be translated into an upper bound for $m_{N_i}$ (and hence the $y_{N_i}$)
\begin{equation}\label{eq:bound1mn}
        m_{N_i} \lesssim \left[ \frac{4\pi^2m^2_{22}v_2^2}{\sum_j m_{\nu_j}|R_{ji}|^2 } \right]^{1/3}.
\end{equation}
While the elements of $R$ are not known, it was argued in Ref.~\cite{Clarke:2015gwa}, that one can still deduce generic upper bounds implied by Eq. (\ref{eq:bound1mn}). In the present two-doublet context these become
\begin{equation}\label{eq:bound2mn}
\begin{split}
    m_{N_1} &\lesssim 4\times 10^7\ \text{GeV} \ \left( \frac{v_2}{v} \right)^{2/3} \left(\frac{m_{22}}{\text{TeV}} \right)^{2/3} \\
    m_{N_2} &\lesssim 7\times 10^7\ \text{GeV} \ \left( \frac{v_2}{v} \right)^{2/3} \left(\frac{m_{22}}{\text{TeV}} \right)^{2/3}
    ,\\
    m_{N_3} &\lesssim 2\times 10^8\ \text{GeV} \ \left( \frac{10^{-3}\ \text{eV}}{m_{\text{min}}}\right)^{1/3} \left( \frac{v_2}{v} \right)^{2/3} \left(\frac{m_{22}}{\text{TeV}} \right)^{2/3}
\end{split}
\end{equation}
where $m_{\text{min}}$ is the lightest $\nu_L$ mass. The mass of the heaviest RHN can thus be considerably larger due to its symmetry-protected decoupling limit. In Ref.~\cite{Sopov:2022bog}, it was additionally argued that the radiative stability of the small inflaton self-coupling in VISH$\nu$ itself implies
\begin{equation}\label{eq:natyn3}
    y_{N_i} < 
    1.4\times10^{-2}\left(\frac{\lambda_S}{2.5\times10^{-10}}\right)^{1/4},
\end{equation} 
which is the stricter naturalness bound on $m_{N_3}$ in lieu of a measured $m_{\text{min}}$.

Our naturalness criteria hence require a (symmetry-protected) suppression of the portal-coupling between the PQ sector to neutrinos ($y_N$), with the weakest bound applying to $y_{N_3}$.  

\paragraph{Lightest $\nu_L$ mass} For our reheating scenario, it will be important that some $N_i$ is long-lived, in the sense that their decay rate, and hence Eq. (\ref{eq:yvyvexp}), are highly suppressed.  However the effective neutrino masses which enter into Eq. (\ref{eq:yvyvexp}) are bounded below by the orthogonality of $R$~\cite{Davidson:2002qv},
\begin{equation}
    \widetilde{m}_i = \sum_j m_j |R_{ji}|^2 \geq m_{\text{min}}.
\end{equation}
Hence the requirement for a highly suppressed $N_i$ decay can be translated to a bound on the lightest $\nu_L$ mass, typically requiring $m_{\text{min}} \ll 10^{-3}$\ eV. 

Let us explain how this can remain consistent with claims in the naturalness discussion above. For example, it may seem, given Eq. (\ref{eq:bound1mn}), that $\widetilde{m}_{1} \geq m_{\text{min}}$ would imply a much larger upper bound for $m_{N_1}$ than given in Eq. (\ref{eq:bound2mn}). However, $m_{N_1}$ is the lightest RHN by assumption, so this is not exactly the case. To see this, consider a candidate $R$ matrix, given in Ref.~\cite{DiBari:2005st}, which realises maximum CP violation, see Eq. (\ref{eq:davidsonibarra}) in the following, with $R_{31}$  small and complex
\begin{equation}
    R = \begin{pmatrix} 
    \sqrt{1 - R^2_{31}} & 0 & -R_{31} \\
    0 & 1 & 0 \\
    R_{31} & 0 & \sqrt{1 - R^2_{31}}
     \end{pmatrix}.
\end{equation}
With normal ordering, it follows that $\widetilde{m}_1 \sim m_{\text{min}}$. However, this form for $R$ gives $\widetilde{m}_3 \sim m_{\text{atm}}$, so that Eq. (\ref{eq:bound1mn}) implies $m_{N_3} < 4 \times 10^7$ GeV, but $m_{N_1} < m_{N_3}$. It follows that $m_{N_1}$ can be bounded by naturalness considerations in a given model, even for small $(y^\dagger_\nu y_\nu)_{11}$, due to the additional RHN generations.

\paragraph{CP-violation in $N_i$ decays}
It can be shown that the CP asymmetry parameter for $N_i$ decays is given, in the real and diagonal basis for $y_N$, by~\cite{Davidson:2002qv}
\begin{equation}
    \epsilon_i = -  \frac{1}{8\pi} \sum_{j\neq i} \frac{\text{Im}[(y_\nu y_\nu^\dagger)^2_{ij}]}{(y_\nu y_\nu^\dagger)_{ii}} \left[ f \left(\frac{M^2_j}{M^2_i}\right) + g\left(\frac{M^2_j}{M^2_i} \right) \right].
\end{equation}
Here, the contributions from interference with the 1-loop vertex and self-energy corrections are respectively~\cite{Covi:1996wh}
\begin{equation}
    f(x) = \sqrt{x} \left[1 - (1+x)\log \frac{1+x}{x} \right], \qquad g(x) = \frac{\sqrt{x}}{1-x}.
\end{equation}
The asymmetry may thus depend on the generation of $N$ and whether assumptions are made about the mass spectrum, which we will take to be hierarchical.
Their sum has two limiting values
\begin{equation}
    f(x)+g(x) \simeq 
    \begin{cases}
    \qquad \ - \frac{3}{2\sqrt{x}}&, \quad x\gg 1 \\
    - \sqrt{x} \left[ \log \left( \frac{1}{x} \right)- 2\right]&, \quad x\ll 1.
    \end{cases}
\end{equation}
For the case of $N_1$ (with $m_{N_1} \ll M_{2,3}$), which is applicable to the thermal leptogenesis scenario and most commonly assumed, one can obtain the well-known Davidson-Ibarra bound~\cite{Davidson:2002qv}
\begin{equation}\label{eq:davidsonibarra}
\begin{split}
    \epsilon_1 &\simeq -\frac{3}{16\pi} \frac{m_{N_1}}{v_2^2} \frac{\sum_j m_j^2\ \text{Im}(R^2_{1j})}{\sum_j m_j |R_{1j}|^2} \lesssim \frac{3}{16\pi} \frac{m_{N_1} m _{\text{atm}}}{v_2^2}.
\end{split}
\end{equation}
An analogous expression follows for $N_2$ in the case of normal-ordering~\cite{DiBari:2005st}. 

As it is easy to realise a scenario in VISH$\nu$ where there is a long-lived population of very heavy $N_i$, it is interesting to examine $N_3$ in the context of the naturalness bounds (\ref{eq:bound2mn}). In order to have $m_{N_3} \gg 10^7$ GeV satisfying (\ref{eq:naturalyukn}), it was assumed that the $R$ matrix satisfies a decoupling limit~\cite{Clarke:2015gwa}, i.e. $|R_{23}|, |R_{33}| \ll 1 $. When this is the case, we find that
\begin{equation}\label{eq:asymmN3}
\begin{split}
    \epsilon_3 &\simeq \frac{1}{16\pi} \frac{m_{\text{min}}}{m_{N_3}v_2^2}\sum_{j=1,2}M_j^2\left[\log \left(\frac{m_{N_3}}{M_j}\right) -2 \right]\frac{\text{Im}[(R^*_{13}R_{1j})^2]}{|R_{13}|^2}\\
    &\sim 0.1 \frac{m_{\text{min}}}{m_{N_3}v_2^2}\sum_{j=1,2}M_j^2 \text{Im}[R_{1j}^2].
\end{split}
\end{equation}
Accordingly, for a large $m_{N_3}$ mass that does not destabilise the electroweak scale, the CP asymmetry is not automatically larger, but can be $m_{\text{min}}$ suppressed.

We will demand that the CP symmetry realised in $N_i$ decays is sufficient for leptogenesis (Section \ref{sec:leptogenesis}). Our implementation of \textit{natural} leptogenesis hence concerns scenarios where the associated lepton asymmetry is generated with $N$'s of mass $< 10^8$ GeV, which is a non-trivial restriction. Of course, the assumption of strict naturalness can be moderately relaxed to allow for more asymmetry, or else a more degenerate neutrino mass spectrum may be considered (as in Ref.~\cite{Ballesteros:2016xejSMASH}). 

\subsubsection{Boltzmann equations} 

We study the thermalisation process at the end of reheating by solving a set of integrated Boltzmann equations (\ref{eq:boltz}) for the energy densities of the relevant baths: inflatons ($\rho_\sigma$), axions ($\rho_a$), SM + Higgs thermal bath ($\rho_R$), the RHNs $\rho_{N_1}$ kinematically accessible from inflaton decay, and their typically much heavier counterparts $N_{2,3}$, 
\begin{equation}\label{eq:boltz}
\begin{split}
    \frac{\mathrm{d}\rho_{\sigma}}{\mathrm{d}N} + 3(1+w_\sigma)\rho_\sigma &\simeq - \frac{\kappa_\sigma(\Gamma_{\sigma\rightarrow N_1 N_1} + \Gamma_{\sigma\rightarrow aa}) + \sum_i\left(\Gamma_{\sigma\sigma \rightarrow N_{i} N_{i}} + \Gamma_{\sigma a\rightarrow N_{i}N_{i}}\right)}{H}\rho_\sigma \\
    &\quad  \ - \frac{\Gamma^{\text{en}}_{\sigma\sigma\rightarrow aa}}{H}\left(\rho_\sigma-\rho_a\right),  
    \\
    \frac{\mathrm{d}\rho_{a}}{\mathrm{d}N} + 4\rho_a  &\simeq - \frac{\sum_i( \Gamma_{aa\rightarrow N_iN_i} + \Gamma_{a\sigma \rightarrow N_i N_i})}{H} \rho_a + \frac{\kappa_\sigma\Gamma_{\sigma\rightarrow aa}}{H}\rho_\sigma + \frac{\Gamma^{\text{en}}_{\sigma\sigma\rightarrow aa}}{H}\left(\rho_\sigma-\rho_a\right), 
    \\
    \frac{\mathrm{d}\rho_{R}}{\mathrm{d}N} + 4\rho_R &\simeq  \frac{\sum_i \kappa_{N_i}\Gamma_{N_i\rightarrow \Phi_2\ell_i}}{H}\left(\rho_{N_i} -\rho^{\text{th}}_{N_i} \right), 
    \\
    \frac{\mathrm{d}\rho_{{N_{i}}}}{\mathrm{d}N} + 3(1+w_{N_{i}})\rho_{N_{i}} &\simeq  \frac{\kappa_\sigma \Gamma_{\sigma \rightarrow N_1N_1} + \Gamma_{\sigma \sigma \rightarrow N_i N_i } + \Gamma_{\sigma a\rightarrow N_i N_i} }{H} \rho_\sigma  + \frac{\Gamma_{a a \rightarrow N_i N_i } + \Gamma_{a \sigma\rightarrow N_i N_i}}{H} \rho_a
    \\
    &\quad  \  - \frac{\kappa_{N_i}\Gamma_{N_i\rightarrow \Phi_2\ell_i}}{H}\left(\rho_{N_i} -\rho^{\text{th}}_{N_i} \right), 
    \\
    H^2 &= \frac{1}{3m_P^2} \left( \rho_\sigma + \rho_a + \rho_R + \sum_{i} \rho_{N_i} \right),
\end{split}
\end{equation}
which are coupled by relevant interaction rates in the collision terms. We also include the Friedmann equation in order to self-consistently model the changing equations of state between radiation and matter-dominated periods. The interaction and decay rates $\Gamma$, the time dilation factors $\kappa$ and the equations of state $w$ are explained below and given in Appendix~\ref{sec:appB}. We use as initial conditions for the subsequent bath evolution, the final values for the energies of the inflatons, axions, together with the assembled thermalised Higgs fluctuations and radiation energies, from the late-time simulation of the turbulent era (\ref{eq:sim3}). The initial conditions for the heavier $N_{2,3}$ baths, which in a certain subset of parameter space can be non-negligible, are inferred from a numerical study of fermionic preheating (see Figures~\ref{fig:fermionicpreheating1}, \ref{fig:fermpreheating} and the discussion in Section~\ref{sec:nureheating}).

Our analysis also assumes several extrapolations of results which we determined empirically from the lattice simulations. These have been described and motivated above in (Section \ref{sec:lattice}). In particular, we extrapolate the non-thermal distribution functions for the inflatons and axions using the assumption of turbulent scaling (\ref{eq:selfsimilar}) with $p =\frac{1}{7}$ and $q \simeq 1.2p$ (Figure~\ref{fig:nkextrap}). We assume that these formulae apply until the particles relax into a thermal distribution due to efficient interactions with the bath, or until the inflaton perturbatively decays into axions or $N_1$ to yield an approximately mono-energetic spectrum in either case (depleting the inflaton number density). Additionally, we model the time-dependence of the effective inflaton, axion and neutrino masses (\ref{eq:effmass}) before the phase transition, and the moment where this occurs, using the turbulent scaling of the two-point function $\langle \sigma^2 \rangle \simeq (f_a^{\text{eff}})^2$ discussed above.

The distribution functions $f_\sigma, f_{\text{ax}}$ are then directly integrated over to furnish non-thermally averaged cross-sections for axion and inflaton (co)annihilations (\ref{eq:sigmaann}) into and absorptions (\ref{eq:qcdabsorption}) by the thermal bath, as well as an equation of state $w_\sigma$ and time dilation factor $\kappa_\sigma$ for the inflaton decays; motivating non-thermal values for the RHN parameters $w_{N_i}$ and $\kappa_{N_i}$. They are also used to model the Bose enhancement of axion and inflaton interactions in the densely populated Bose sphere. 

We include in the collision terms all relevant decay rates $\Gamma_{1\rightarrow 23}$ (\ref{eq:decayrates}). For computational expedience, we include only the leading number-changing 2-2 processes when they are relevant to the outcomes of the reheating process. As we discuss below, we find that the subset of these which could, in principle, thermalise the axions and inflaton do not become efficient in general. Nevertheless, we find a certain region of parameter space where annihilations can result in a sizeable, yet sub-dominant population of $N_{i}$'s after the phase transition through a process of non-thermal freeze-in. We also assume, for the purposes of calculating interaction rates, the instantaneous thermalisation of SM final states, and any $N_i$ once $\Gamma_{N_i \rightarrow \Phi_2 \ell} > H$ with $T >N_{i}$. Production of inflatons (and axions) through inverse decays (or annihilations) of $NN \rightarrow \sigma$ and $aa \rightarrow \sigma$ is almost always negligible unless the latter is Bose enhanced. We therefore do not include the former due to the low reheating temperatures considered, while the latter is effectively modelled through $\Gamma^\mathrm{eff}_{\sigma\sigma \rightarrow aa}$ which resembles a step function.\footnote{In the end, including this rate did not modify our results in an important way, so a more precise integration over final state distribution functions is not needed.} 

\subsubsection{Scenarios for viable reheating and leptogenesis} 

\begin{figure*}[t]
\begin{center}
\includegraphics[width = 0.51\textwidth]{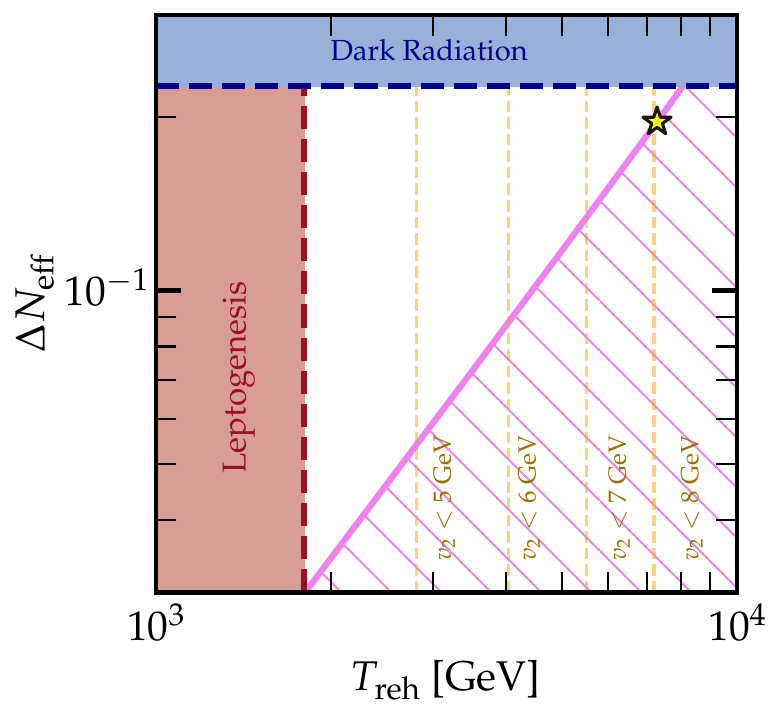}
\includegraphics[width = 0.3\textwidth]{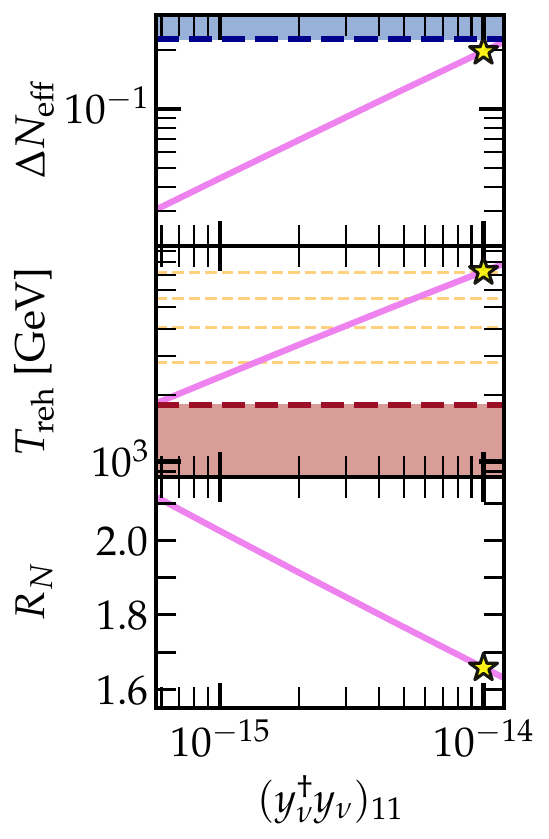}
\caption{\label{fig:N1fromdecay} In this plot we characterise, for $\xi_S = 0.5$, the available parameter space for the first scenario where reheating and leptogenesis is performed by long-lived $N_1$ inflaton decay products. This corresponds to the unshaded, un-hatched region of the left panel (see text for details). The red leptogenesis bound is illustrated for $v_2 = 4$ GeV, while its location for larger $v_2$ GeV is indicated by the labelled orange dashed lines. In the right panel, we illustrate the dependence of $\Delta N_{\text{eff}}$, $T_{\text{reh}}$ and the reduced number of inflationary e-folds $R_N$ (\ref{eq:matterefolds})  on  $(y_\nu^\dagger y_\nu)_{11}$ along the violet border of the first plot. The starred point is assumed in Figures~\ref{fig:bathplots} and  \ref{fig:fig:vishnugwomega}. }
\end{center}
\end{figure*}

\begin{figure*}[t]
\begin{center}
\includegraphics[width = 0.33\textwidth]{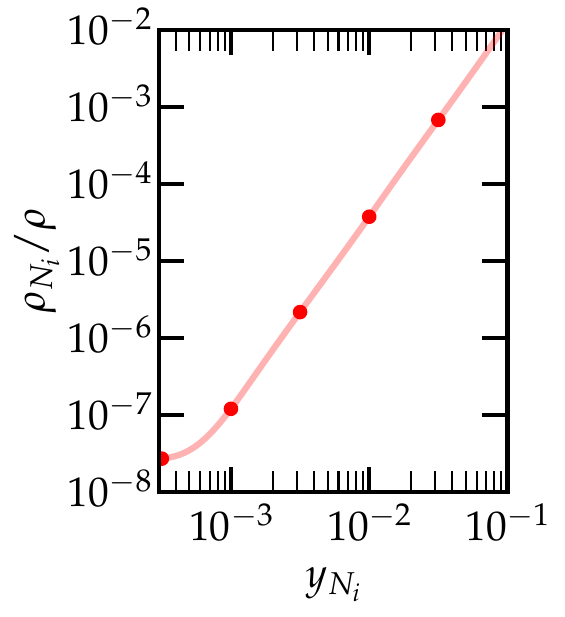}
\includegraphics[width = 0.39\textwidth]{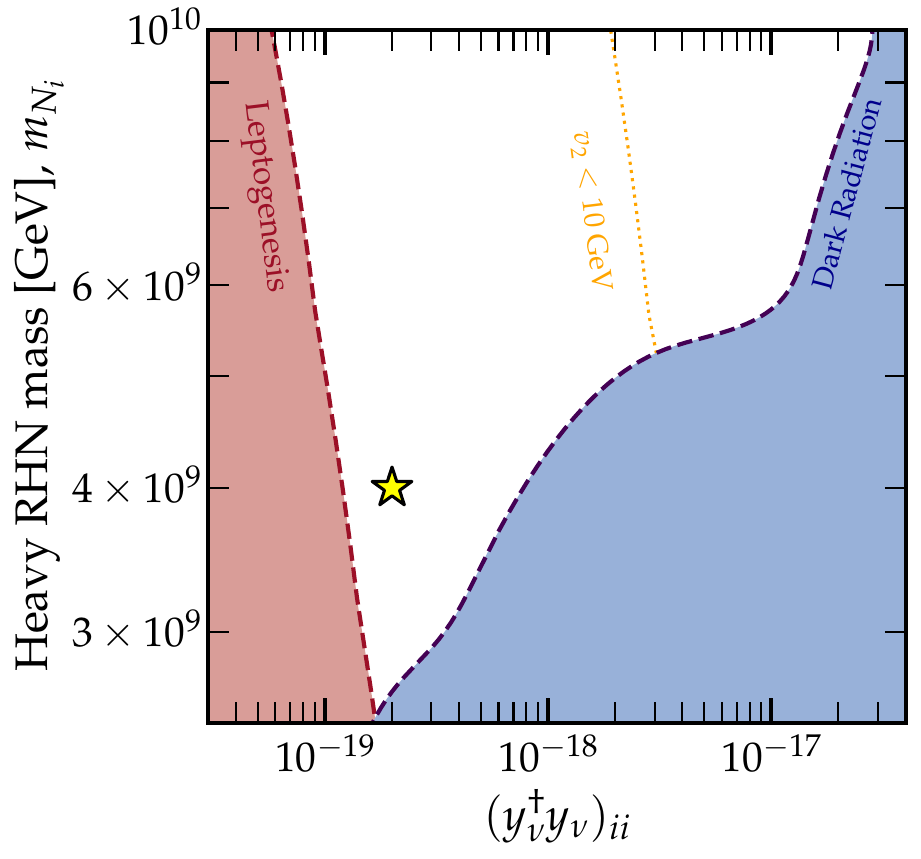}
\includegraphics[width = 0.26\textwidth]{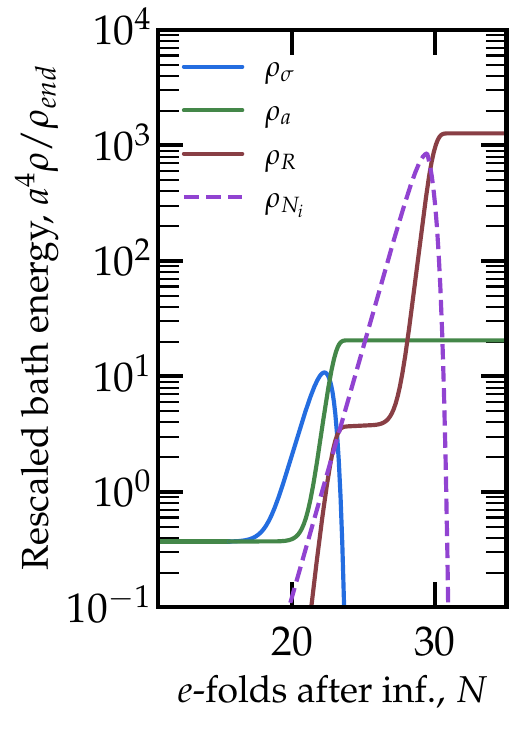}
\caption{\label{fig:fermpreheating}  In the left plot estimate the dependence, on the Yukawa coupling to the inflaton, of the initial relic neutrino bath energy density produced in fermionic preheating on the Yukawa coupling to the inflaton (as a fraction of the total energy). The energy fraction is approximately preserved during the initial radiation dominated era for stable $N_i$.  In the centre, we plot the available parameter space for a fermionic preheating scenario in which there is a long-lived heavy neutrino, $N_i$. This was argued to be unnatural in the text. The orange dashed contour illustrates the non-thermal leptogenesis bound for $v_2 <10$ GeV. On the right, we illustrate the bath evolution for the starred point on the central plot. }
\end{center}
\end{figure*}

We now detail the three scenarios for viable reheating and leptogenesis we have identified in VISH$\nu$, summarising parameter regions, for benchmark choices, in Figures~\ref{fig:N1fromdecay}, \ref{fig:N3fromann} and~\ref{fig:fermpreheating}. Of these, only the first two are found to respect that naturalness criteria we have explained above. The leptogenesis and dark radiation bounds discussed in this section are explained in Sections \ref{sec:leptogenesis} and \ref{sec:darkrad}, respectively.

\label{sec:nureheating}

\paragraph{Long-lived $N_1$ from inflaton decay (\textit{$N_1$-from-decay})}
In the simplest of the three reheating scenarios, we assume a general parameter space 
where $N_2, N_3$ production is negligible, but
the inflatons can decay via $\sigma \rightarrow N_1 N_1$. As this requires $y_{N_1} < \sqrt{\lambda_S/2}$ it follows automatically that
\begin{equation}
    \frac{\Gamma_{\sigma \rightarrow N_1 N_1}}{\Gamma_{\sigma \rightarrow aa}} \simeq \left( \frac{y_{N_1}}{\sqrt{\lambda_S/2}}\right)^3 < 1
\end{equation}
so the universe becomes initially axion-dominated.\footnote{One could in principle consider a parameter point where the kinematically allowed decays to all three generations overcome this bound, but this is difficult to achieve in general.} Hence, if $ \Gamma_{N_1 \rightarrow \Phi_2 \ell} \gtrsim \Gamma_{\sigma\rightarrow N_1 N_1} $
the reheating mechanism fails. This is because the $N_1$ decays realise only a sub-dominant thermal bath which is decoupled from the hot axions, which simply red-shift.

\begin{figure*}[t]
\begin{center}
\includegraphics[width = 0.4\textwidth]{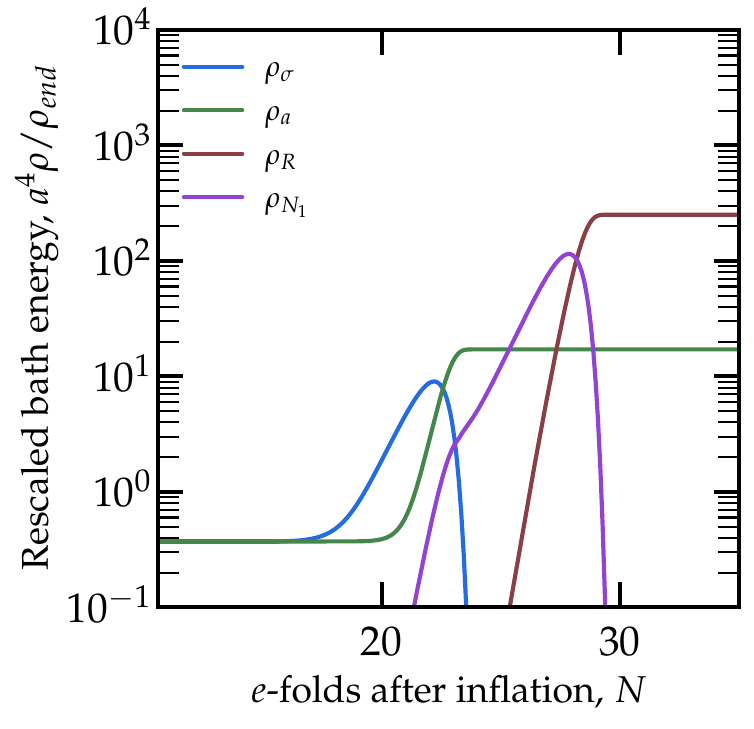}  \qquad \includegraphics[width = 0.4\textwidth]{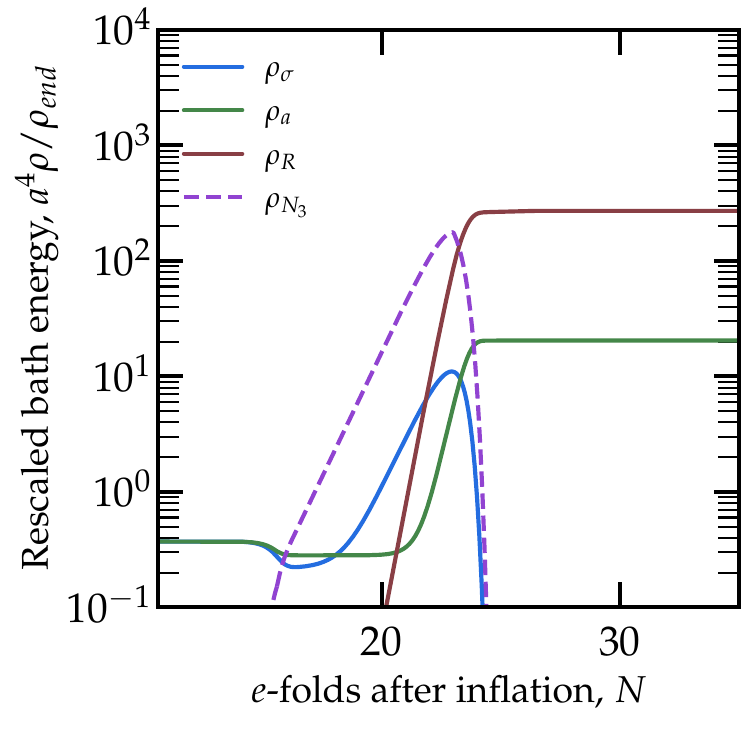}
\caption{\label{fig:bathplots} In this figure, we depict the solutions to Eq. (\ref{eq:boltz}) corresponding to the starred points in Figures~\ref{fig:N1fromdecay} and \ref{fig:N3fromann} (and the SGWB amplitudes in Figure~\ref{fig:fig:vishnugwomega}). Specifically, the left plot depicts a viable $N_1$-from-decay scenario, while the right plot gives an example of $N_3$-from-annihilation.    }
\end{center}
\end{figure*}

\begin{figure*}[t]
\begin{center}
\includegraphics[width = 0.6\textwidth]{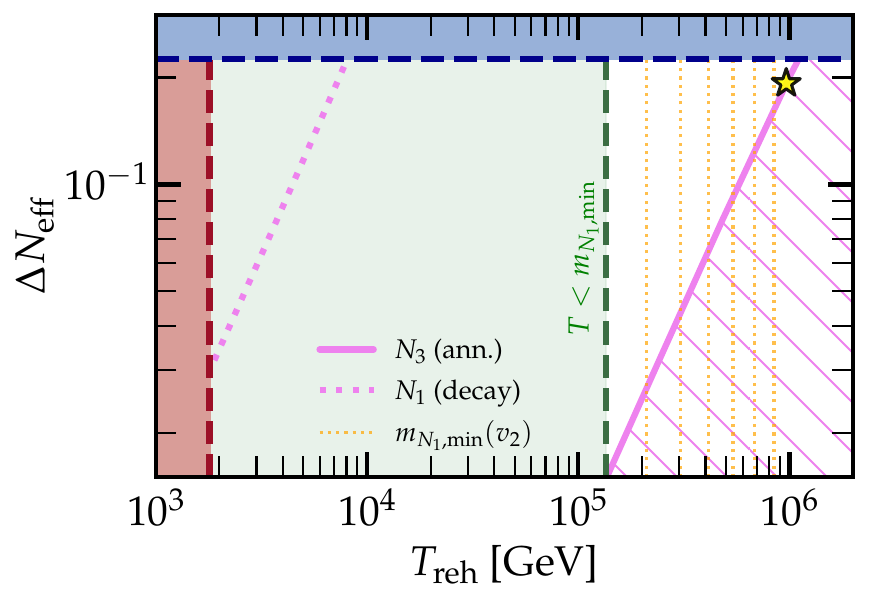}\includegraphics[width = 0.26\textwidth]{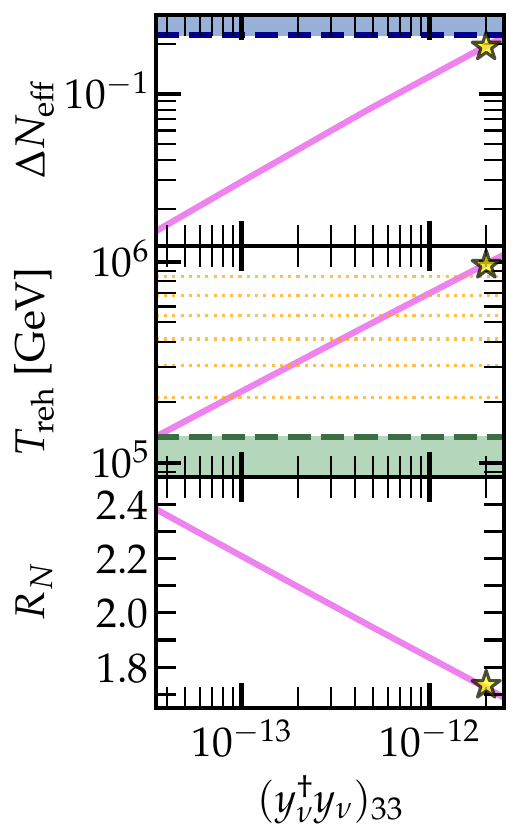}
\end{center}
\caption{\label{fig:N3fromann} In this figure we consider, for $\xi_S = 0.5$, the second reheating scenario, where a long-lived freeze-in population of $N_3$ produce a bath hot enough for thermal leptogenesis ($T_{\mathrm{reh}} > M_{1,\text{min}}$~\ref{eq:thermalDI}). On the left panel, this is the green bound for $v_2 = 4$ GeV, which together with the solid violet contour, and the blue-shaded dark radiation bound, encloses the available parameter space. For $v_2 = 5,6,7,8,9$ GeV, left to right, the leptogenesis bound is indicated by the orange dotted lines. The regions outside are hatched because they may be accessible to other benchmark choices. We also include the $N_1$-from decay parameter region from Figure~\ref{fig:N1fromdecay}, bounded by the red-shaded leptogenesis bound, for comparison. On the right panel, we illustrate the dependence of the dark relativistic degrees of freedom, reheating temperature and reduction in e-folds due to matter domination ($R_N$) on $(y_\nu^\dagger y_\nu)_{33}$, which sets the decay rate of $N_3$. The starred point is used in Figures~\ref{fig:bathplots} and \ref{fig:fig:vishnugwomega}.}
\end{figure*}

On the other hand, if the $N_1$ are long-lived, with $ \Gamma_{N_1 \rightarrow \Phi_2 \ell} \ll \Gamma_{\sigma\rightarrow N_1 N_1} $, then they will come to dominate as a matter-like component. So the situation can be summarised as $\Gamma_{N_1} \ll \Gamma_\sigma,\Gamma_{N_{2,3}}$ when the non-thermal population of $N_{2,3}$ can be neglected. This scenario is depicted in the left panel of Figure~\ref{fig:bathplots}, which corresponds to the starred point in Figure~\ref{fig:N1fromdecay}.  As a result, the energy fraction in dark radiation is suppressed for sufficiently long-lived $N_1$, 
reducing the axion contribution to the stringently constrained dark relativistic degrees of freedom, $\Delta N_{\text{eff}}$ (Section \ref{sec:darkrad}). This is represented by the blue constraint in Figure~\ref{fig:N1fromdecay}. We may assume that the $N_{2,3}$ have thermalised much earlier, since $\widetilde{m}_{2,3} \sim \widetilde{m}_{\text{atm}}$ realises $(y_\nu^\dagger y_\nu)_{22,33} \sim 10^{-8}$-$10^{-6}$ with (\ref{eq:yvyvexp}), for $m_{N_2}, m_{N_3} \sim 10^7$ GeV. 

As explained in Section \ref{sec:leptogenesis}, when the $N_1$ decay to realise the SM bath, they can produce enough lepton asymmetry for successful non-thermal leptogenesis only if the reheating temperature is sufficiently high. The lower bound on $T_{\text{reh}}$ (\ref{eq:trehbound}), the red region in Figure~\ref{fig:N1fromdecay}, has been drawn for $v_2 \gtrsim 4$ GeV~\cite{nuDFSZ}. This choice is consistent with (\ref{eq:bound2mn}) and  $y_{N_1} < \sqrt{\lambda_S/2}$.\footnote{This limit also aligns with the small Higgs vev hierarchy explaining a large top mass in the top-specific VISH$\nu$.} The largest possible reheating temperature which may be achieved ($\sim 10^4$ GeV) is an order of magnitude too small for thermal leptogenesis.

We can account for the smallness of the region by observing that the $N_1$ abundance must be generated before the inflaton number density is significantly reduced. (The $N_1$ at this stage are non-thermal, and so inverse decays are negligible.) This constrains the branching ratio of the decays, which cannot vary by more than an order of magnitude. As we depict in Figure~\ref{fig:N1fromdecay}, the parametric freedom owing to the second Higgs doublet of the underlying DFSZ sector ($v_2 \ll v_1$) realises a small amount of viable parameter space, consistent with
\begin{equation}
    5.7 \times 10^{-16}\lesssim \ (y^\dagger_\nu y_\nu)_{11} \lesssim 1.2 \times 10^{-14} \implies  6.5 \times 10^{-21}\ \text{eV} \lesssim \ \widetilde{m}_1 \ \lesssim 5.5 \times 10^{-19}\ \text{eV}
\end{equation}
encompassed by the violet line for $y_{N_1} \simeq 7\times10^{-6}$. This choice gives a close to maximal branching fraction for $\lambda_S \simeq 2.5 \times 10^{-10}$. We hence \textit{hatch} the region outside in Figure~\ref{fig:N1fromdecay} which, while inaccessible for this benchmark point, may be partly available for the order of magnitude larger $\lambda_S$ (hence larger reheating temperatures and $m_{N_1}$) allowed by $\xi_S \lesssim 1$. We will also see how some of the region to the right can be populated by freeze-in by (co)annihilations at higher temperatures, Figure~\ref{fig:N3fromann}.

This parameter regime clearly satisfies the Vissani-like~\cite{Vissani:1997ys} naturalness bound (\ref{eq:naturalyukn}).\footnote{The required tiny value of $\widetilde{m}_1$ can also be stable under radiative corrections~\cite{nuDFSZ}.}
Smaller values of $y_{N_1}$ correspond to the region enclosed by the green line together with the dark red and blue bounds, with a reduced range of compatible $(y^\dagger_\nu y_\nu)_{11}$. 

\begin{figure*}[t]
\begin{center}
\includegraphics[width = 0.4\textwidth]{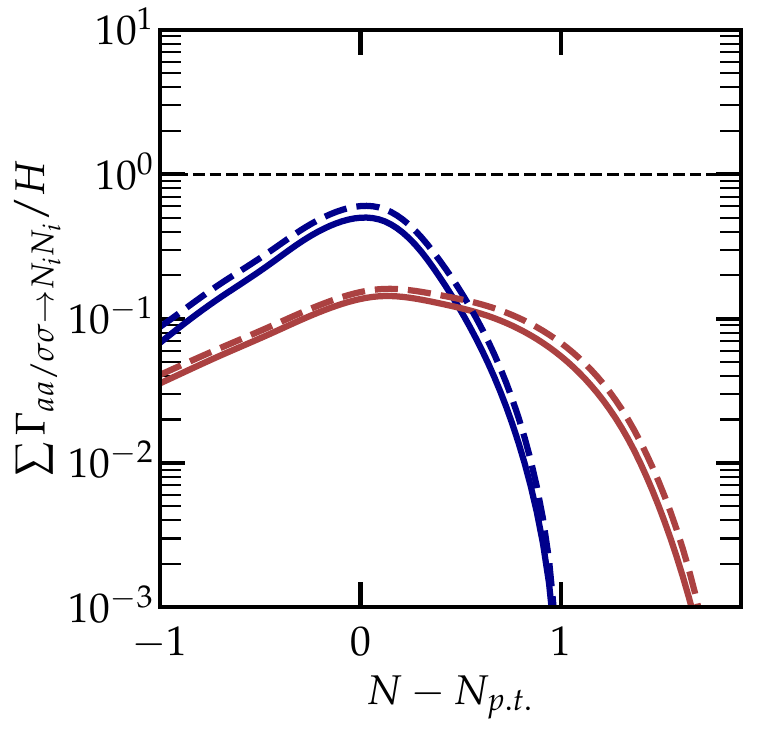}\ \ \
\includegraphics[width = 0.4\textwidth]{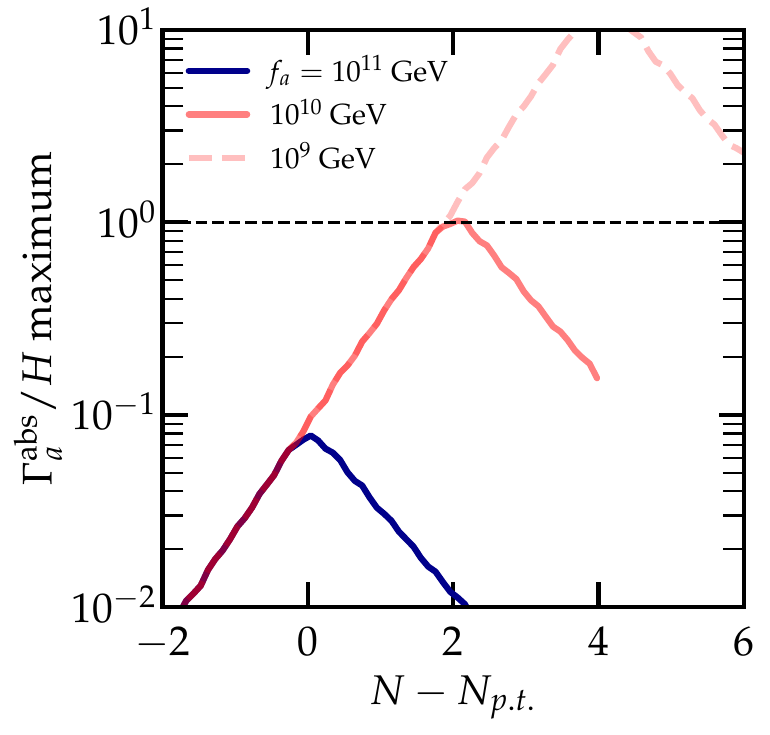}
\end{center}
\caption{\label{fig:rates} In the left panel, we compare to the expansion rate the total annihilation rates of the axions (solid) and inflatons (dashed) into the $\rho_{N_2} + \rho_{N_3}$ bath for $y_{N_3} = 7.4\times10^{-4}$ (blue) and $y_{N_3} = 3.7 \times10^{-4}$ (red), corresponding to masses of $2.6 \times 10^7$ GeV and $5.2 \times 10^7$ GeV, respectively. The plot illustrates that the rates reach a maximal, but only marginally efficient value at the phase transition, with $7.4\times10^{-4}$ resulting in the larger rate. In the right panel we display the absorption rate $\Gamma^{\text{abs}}_{a}$, for when the temperature, and hence the rate, is maximised (full thermalisation of the non-axion component).  The rate peaks at the phase transition. The plot illustrates that the axion absorption rate by the radiation bath is always subdominant for our benchmark $f_a$, however, for $f_a < 10^{10}$ GeV, thermalisation through the QCD portal could in principle be achieved.  }
\end{figure*}

\paragraph{Reheating through axion/inflaton (co)annihilations to $N_iN_i$}
Each of the dominant absorption (\ref{eq:qcdabsorption}) and (co)annhilation (\ref{eq:sigmaann}) processes changing the inflaton and axion number densities are at most marginally efficient (see Figure~\ref{fig:rates}). Nonetheless, we find, using the extrapolated distribution functions from lattice estimates, that the non-thermally averaged (co)annihilation rates are maximised for $y_{N_i} \sim \text{few}\times10^{-4}$ in the vicinity of the phase transition, and can result in a sizeable freeze-in population of $N_i$. These Yukawa couplings imply $M_{1,2}$ masses which can satisfy the naturalness mass bounds (\ref{eq:bound2mn}) only for $\tan \beta \sim \mathcal{O}(1)$ or an $m_{N_3}$ mass in a decoupling limit allowing $\tan \beta \gg 1$ . We only sketch the first case, but consider the second in more detail, exhibiting a viable parameter region in (Figure~\ref{fig:N3fromann}). The leptogenesis bounds mentioned may be found later in Section~\ref{sec:leptogenesis}.

\underline{\textit{$N_1$ from annihilations:}} We will argue that the first case is difficult, but not impossible to achieve, considering a detailed analysis beyond the scope of this paper. The smaller values of $\tan \beta \sim \mathcal{O}(1)$ result in more stringent leptogenesis bounds on $m_{N_1}$ (thermal: \ref{eq:thermalDI}) and on $T_{\text{reh}}$ (thermal and non-thermal: \ref{eq:trehbound}). It is clear, then, that thermal leptogenesis with $\tan \beta \sim 1$ cannot be achieved in this case, which would require $m_{N_1} \sim 10^8$-$10^9$ GeV, much heavier than the masses where the annihilation rates are efficient (so $N_1$ would not be the lightest RHN). In this context, $N_1$ must then produce the lepton asymmetry non-thermally as it reheats the universe, but, for $\tan \beta \sim \mathcal{O}(1)$, this requires $T_{\text{reh}} \gtrsim 10^7$ GeV, see Eq. (\ref{eq:trehbound}). This requires reheating to end well before $\Gamma_{\sigma \rightarrow aa} > H$. Hence, this scenario only remains viable if the inflatons have disappeared before their decays become efficient, because $N_1$ cannot otherwise suppress the energy fraction in hot axions. Clearly, the scenario is only rescued if the inflatons thermalise at the phase transition and subsequently get a Boltzmann-suppressed number density. It appears to be possible to realise this situation when all $N_{1,2}$ saturate their naturalness bounds, and $N_3$ is chosen such that the summed (co)annihilation rate can be efficient. However, the parameter space is sufficiently marginal to require a dedicated study, particularly due to the non-trivial particle distributions being assumed.

\underline{\textit{$N_3$-from-annihilations:}} In this case we have the parametric freedom to consider $N_3$ heavy enough that its non-thermal freeze-in production rate is enhanced (we fix $y_{N_3} = 7 \times 10^{-4}$ GeV), but consider $N_{1,2}$ to take values at large $\tan \beta$ (small $v_2$) that permit lower $m_{N_1}$ masses for thermal leptogenesis, see Eq. (\ref{eq:thermalDI}), and hence a lower reheating temperature than for $\tan \beta \sim 1$.  This $y_{N}$ parameter regime, which is consistent with electroweak radiative stability,  Eq. (\ref{eq:bound2mn}), results in long-lived $N_3$. $N_1$ and $N_2$, whose freeze-in production is negligible, thermalise much earlier through (inverse) decays with the radiation bath. RHN matter-domination is then much earlier than in the $N_1$-from-decay scenario and the possible reheating temperatures are correspondingly higher, sufficient to keep the $N_1$ thermal at the time of the $N_3$ decay to populate the thermal radiation bath for a subset of the parameter space.
Here, the relic dark radiation is suppressed by the larger abundance of $N_3$ relative to inflatons at the time of their decay.

The available parameter is plotted in Figure~\ref{fig:N3fromann}. As (co)annihilations are $\propto y^4_{N_1}$, smaller $m_{N_3}$ masses will be enclosed by the corresponding contour on the $T_{\text{reh}}, N_{\text{eff}}$ plane by considering this bound to be saturated. For $v_2 \ll v_1$, the range of possible $N_3$ decay rates is then set by
\begin{equation}
    3.5 \times 10^{-14}\lesssim \ (y^\dagger_\nu y_\nu)_{33} \lesssim 2.5 \times 10^{-12} 
\end{equation}
which, as in the $N_1$ from decay scenario, requires $m_{\text{min}} \lll 10^{-3}$ eV.

\paragraph{Decays of long-lived $N_{i}$ from preheating (unnatural) 
}

\label{sec:nureheating3}

As we illustrate in Figure~\ref{fig:fermpreheating}, non-trivial initial energy fractions of $N_i$ can result from fermionic preheating during the initial oscillations of the inflaton background. The strength of this effect is enhanced for larger $y_{N_i}$, and hence larger RHN masses, being most important for $N$'s which are otherwise kinematically inaccessible to perturbative decays or scatterings. This production saturates~\cite{Fermions1} well before the inflaton fragmentation (see Figure~\ref{fig:fermionicpreheating1}), which we expect to disrupt the coherence and periodicity of the zero-mode required for the mechanism to be efficient. 

It is tempting to imagine the $N$'s, which are produced with a non-relativistic $k \ll m_{N_{i}}$ spectrum, will quickly come to dominate the energy density after preheating as a matter-like component. However, it should be remembered that, even though in this case $\rho_{N_i} \simeq m_{N_{i}}n_{N_{i}}$, the masses red-shift as $m_{N_{2,3}} \propto a^{-1}$ until the PQ phase transition, so that their energy density is $\propto a^{-4}$, and is thus radiation-like.
Nevertheless, after the PT, the $N_i$ from preheating can play the same role as $N_1$ in the first reheating scenario provided their decays to the SM are sufficiently suppressed. That is, they can become a dominant energy component before realising the SM bath through decays. 

\label{sec:nureheating2}

Due to the large value of $y_{N_i}$ required, this mechanism can only be achieved with a stable electroweak scale for heavy $N_3$, see Eq. (\ref{eq:bound2mn}), or with large $v_2$ (small $\tan \beta$). Let us consider the first option. In order to suppress the axion dark radiation, the $N_3$ decays would realise a reheating temperature which is much too low for thermal leptogenesis with $N_1$, meaning the asymmetry must be generated directly by non-thermal $N_3$ decays. Keeping the $N_3$ sufficiently long-lived requires a very tiny $m_{\text{min}}$. This is then a problem, because the asymmetry generated in naturally heavy $N_3$ decays, Eq. (\ref{eq:asymmN3}), is suppressed by $m_{\text{min}}$. In the second option, the correspondingly large $v_2 \gg v$ requires a much higher reheating temperature, see Eq. (\ref{eq:trehbound}), for successful non-thermal leptogenesis than what can be achieved while suppressing dark radiation.

Hence, we consider the fermionic preheating scenario, though viable, to fail our criteria for naturalness. It requires $N_1$ and/or $N_2$ to have a mass $\gtrsim 10^ 9$ GeV, leading to radiative corrections to $m^2_{22}$ which are $\gtrsim 10^2$ larger than TeV scale, thus requiring fine tuning. We illustrate a viable (but unnatural) region of parameter space for $y_{N_i} < 0.1$ and a range of $(y_\nu^\dagger y_\nu)_{ii}$ in Figure~\ref{fig:fermpreheating}. The former is not a physical limit, but is adopted to avoid an analysis of early backreaction effects on the bosonic preheating. On the other hand, as short-lived $N_i$ cannot satisfy an instantaneous decay approximation in this context during preheating, we do not expect this to realise an instant preheating mechanism. 

\section{Further discussion}
\label{sec:part2}

\begin{figure}[t!]
\centering
\includegraphics[width=0.8\textwidth]{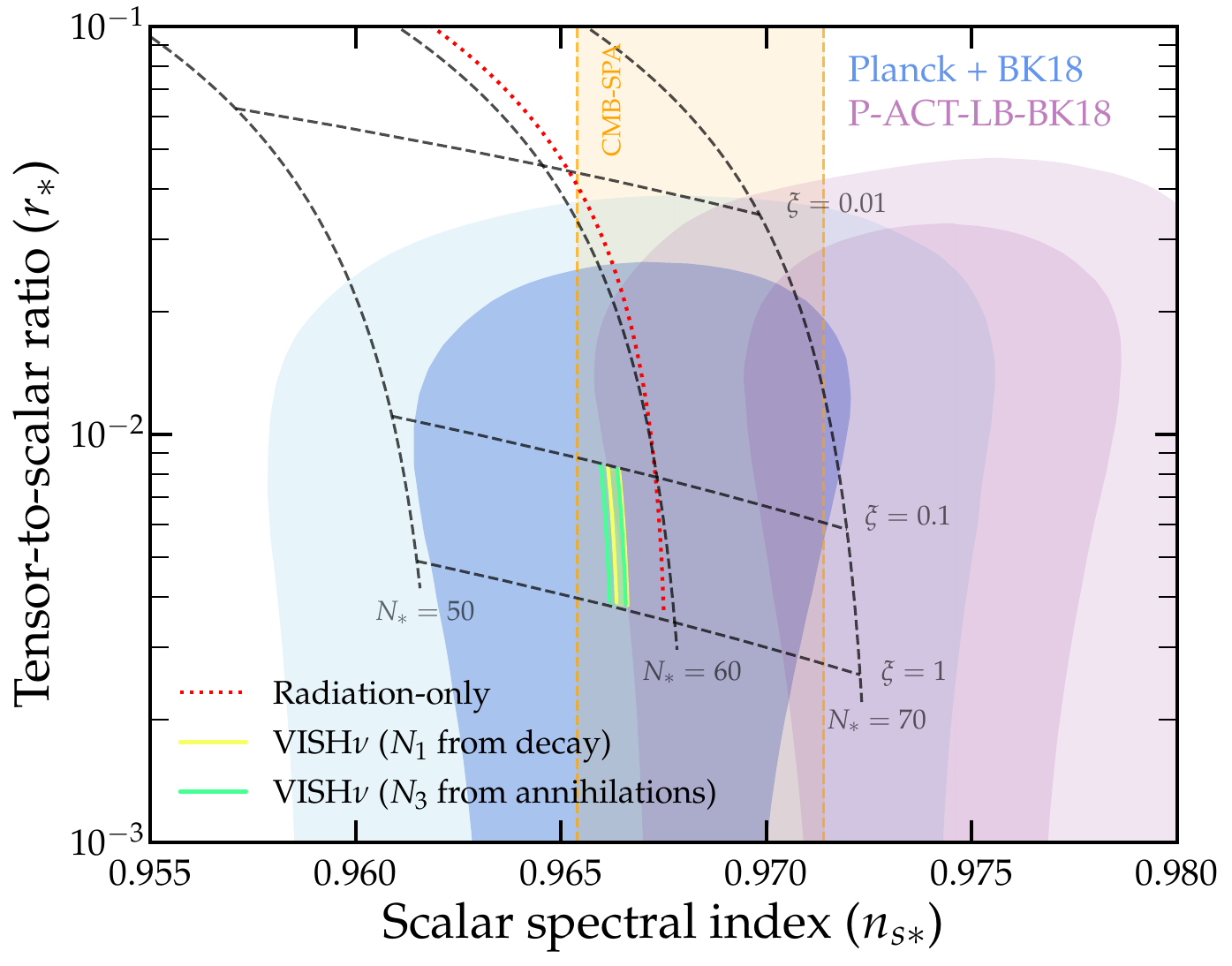}
\caption{\label{fig:rnsplot} In this plot we include the approximate range of $(r, n_s)$ values for the two candidate VISH$\nu$ reheating scenarios we have identified in Section~\ref{sec:reh}, bounded by the yellow and green solid lines for $N_1$-from-decay and $N_3$-from-annihilation, respectively. These are compared to the inflationary predictions for the radiation-only expansion (red, dotted), as assumed in the SMASH model~\cite{Ballesteros:2016xejSMASH}. See text for details.}
\end{figure}

As we discuss below, each VISH$\nu$ reheating scenario can lead to different predictions for the scalar spectral index, successfully realises leptogenesis, obeys dark radiation constraints and leads to a post-inflationary axion. 

\subsection{e-folds of inflation}
\label{sec:efolds}

The large-scale curvature perturbation ($\mathcal{R}$) and tensor ($h$) perturbation power spectra may be characterised as power laws, about a fiducial pivot scale $k_*$, with associated spectral amplitudes ($A_s$ and $A_t = rA_s$) and spectral tilts ($n_s$ and $n_t$)
\begin{equation}
    \Delta_\mathcal{R}(k) = A_s \left( \frac{k}{k_*}\right)^{n_s - 1},\qquad \Delta^2_t (k)=rA_s\left( \frac{k}{k_*}\right)^{n_t}
\end{equation}
Here, $r$ is the tensor-to-scale amplitude ratio, proportional to the inflationary energy scale. As mentioned in Section~\ref{sec:vishnu},  Figure~\ref{fig:lxiplot} the fit to the scalar amplitude $A_s \simeq 2.1 \times 10^{-9}$~\cite{Planck:2018vyg} results in a parameter space where $\lambda_S \ll 1$ for $\xi_S \lesssim 1$.

The remaining constraints on the inflationary power spectrum at the current level of precision are typically depicted in the $r$-$n_s$ plane. In Figure~\ref{fig:rnsplot}, we include the well-known Planck 2018~\cite{Planck:2018jri} and Bicep/Keck (BK18~\cite{BICEP:2021xfz}) fit (blue region) as well as new combined fits with Atacama Cosmology Telescope (ACT-DR6~\cite{ACT:2025tim}) and South Pole Telescope (CMB-SPA~\cite{SPT-3G:2025bzu}) data.  (We take from the latter the $95\%$ CL scalar spectral index constraint resulting in the yellow band of Figure~\ref{fig:rnsplot}.) With respect to CMB-only constraints~\cite{German:2025ide}, the inflationary predictions consistent with the VISH$\nu$ parameter space are strongly favoured.
As discussed in Ref.~\cite{ACT:2025fju}  (see also Ref.~\cite{SPT-3G:2025bzu}), combined fits of these with BAO data from the DESI  collaboration~\cite{DESI:2025zgx}, which does not measure $n_s$ directly, currently introduces a tension with the smaller values of $n_s$ favoured by the CMB-only datasets. The extent to which this result challenges the inflation model is unclear given the inconsistency of the datasets being combined (see comments in Ref.~\cite{Kallosh:2025rni,Ferreira:2025lrd, SPT-3G:2025bzu}). We include the constraint for later reference as the purple region of Figure~\ref{fig:rnsplot}.

\paragraph{Predicting the scalar spectral index} In the single-field inflation model, the large-scale scalar and tensor perturbation power spectra constrained by CMB data are directly related to the scalar potential (in our case the valley close to the $|S|$-direction) in the vicinity of a particular inflaton background field-value. As the inflaton introduces a preferred time-slicing of the spacetime, this corresponds to a moment in time, namely when the relevant scale left the horizon ($k=a_kH_k$). In particular, the scalar spectral index $n_s$ is sufficiently well-constrained that there is a measurable dependence on the remaining duration of inflation. 

Unless fixed by a post-inflationary expansion history pre-BBN, the exact predictions of a given inflation model are then subject to a degree of uncertainty. Hence, our analysis of the reheating epoch can now make precise the predictions of Ref.~\cite{Sopov:2022bog} regarding the VISH$\nu$ inflation model. 

Let us define, for a relevant co-moving scale $k$, and a particular inflation model, the number of e-folds of scale factor expansion which are predicted to have elapsed after horizon-crossing ($k = aH$) and the end of inflation,
\begin{equation}
    N(k) \equiv \log \frac{a_{\text{end}}}{a_{k}} = \int^{t_{\text{end}}}_{t_k} \mathrm{d}t\ H,
\end{equation}
where $a_{\text{end}} > a^{\text{inf}}_k = k/H^{\text{inf}}_k$.\footnote{In this context, the calculation is done in the Einstein frame due to the non-minimal coupling $\xi$, which is related by a well-known conformal transformation (given in~\cite{Sopov:2022bog}) to the Jordan frame model given in Section~\ref{sec:vishnu}.  } We denote by $N_*$ the value corresponding horizon exit of the CMB pivot scale in Figure~\ref{fig:rnsplot}, which, following Refs.~\cite{Planck:2013jfk, Liddle:2003as} may be inferred by the relationship to the present horizon size, so that using measured values 
\begin{equation}
    N_* \equiv N(k_*) \simeq 67 - \log\frac{k_*}{a_0H_0} + \frac{1}{4} \log \frac{V_k}{m_P^2} + \frac{1}{4} \log \frac{V_k}{\rho_{\text{end}}}  - R_N,
\end{equation}
where ``eq'' denotes matter-radiation equality and instantaneous epochal transitions have been assumed. $V_k$ denotes (in the Einstein frame) the inflationary potential energy at horizon exit of $k$.
Here, $R_N$ accounts for deviations from radiation domination before the thermal era, which amount to a reduction in e-folds. In the VISH$\nu$ context, it is given by 
\begin{equation}\label{eq:matterefolds}
    R_N = \frac{1}{12}\left[\log \frac{g_{*, \text{reh}}}{106.75} + \sum_{i=0}^{n_{\text{MD}}} \log \frac{\rho_{\text{md},i}}{\rho_{\text{rd},i}} \right]
\end{equation}
allowing for $n_{\text{MD}}$ matter-dominated periods before the end of reheating, beginning and ending at $a_{\text{md},i}$ and $a_{\text{rd},i}$, the last being $a_{\text{reh}}$. Larger $R_N > 0$ decreases $n_s$. This raises the possibility that the reheating scenarios of VISH$\nu$, and hence the SMASH model, may be distinguished through future CMB experiments, as each realises a different $R_N$.

In Section~\ref{sec:reh}, it was found that the $N_1$-from-decay reheating and leptogenesis scenario (Figure~\ref{fig:N1fromdecay}) results in a range $1.6  \lesssim R_N \lesssim 2.1$ which is bounded by the yellow contours in Figure~\ref{fig:rnsplot}. The second scenario we considered, $N_3$-from-annihilations, results in the slightly broader range $1.7 \lesssim R_N \lesssim 2.4$ corresponding to the region bounded by the green contours in Figure~\ref{fig:rnsplot}. The maximal value of $N_*$, and hence $(n_s)_*$, achievable in the inflation model corresponds to no matter domination, which is the expectation for reheating in the SMASH axion-majoron model~\cite{Ballesteros:2016xejSMASH}. However, all reheating scenarios we have identified involve $R_N > 0$ resulting in a deviation in $(n_s)_*$ at the $\sim 0.1\%$ level. 

Our analysis was performed with a benchmark choice of $\xi_S = 0.5$. The reader may wonder how significantly our results change for different values of this coupling. There are two $\xi$-dependent parameters of interest, $\lambda_S(\xi_S)$, $\phi_{\text{end}}(\xi_S)$, which together coordinate the moment of the phase transition, and the time where inflaton to axion decays become efficient. We observe that, for $\xi_S$ in the range $0.01$-$1$, $\phi_{\text{end}}$ varies by a factor $\sim 3$, while $\lambda_{S}$ varies by $10^3$. Hence, $a_{PQ}$, which is $\propto \phi_*$, does not vary by much, while $\Gamma_{\sigma \rightarrow aa}/H \propto \sqrt{\lambda_S}/\phi_{\text{end}}$ can change by $\sim 10$ in this range. Taking the example of the $N_1$-from-decay reheating scenario, we can expect the extra inflaton matter domination before decay to increase $R_N$ by up to  $+0.5$ for $\xi_S = 10^{-2}$, while the maximum possible reheating temperature will decrease by a factor $5$, which challenges the leptogenesis mechanism. Hence, we do not extend our predictions beyond $\xi < 0.1$ in Figure~\ref{fig:rnsplot}, where we expect the results are a good approximation for $\xi_S \neq 1$.

\subsection{Dark radiation}
\label{sec:darkrad}

As a consequence of the reheating epoch in VISH$\nu$ models, a relic axion bath is produced which ultimately contributes a BSM fraction $\Delta N_{\text{eff}}$ to the number of dark relativistic degrees of freedom, parameterised by $N_{\text{eff}}$. In particular, for non-thermal relic axions, the contribution can be larger than the value expected from thermal axion production~\cite{Mazumdar:2016nzr}, constraining a significant part of the parameter space, as seen in Figures \ref{fig:N1fromdecay}, \ref{fig:N3fromann} and \ref{fig:fermpreheating}.

This quantity is stringently constrained at CMB decoupling by \textit{Planck} and BAO data, \textit{viz.} $N_{\text{eff}} = 2.99{}^{+0.34}_{-0.33}$~\cite{Planck:2018vyg}, but the impact of additional radiation density on the expansion rate is also imprinted in the abundances of light elements at BBN. Assuming three light neutrino species, a recent joint analysis of these constraints in Ref.~\cite{Yeh:2022heq} limits any additional contribution to
\begin{equation}\label{eq:delneff}
    \Delta N_{\text{eff}} < 0.226.
\end{equation}
The contribution from relic axion radiation is given as follows 
\begin{equation}
    \Delta N_{\text{eff}} = \frac{\rho_{a,0}}{\rho_{\nu,0}} = \frac{\rho_{a,0}}{\rho_{\text{SM},0}} \frac{\rho_{\text{SM},0}}{\rho_{\nu,0}} = \frac{g_{*}(T_{\text{reh}})}{g_{*}(T_0)} \left[ \frac{g_{*S}(T_0)}{g_{*S}(T_{\text{reh}})}\right]^{4/3} \left[\frac{8}{7} \left(\frac{11}{4}\right)^{4/3} + N^{\text{SM}}_{\text{eff}} \right]\frac{\rho_{\text{a,rh}}}{\rho_{\text{SM,rh}}}
\end{equation}
Hence, for $N^{\text{SM}}_{\text{eff}} \simeq 3.044$~\cite{Bennett:2020zkv}, we have that  
\begin{equation}
    \Delta N_{\text{eff}} \simeq 7.4\ \frac{g_{*}(T_{\text{reh}})}{g_{*}(T_0)} \left[ \frac{g_{*S}(T_0)}{g_{*S}(T_{\text{reh}})}\right]^{4/3}\frac{\rho_{\text{a,rh}}}{\rho_{\text{SM,rh}}}
\end{equation}
which corresponds to the dark radiation constraints in Section \ref{sec:reh}. Note that $g_*(T_{\text{reh}})$ is slightly larger than the SM value due to the additional Higgs states in the radiation bath. 

\subsection{(Non)-thermal leptogenesis}
\label{sec:leptogenesis}

In the VISH$\nu$ model, the PQ scalar must reheat the universe, but at the same time carries a lepton number. It thus becomes possible to produce a large abundance of $N_i$'s non-thermally from the inflaton energy density. This can occur through freeze-in decays and annihilations, or else by fermionic preheating, as we have seen in Section~\ref{sec:reh}. In fact, for the majority of the available parameter space, and in the desired region which respects the VISH$\nu$ naturalness criteria, the reheating is ultimately performed by the late-time decays of the $N_i$, which must be sufficiently long-lived to suppress relic axion dark radiation (see Section \ref{sec:darkrad}).

In these scenarios, reheating temperatures may be below the lightest $N$ respecting the  Davidson-Ibarra bound, Eq. (\ref{eq:davidsonibarra}), in the context of thermal leptogenesis~\cite{Davidson:2002qv,nuDFSZ}, 
\begin{equation}\label{eq:thermalDI}
    M_{1} \gtrsim 5\times10^8\ \text{GeV}\ \left( \frac{v_2}{v} \right)^2 \simeq 1.3 \times 10^5\ \text{GeV}\ \left(\frac{v_2}{4\ \text{GeV}} \right)^2,\qquad T_{\text{reh}} > m_{N_1}.
\end{equation}
However, the large initial abundance of $N$'s can still produce the required lepton asymmetry directly through their out-of-equilibrium decays. 

There are then two scenarios for non-thermal leptogenesis, namely long-lived $N_1$ decays produced from $\sigma \rightarrow N_1N_1$~\cite{Lazarides:1990huy,Murayama:1992ua}, or decays from heavier neutrinos (including $N_{2,3}$~\cite{Asaka:2002zu}) produced in this context by freeze-in from annihilations or fermionic preheating. 
Washout can be automatically suppressed in the long-lived $N$ scenario, 
typically realising $T_{\text{reh}} \ll m_{N_1}$ so the $N_1$ cannot alter or thermalise the asymmetry produced after reheating ends.\footnote{This is in contrast to thermal leptogenesis, where the lepton asymmetry from $N_{2,3}$ processes is washed out by the still thermal $N_1$'s, before these yield the asymmetry.} 
The resulting baryonic fraction, Eq. (\ref{eq:baryonyield}), may thus be estimated without including the explicit Boltzmann equation for $n_{B-L}$ in Eq. (\ref{eq:boltz}).

The baryon-to-entropy ratio, measured to be $Y_B = (8.72\pm 0.08)\times10^{-11}$~\cite{PlanckCosParam}, can then directly estimated from the full conversion of $N$'s to radiation ($\rho_R = \rho_N$)~\cite{Zhang:2023oyo} as follows 
\begin{equation}\label{eq:baryonyield}
    Y_B \equiv \frac{n_B}{s} =  c_{\text{sph}}\epsilon \frac{n_N}{s} \simeq \frac{3}{4}c_{\text{sph}}\epsilon\frac{T_{\text{reh}}}{M_{i}}
\end{equation}
where $c_\text{sph} = \frac{32}{92}$ is the two electroweak doublet sphaleron conversion factor for the $B$-$L$ asymmetry to baryon number, $B$; and $\epsilon_i$ is the amount of CP violation per $N_i$ decay. The reheating temperature is then bounded from below from the requirement that Eq. (\ref{eq:baryonyield}) is achieved for a maximal value of $\epsilon_i$, see Eq. (\ref{eq:davidsonibarra}), which for $N_{1,2}$ is given by
\begin{equation}\label{eq:epsmax}
\begin{split}
    |\epsilon_i| &\lesssim \frac{3}{16\pi}\frac{m_{N_i}m_{\text{atm}}}{v^2_2}\\
    &\simeq  0.0030 \times y_{N_i}  \left(\frac{m_{\text{atm}}}{0.05\ \text{eV}} \right)\left( \frac{f_a}{10^{11}\ \text{GeV}}\right)\left(\frac{10\ \text{GeV}}{v_2} \right)^2
\end{split}
\end{equation}
Note that $v_2 \gtrsim 4$~GeV is possible in the VISH$\nu$ model framework~\cite{nuDFSZ}, which weakens the constraint given for the benchmark parameter point by an order of magnitude. 

Hence, we can combine Eqs. (\ref{eq:epsmax}) and (\ref{eq:baryonyield}) to obtain a lower bound on the reheating temperature for non-thermal leptogenesis,
\begin{equation}\label{eq:trehbound}
    T_{\text{reh}} \gtrsim 1.8 \times 10^3\ \text{GeV}\  \left(\frac{v_2}{4\ \text{GeV}}\right)^2\left(\frac{32/92}{c_{\text{sph}}} \right) \left(\frac{0.05\ \text{eV}}{m_{\text{atm}}}  \right) \left( \frac{Y_B}{8.72\times10^{-11}}\right).
\end{equation}
This bound is the dark red shaded region in Figures~\ref{fig:N1fromdecay}, \ref{fig:N3fromann} and \ref{fig:fermpreheating}. 

Note that, for the purposes of this paper, we do not assume that the heavy neutrino masses are degenerate, which would further relax these bounds~\cite{Pilaftsis:2003gt}.

\subsection{Post-inflationary axion dark matter}

The VISH$\nu$ model features a QCD axion dark matter candidate (which is at the same time a majoron). Let us mention briefly about what is assumed for this to be the case, in the context of the reheating analysis we have performed.\footnote{The topic will be revisited in more detail elsewhere Ref.~\cite{Companion}.}

As explained in Section~\ref{sec:lattice}, despite a low reheating temperature, the growth of axion fluctuations during the preheating epoch results in a non-thermal restoration of the PQ symmetry after inflation for $f_a \ll 10^{17}$ GeV. The alternative, to have $f_a \gtrsim 10^{17}$ GeV, has been shown to result in an over-enhancement of axion isocurvature~\cite{Ballesteros:2021bee} through the tachyonic instability reviewed in Section~\ref{linearanalysis}. Consequently, the axion in VISH$\nu$ is \textit{post-inflationary}, fulfilling the role of dark matter for a fixed value of $f_a \simeq 10^{11}$ GeV (up to theoretical uncertainties). The non-thermal phase transition, which was observed in Section~\ref{sec:latticept} to result in the formation of axion strings, further confirms this picture. The non-thermal axions produced from preheating and inflaton decay do not contribute appreciably to the axion dark matter density, but are an early dark radiation component, as we have discussed. 

The axion dark matter fraction is thus both populated by the usual misalignment mechanism~\cite{Preskill:1982cy, Abbott:1982af, Dine:1982ah} (with $\langle \theta \rangle = \frac{\pi^2}{3}$) as well as by defects. We will assume that in both cases the non-thermal, non-linear context of the PQ PT does not modify the usual predictions, though this can be put to the test by future simulations. In the case of axion strings, there is some evidence that the attractor solution exists for random initial conditions~\cite{Gorghetto:2018myk}. 

As discussed in Section~\ref{sec:vishnu} and Ref.~\cite{Sopov:2022bog}, we take an $N_{DW} =1$ axion to be realised through flavoured PQ couplings in the DFSZ sector, so that the string-domain wall network is short-lived. This leads to a DFSZ-like sensitivity for the axion-photon coupling ($\frac{E}{N} = \frac{8}{3}$), but realises a KSVZ-like ($N_{DW} =1$) mass window for the DM axion mass. Hence, despite interesting low energy phenomenological departures, the mass of the axion cannot be used to distinguish the model from SMASH, which is set cosmologically in the same way. 

\section{Gravitational waves }
\label{sec:part3}

The cosmological history we have outlined above additionally  leads to specific primordial gravitational wave (GW) sources contributing to a present-day cosmological stochastic gravitational wave background (SGWB); these include not only inflationary and thermal fluctuations, but more particularly the primordial anisotropic stress of large field gradients generated by preheating, turbulence and cosmic string evolution (the latter being considerably suppressed for the post-inflationary regime of $f_a$).
Importantly, in the context of an explicit model, all these spectra are inter-related and dependent on the expansion history. Indeed, as relic GWs remain the only direct carrier of information about the pre-thermal era, due to their decoupling from other energy components, they provide an important complement to the phenomenology we have discussed so far; particularly so in the context of a ``hidden sector'' model, where the portal coupling with the visible sector is automatically small, while the gravitational coupling remains universal. 

\begin{figure}[t!]
\centering
\includegraphics[width=0.75\textwidth]{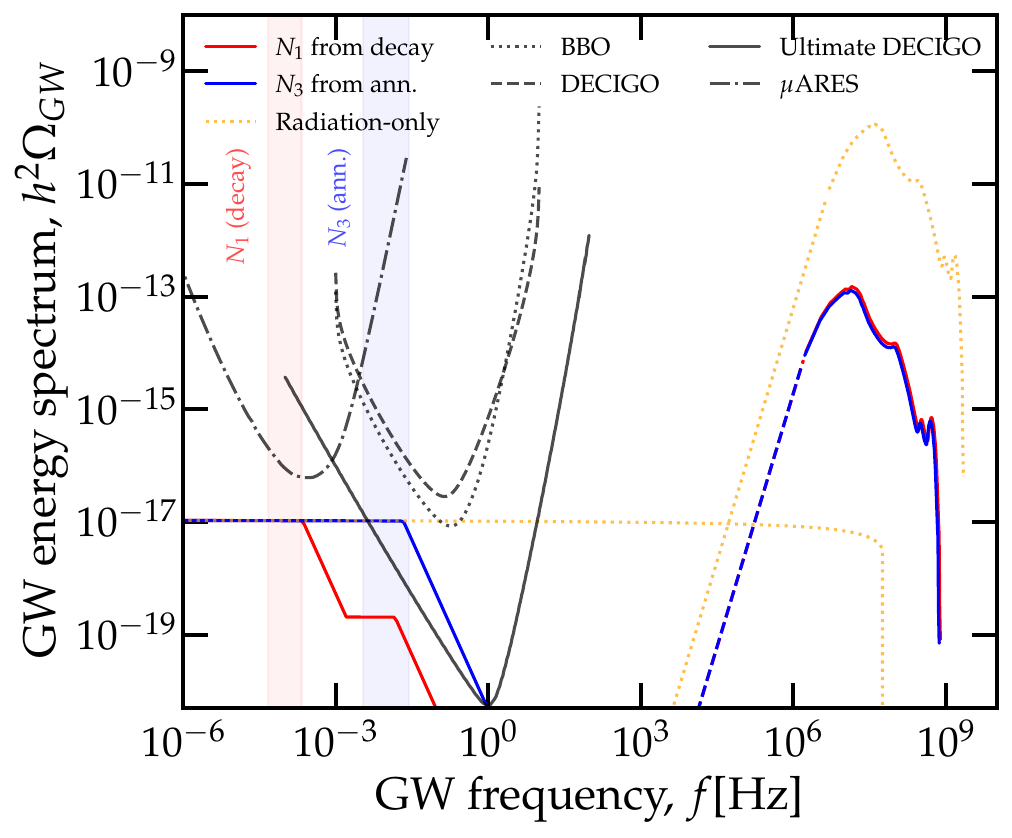}
\caption{\label{fig:fig:vishnugwomega}{In this plot, we provide estimates for the combined SGWB contributions from inflation and preheating for the two starred parameter points in Figures~\ref{fig:N1fromdecay} and~\ref{fig:N3fromann}, corresponding to the two viable reheating scenarios discussed in Section~\ref{sec:nureheating}. The band of frequencies associated to the viable reheating temperatures in each case is shaded, representing the approximate location of a kink in the spectrum. These are compared to sensitivity curves for proposed space-borne laser interferometers: 
BBO~\cite{Crowder:2005nr}, DECIGO~\cite{Seto:2001qf} (fits from Refs.~\cite{Schmitz:2020syl,schmitz_new_2020}),  and the ultimate DECIGO~\cite{Kuroyanagi_2015} sensitivity, together with
$\mu$Ares~\cite{Sesana_2021}. We also plot a benchmark contour for a radiation-only equation of state consistent with the inflation model (which facilitates a comparison with SMASH~\cite{Ringwald:2022xif}). 
}}
\end{figure}

In this section, we estimate the SGWB contributed by inflation and preheating in VISH$\nu$, facilitating comparison with a similar study of the gravitational wave spectrum of the SMASH model~\cite{Ringwald:2020vei,Ringwald:2022xif}. We also discuss a contribution from the initially high temperature radiation bath. As can be seen in Figure~\ref{fig:fig:vishnugwomega}, future space-borne laser interferometers, particularly in the $\mu$Hz-Hz range, may not only lead to indirect evidence 
of our proposal, but observationally distinguish the reheating scenarios through features imprinted in the inflationary spectrum. For example, the reheating temperatures of the $N_3$-from-annihilation reheating are high enough for the transition away from $N_3$ domination to fall within an ultimate DECIGO sensitivity~\cite{Kuroyanagi_2015}, leaving a distinctive kink. 

Nonetheless, the direct detection of the SGWB at the ultra high frequencies  relevant to preheating remains challenging in view of  strain sensitivity and detector noise. Despite this it also lacks known astrophysical foregrounds~\cite{Aggarwal:2025noe}, which pose a difficulty for $<$ kHz frequencies. 

\paragraph{Inflation}

As discussed earlier, the inflationary era contributes a SGWB \cite{Grishchuk:1974ny,Starobinsky:1979ty} in the form of initial super-horizon tensor perturbation modes ($h_k$) with a primordial power spectrum
\begin{equation}
    \Delta^2_{t} (k, \tau_i) \simeq \frac{2}{\pi^2}\frac{H^2_*}{m_P^2}\qquad \text{where} \qquad k=a_* H_*,
\end{equation}
which then corresponds to a present-time energy fraction per logarithmic frequency interval
\begin{equation}
     h^2\Omega_{\text{GW}}(f) = \frac{h^2}{\rho_c}\frac{\mathrm d\rho_{\text{GW}}}{\mathrm d \log f}  =   \underbrace{\left|\frac{h_k(\tau_0)}{h_k(\tau_i)} \right|^2}_{\equiv \mathcal{T}_0(f)} \frac{h^2\Delta^2_t(f,\tau_i)}{12a_0^2H_0^2} \qquad \text{with} \qquad f = \frac{k}{2\pi a_0},
\end{equation}
in units of the critical density $\rho_c = 3H_0^2m_P^2$. (Here, $*$ again denotes a quantity evaluated at the horizon exit of the corresponding scale $k$, `$0$' denotes evaluation at the present time and $\tau_i$ is a time just after the end of inflation.) 

While the amplitude of the primordial tensor power spectrum is proportional to the energy-scale of inflation, the frequency-dependence (beyond the very weak inflationary spectral tilt) may be summarised in the transfer function, $\mathcal{T}_0(f)$, which carries information about the subsequent expansion and thermal history of the universe. When the GW modes re-enter the horizon during the intervening eras after inflation, having been frozen on super-horizon scales, they oscillate and red-shift as contributions to a radiation component, and may be enhanced or suppressed (as a fraction of the critical density) according to the expansion rate and equation of state. In particular, for frequencies which enter during a radiation-dominated era, the approximate flatness of the inflationary spectrum is preserved;while  for those which enter during a matter-dominated era there is a $\propto k^{-2}$ suppression of higher-frequency modes (see \ref{eq:mattergw}).

Indeed, for frequencies corresponding to modes that entered the horizon during the radiation era after the extended reheating epoch, namely $ a_{\text{eq}}H_{\text{eq}}\ll 2\pi a_0f \ll a_{\text{reh}}H_{\text{reh}}$, it can be shown that~\cite{Saikawa:2018rcs,Ringwald:2020vei}
\begin{equation}\label{eq:radiationgw}
\begin{split}
    h^2 \Omega^{\text{inf,RD}}_{\text{GW}} (f) &= h^2\Omega_{\text{rad}} \frac{g_{*\rho}(T_{\text{hc}})}{g_{*\rho,0}} \left[\frac{g_{*s,0}}{g_{*s}(T_{\text{hc}})} \right]^{4/3} \Delta^2_t (f,\tau_i) \\
    &\simeq 1 \times 10^{-17}\times g_{*\rho}(T_{\text{hc}})\left[ g_{*s}(T_{\text{hc}})\right]^{-4/3}  \left [ \frac{H_*(f)}{10^{13}\ \text{GeV}}\right]^{2}
\end{split}
\end{equation}
where the relativistic degrees of freedom in entropy $g_{*s}$ and in energy density $g_{*\rho}$ are dependent on the temperature $T_{\text{hc}}$ of the dominant radiation bath at the horizon-crossing of the modes of frequency $f$, with
\begin{equation}\label{eq:freq}
    f \simeq 1\times10^{-4}\ \text{Hz}\ \times [g_{*\rho}(T_{\text{hc}})]^{1/2}\left[ g_{*s}(T_{\text{hc}})\right]^{-2/3} \left[\frac{T_{\text{hc}}}{10\ \text{TeV}} \right].
\end{equation}
Note that the thermodynamical pre-factors, which arise from the conservation of comoving entropy, result in  dips in the spectrum as species decouple~\cite{Schwarz:1997gv,Seto:2003kc} $O(1)$ at the frequencies of interest, which are consistent with those imprinted by the SM~\cite{Saikawa:2018rcs}, since the dark sector is decoupled and sub-dominant during the radiation era for  $T\ll T_{\text{reh}}$. 

Let us now assume an instantaneous approximation for the transition between the extended reheating epoch (ending with $N_{\text{MD}}$ e-folds of matter domination) and the thermal radiation-dominated epoch after inflation. For frequencies corresponding to modes which entered the horizon before the end of reheating, assuming a general equation of state $w$, it can be shown that (see e.g. Refs.~\cite{Zhao:2011bg,Liu:2015psa}) 
\begin{equation}\label{eq:mattergw}
\begin{split}
    h^2 \Omega^{\text{inf, pre-RD}}_{\text{GW}} (f) &= \left(\frac{a_{k}}{a_{\text{reh}}} \right)^{1-3w} \times h^2\Omega_{\text{rad}} \frac{g_{*\rho}(T_{\text{reh}})}{g_{*\rho,0}} \left[\frac{g_{*s,0}}{g_{*s}(T_{\text{reh}})} \right]^{4/3} \Delta^2_t (f,\tau_i) \\ 
    \text{where}\quad &\left(\frac{a_{k}}{a_{\text{reh}}} \right)^{1-3w}  = \left( \frac{k_{\text{reh}}}{k}\right)^{2\beta} \qquad \text{with}\quad \beta = \frac{1-3w}{1+3w}
\end{split}
\end{equation}
We may then set $w=0$ ($\beta=1$) for early matter-domination, which corresponds to $\propto k^{-2}$ suppression. Accordingly, beyond the tensor amplitude dependence on $N$ and $\xi$, the model-dependence is mostly imprinted in the higher-frequency regime $ $ (this will also be the case for preheating). Using Eqs. (\ref{eq:radiationgw}), (\ref{eq:freq}) and (\ref{eq:mattergw}), we estimate the effective contribution from inflation in Figure~\ref{fig:fig:vishnugwomega}. Note that the sharp corners in the amplitude spectra we have plotted are not physical and would be rounded (with small oscillatory features) if we have solved the equations of motion directly. Nonetheless, the piecewise power laws give a good guide to the transferred spectral amplitude. 

\paragraph{Preheating and turbulence}
During preheating, as we saw in Section \ref{sec:latticeprefragturb}, sizeable gradient energies are generated in the PQ scalar field configuration, resulting in considerable anisotropic stress and, hence, efficient GW production~\cite{Khlebnikov:1997di,Easther:2006gt,Easther:2006vd,Easther:2007vj,Garcia-Bellido:2007nns,Garcia-Bellido:2007fiu,Dufaux:2007pt,Dufaux:2008dn,Figueroa:2017vfa}. Indeed, the emission of gravitational waves continues in the non-linear regime while re-scattering brings the system into a stationary state, after a moderate growth of UV modes during the initial energy cascade, the production of GW is expected to become negligible.

This was recapitulated in our simulations. In particular, we obtain an approximation using the lattice-averaged gravitational wave energy spectrum computed at various time-steps (see Figure ). We may then obtain a prediction for the resulting contribution to the present SGWB as
\begin{equation}
    h^2 \Omega^{\text{preh}}_{\text{GW}} (f) = h^2\Omega_{\text{rad}} \frac{g_{*\rho}(T_{\text{reh}})}{g_{*\rho,0}} \left[\frac{g_{*s,0}}{g_{*s}(T_{\text{reh}})} \right]^{4/3}e^{-4R_N}\underbrace{\left\langle\frac{1}{\rho(\tau_\text{prod})} \frac{\mathrm d \rho_{\text{GW}}}{\mathrm d \log k}(\tau_{\text{prod}}) \right\rangle}_{\text{lattice simulations}}\Bigg|_{k=\frac{f}{2\pi a_o}}
\end{equation}
where, as before, there is a dependence on the e-folds of matter domination, which dampens the spectrum (as the modes propagate through the entire matter-dominated era after production). We also assume that GWs are entirely sourced during the initial radiation-like epoch immediately after inflation, and again approximate the transition to primordial matter-domination as instantaneous.  

\paragraph{Thermal bath} 

A primordial thermal radiation bath of sufficiently high temperature can source a non-negligible contribution to the expected SGWB amplitude via thermal fluctuations with a peak frequency generally at $\sim 80$ GHz~\cite{Ghiglieri:2015nfa,Ghiglieri:2020mhm,Ringwald:2020ist}, hence dubbed the Cosmic Gravitational Microwave Backround (CGMB). While the reheating temperature is ultimately low in VISH$\nu$, a large maximum thermalisation temperature ($T_{\text{max}}\sim10^{13}$ GeV) is still realised in the spectating radiation bath formed from Higgs decays during preheating. The amplitude of the spectrum nonetheless comes to be additionally suppressed by the primordial intervals of matter domination. Following Ref.~\cite{Ringwald:2020ist}, but accounting for the equation of state during reheating, we have
\begin{equation}
\begin{split}
    h^2 \Omega^{\text{thermal}}_{\text{GW}} &\simeq 1.7e^{-4R_N}\times 10^{-17} \times \left[\frac{T_{\text{max}}}{10^{13}\ \text{GeV}}\right] \left[ \frac{g_{*s}(T_{\text{max}})}{106.75}\right]^{-5/6} \left[ \frac{f}{\text{GHz}}\right]^3 \\
    &\qquad \times \ \ \hat{\eta} \left(T_{\text{max}}, 2\pi \left[ \frac{g_{*s}(T_{\text{max})}}{g_{*s}(T_{\text{fin}})}\right]^{1/3} \frac{f}{T_0} \right)
\end{split}
\end{equation}
where 
$\hat{\eta}$ is a dimensionless source term defined in Ref.~\cite{Ringwald:2020ist} and ``fin'' refers to the epoch after neutrino decoupling. Through its amplitude dependence on $T_{\text{max}}$, and the relativistic degrees of freedom, the CGMB may thus be used to probe the size of the Higgs misalignment in the VISH$\nu$ inflationary trajectory, which is $\propto \sqrt{2\lambda_S/\lambda_1}$.

\section{Summary and conclusion}
\label{sec:con}

In this paper, we have analysed the post-inflationary reheating epoch of the VISH$\nu$ axion-majoron model~\cite{Sopov:2022bog}, an $N_{DW} = 1$ flavour-variant of the $\nu$DFSZ~\cite{Volkas1988,nuDFSZ}. Such a model provides a minimal way of realising solutions to five of the fundamental issues (dark matter, neutrino mass, BAU, inflation and the strong CP problem) which must be addressed exclusively BSM, \textit{while introducing no physics that destabilises the electroweak scale}. The latter requirement translates to tiny, but symmetry-protected, inter-sector couplings given the much heavier Peccei-Quinn/RHN mass-scale.  

Inflation is assumed to be driven by the slow-roll of a CP-even scalar combination which is mostly the PQ scalar modulus or ``saxion'', an arrangement which avoids a low-scale violation of unitarity. Hence, the demand for naturalness, which suppresses the interaction rate with the thermal bath, sets up a very long, non-thermal reheating epoch driven largely by the non-linear inhomogeneous oscillations of the dark bosonic fields. This regime is provoked by the non-perturbative production of non-thermal inflatons and axions through parametric resonance in the immediate aftermath of inflation, replacing the homogeneous inflaton condensate as the dominant energy component.\footnote{During this initial process of preheating, RHNs can also be produced non-perturbatively}
The dominant portal to the SM then becomes the RHNs, rather than the Higgs, from which the PQ scalar is effectively decoupled at tree-level to suppress high mass-scale corrections to the Higgs mass parameters. 

To ensure that the non-linear dynamics are modelled accurately, we studied the initial stages of bosonic reheating with lattice simulations using a modified version of CLUSTEREASY~\cite{Felder:2000hq,Felder:2007kat}; encompassing preheating and the realisation of a post-inflationary axion (Section~\ref{sec:latticeprefragturb}), inflaton fragmentation and field turbulence (Section~\ref{sec:latticefragturb}), as well as the non-thermal phase transition (Section~\ref{sec:latticept}). The results of the lattice analysis then informed the perturbative study of the end of reheating and axion thermalisation (Section~\ref{sec:reh}), where we solved integrated Boltzmann equations for the baths, Eq. (\ref{eq:boltz}), using well-motivated lattice extrapolations (Appendix~\ref{sec:appB}).

We showed that, quite generally, the DFSZ-like axions produced from preheating, rescattering, primordial string radiation and inflaton decay do not thermalise through the QCD portal for $f_a \sim 10^{11}$ GeV, nor through the RHN portal for ``natural'' parameter choices set out in Section~\ref{sec:neutrinos} that simultaneously achieve thermal leptogenesis. This primordial axion bath thus constitutes an early dark radiation component, which must be suppressed in order to respect well-known constraints on $\Delta N_{\text{eff}}$. We then showed how long-lived RHNs, produced via freeze-in during reheating, can act as a bath of fermionic reheatons, inducing primordial RHN matter domination before decaying to populate the thermal SM bath. In suppressing the energy fraction in axions, this mechanism renders the reheating viable. This may be achieved for arbitrarily small Higgs portal couplings of the PQ scalar, 
which, if sizeable, would otherwise make the reheating much more efficient.  

We further demonstrated how the RHNs may thus produce sufficient asymmetry for successful leptogenesis in at least three different regions of parameter space. In two, which have called $N_1$-from-decay (see Figure~\ref{fig:N1fromdecay}) and $N_3$-from annihilation (Figure~\ref{fig:N3fromann}), this could be achieved while the electroweak scale remains radiatively protected from the RHN mass scale.  We additionally analysed the case where RHNs are produced during fermionic preheating (Figure \ref{fig:fermpreheating}), which we argued to lead to an unnatural, though viable, scenario for both reheating and leptogenesis. All scenarios are, furthermore, realisable without enforcing a degeneracy in the neutrino masses, nor introducing PQ violating non-minimal couplings, both considered in the analysis of Ref.~\cite{Barenboim:2024xxa}.

In $N_1$-from-decay, the lepton asymmetry is produced directly in the non-thermal decays of $N_1$ which populate the bath at the end of reheating (with $T_{\mathrm{reh}} < m_{N_1}$). The $N_1$ bath is seeded by inflaton decays, and thus requires a sufficiently large branching fraction. In $N_3$-from-annihilation, freeze-in production of $N_3$ is enhanced around the non-thermal PQ phase transition, and the thermal bath produced by $N_3$ decays can have  $T_{\mathrm{reh}} > m_{N_1}$, hence permitting conventional thermal leptogenesis. 

In each case, the requirement of both sufficient matter domination and reheating temperature enclose the viable parameter region.  We find that the demand for radiative stability is in both cases very stringent, but can be achieved with (radiatively stable) $\tan \beta \gg 1$ enhancing the CP violation resulting from decays. This corresponds to the requirement that $v_2 \ll v_1$, which, in the top-specific VISH$\nu$ model, leads to a naturally heavier top quark~\cite{Sopov:2022bog}.

In assembling a complete cosmic history for VISH$\nu$, the inflationary predictions of Ref.~\cite{Sopov:2022bog} could be made more precise. In particular, we have shown how each reheating scenario, in contrast to a radiation-only reheating (as in SMASH), leads to a slightly lower range for the spectral index of the scalar perturbations due to additional e-folds of matter domination (Figure~\ref{fig:rnsplot}). This may be challenged by future CMB data from, e..g EUCLID~\cite{EUCLID:2011zbd}, LiteBIRD~\cite{LiteBIRD:2022cnt}, the SKA~\cite{Maartens:2015mra}.  Indeed, the tensor-to-scalar ratio $r \sim \text{few}\times 10^{-3}$ predicted by the model will likewise confront new measurements at the South Pole~\cite{Moncelsi:2020ppj} and the Simons Observatory~\cite{SimonsObservatory:2018koc}.

Additionally, the relic axions from the VISH$\nu$ reheating scenarios we have identified lead to $\Delta N_{\text{eff}} > 0.01$, encompassing values which may be probed by future CMB experiments, as would have been the case with CMB-S4~\cite{CMB-S4:2016ple}. Moreoever, each of the reheating scenarios require long-lived RHNs consistent with a highly suppressed absolute neutrino mass scale $m_{\text{min}} \lll 10^{-3}$ eV. Hence, the scenarios we have identified may also be falsified by future neutrino mass measurements from e.g. tritium decay kinematics, neutrinoless double beta decay and cosmological probes.\footnote{In this case, thermalisation processes consistent with much larger radiative corrections to the electroweak scale would be required to keep the model viable, given our parameter assumptions.}

We also estimated the most distinctive contributions of the model to the SGWB (Figure~\ref{fig:fig:vishnugwomega}). That is, from inflation, preheating and turbulent eras, as well as the imprint left on these by the expansion history. It was demonstrated how the intervals of matter-domination required in each scenario, which suppress the amplitude for $f\gtrsim$ mHz frequency ranges, may lead to experimentally testable departures from a radiation-only (SMASH-like) reheating epoch. In the case of the $N_3$-from-annihilation scenario, this may realise a dip within the sensitivity of an ultimate DECIGO~\cite{Kuroyanagi_2015}. While a null result for an inflationary spectrum at $f\gtrsim$ mHz in conjunction with a measurement of $r \sim \text{few}\times 10^{-3}$, would also present compelling indirect evidence for an expansion history consistent with an $N_1$-from-decay reheating epoch. 

As we have seen, the VISH$\nu$ model, despite its weak portal couplings between the visible and dark sectors, must not only carry a significant burden in realising a self-consistent cosmic history, but, in so doing, may be constrained or excluded by future observations in a number of ways (in its most natural form). The model also realises  and motivates post-inflationary QCD axion dark matter, which lies within the discovery potential of a number of haloscope experiments, such as MADMAX~\cite{Caldwell2017}.

While VISH$\nu$ is not the only proposal for a benchmark BSM extension which can be a complete BSM particle-cosmological description from inflation to the present (see also Refs.~\cite{Ballesteros:2016xejSMASH, Ballesteros:2016euj,Ballesteros:2019tvf}), it may be distinguished both by the smallness of the high-scale BSM corrections to the electroweak scale it introduces, together with interesting phenomenological implications. Of course, it would also be very interesting to consider minimal ways in which an explanation of dark energy can arise in this context, or to consider well-motivated  completions which can resolve the grander hierarchy and axion quality problems arising from the Planck scale, which we have not considered here. 

\acknowledgments
This work was supported
in part by the Australian Research Council through the ARC Centre of Excellence for Dark Matter Particle Physics, CE200100008. AHS is supported by the Australian Government Research Training Program Scholarship initiative, and thanks the Mainz Institute for Theoretical Physics (MITP) for hospitality and a stimulating environment during the ``Crosslinks of early universe cosmology'' summer school. CT acknowledges support from the Cluster of Excellence “Precision Physics, Fundamental Interactions, and Structure of Matter” (PRISMA+ EXC 2118/1) funded by the Deutsche
Forschungsgemeinschaft (DFG, German Research Foundation) within the German Excellence Strategy (Project No. 390831469).

\appendix

\section{Benchmark parameters and lattice simulation details}
\label{sec:appA}

In this Appendix, we provide some benchmark values fixed for the purposes of our lattice simulations (Table \ref{tab:table1}), provide the lattice parameters used in the simulations discussed in Section \ref{sec:lattice}, (\ref{eq:sim1}-\ref{eq:sim6}), give the partial decay widths to fermions for the 7 real scalar components modelled by our lattice simulation, Eq. (\ref{eq:decays}), and explain a subtletly related to the gauge choice.

Our benchmark parameters (Table \ref{tab:table1}) include $\lambda_S \simeq 2.5 \times 10^{-10}$, which follows from the inflation fit to $A_s$ (Figure \ref{fig:lxiplot}) with $\xi \simeq 0.5$. In the lattice simulation, we choose $y_{N_1}  \simeq 7.4\times10^{-6}$ to realise a kinematically accessible $\sigma \rightarrow N_1 N_1$ decay, with $m_{N_1}$ above the Davidson-Ibarra bound, Eq. (\ref{eq:davidsonibarra}), for the benchmark choice of $v_S \sim 10$ GeV. The eigenvalues $y_{N_2}$ and $y_{N_3}$, as well as the matrix $y_\nu$ are all free for the purposes of the simulation.\footnote{In practice, the dependence on $y_{N_1}$ is weak as the decays are not efficient in the lattice integration time. }

We explicitly consider results of simulations with the following lattice simulation parameters. Here, $N^3$ is the number of lattice sites in the three-dimensional box, $L$ is the lattice box side length in units of $(\sqrt{\lambda_S}\sigma_{\text{end}})^{-1}$, from which one obtains $\kappa_{\text{IR}} =\frac{2\pi}{L}$ and $\kappa_{\text{UV}} = \frac{\sqrt{3}N }{2}\kappa_{\text{IR}}$ as the corresponding IR and UV momentum cut-offs (in units of $\sqrt{\lambda_S}\sigma_{\text{end}}$), $\Delta \tau$ is the time-step, in units of $(\sqrt{\lambda_S}\sigma_{\text{end}})^{-1}$, chosen to be $\ll \frac{L/N}{\sqrt{3}}$ for numerical stability.
\begin{align}
    N^3=512^3,\ L=75\pi,\ \Delta \tau = 0.018,\ f_a = 1\times10^{11}\text{GeV} \qquad &\text{(Preheating,\  Figs.~\ref{fig:spectrapreheating},\ref{fig:varenpreheating}}), 
    \label{eq:sim1}\\
    N^3=256^3,\ L=15\pi,\ \Delta \tau = 0.0048,\ f_a = 1\times10^{11}\text{GeV} \qquad &\text{(Turbulence 1,\  Fig.~\ref{fig:nkevol1}}) ,
    \label{eq:sim2}\\
    N^3=128^3,\ L=8\pi,\ \Delta \tau = 0.009,\ f_a = 1\times10^{11}\text{GeV} \qquad&\text{(Turbulence 2,\  Fig.~\ref{fig:nkevol2}}) ,
    \label{eq:sim3}\\
    N^3=128^3,\ L=5\pi,\ \Delta \tau = 0.0011,\ f_a = 2\times 10^{16}\text{GeV} \qquad&\text{(PQPT 1,\  Fig.~\ref{fig:afterpt}}) ,
    \label{eq:sim4}\\
    N^3=128^3,\ L=9\pi,\ \Delta \tau = 0.00199,\ f_a = 5 \times 10^{14} \text{GeV} \qquad&\text{(PQPT 2,\  Fig.~\ref{fig:afterpt}}) ,
    \label{eq:sim5}\\
    N=128^3,\ L=11\pi,\ \Delta \tau = 0.0024,\  f_a = 1\times10^{16}\text{GeV} \qquad&\text{(PQPT 3,\  Figs.~\ref{fig:strings1},\ref{fig:strings2}}) \label{eq:sim6}.
\end{align}

For definiteness, we consider a $\Phi_1 S$-Inflation scenario in the simulation. After inflation ends, this can only be approximate, as the small-field potential no longer stabilises the negligible vacuum angle and the Higgs basis becomes misaligned, which would spoil our choice of unitary gauge. Nonetheless, it is a good approximation for the scalar field configuration during the early stages of reheating when the Higgs components promptly thermalise, and so the zero-modes of the $\Phi_2$ components may be neglected in what follows. 

\begin{table*}
\centering
\def\arraystretch{1.1}
\begin{tabular}{|c||c|c|c|}
\cline{2-4}
\multicolumn{1}{c|}{} & $\mu = m_Z, \mathbf{m_{22}}$ & $\mu = \sqrt{\lambda_S}\sigma_f$ & $\mu = m_P$ \\
\hline 
$\lambda_1(\mu)$ & $0.259$ & $0.0978$ & $0.240$\\
\hline 
$\lambda_2(\mu)$ & $0.28$ & $0.62$ & $1.2$\\
\hline 
$\lambda_3(\mu)$ & $\mathbf{0.33}$ & $0.58$ & $0.27$ \\
\hline 
$\lambda_4(\mu)$ & $\mathbf{ -0.10 }$ & $ -0.033 $ & $ -0.029 $ \\
\hline
$\lambda_{1S}(\mu)$ & $ 5.7 \times 10^{-20} $ & $ -4.9 \times 10^{-16}$ & $- 1.4 \times 10^{-15}$ \\
\hline 
$\lambda_{2S}(\mu)$ & $1.4 \times 10^{-17}$ & $-3.3 \times 10^{-15}$ & $- 5.5 \times 10^{-15}$ \\
\hline 
$\lambda_{12S}(\mu)$ & $ 5.6 \times 10^{-7}$ & $ 5.9 \times 10^{-7}$ & $6.0\times 10^{-7}$ \\
\hline 
$y_t(\mu)$ & $ 0.961$ & $ 0.450 $ & $0.382 $ \\
\hline 
$y_b(\mu)$ & $ 0.44$ & $ 0.17 $ & $ 0.14 $ \\
\hline 
$g_1(\mu)$ & $0.357 $ & $ 0.429 $ & $ 0.473 $ \\
\hline 
$g_2(\mu)$ & $ 0.652 $ & $ 0.546 $ & $ 0.515$ \\
\hline 
$g_3(\mu)$ & $ 1.22 $ & $ 0.568 $ & $ 0.496 $ \\
\hline 
$[M_{11}(\mu)]^2$ & $\mathbf{- [170\ \textbf{GeV}]^2}$ & $ [609\ \text{GeV}]^2 $ & $ [863\ \text{GeV}]^2 $  \\
\hline 
$[M_{22}(\mu)]^2$ & $\mathbf{[1.1\ \textbf{TeV}]^2}$ & $ [1.2\ \text{TeV}]^2$ & $ [1.5\ \text{TeV}]^2 $ \\
\hline 
\end{tabular}
\caption{
\label{tab:table1} Values for the couplings used in the simulation (central column) obtained by renormalisation group evolution (using 2-loop $\beta$ functions generated in PyR@TE~\cite{Sartore:2020gou}), to an appropriate mass-scale for the reheating analysis (viz. the order of the effective masses (and $H_{inf}$) $\mu \sim \lambda_S^{1/4}\sigma_{f} \sim 5 \times 10^{13}$ GeV), from benchmark inputs consistent with stability and phenomenological criteria set out in Ref.~\cite{Sopov:2022bog}. Note that running in $\lambda_S \simeq 2.5 \times 10^{-10}$, $(M_{SS})^2 \simeq - (1.2 \times 10^6 \ \text{GeV})^2$, $y_{N_i}$ and $y_\nu$ is negligible, and we have fixed $\tan \beta \sim 26$ in the low-energy model. }
\end{table*}

We use a relativistic fermion bath as an adequate substitute for the full extended SM thermal plasma, augmented by a tiny non-thermal population of releativstic $N$'s which may thermalise later. Specifically, we implement the growth and evolution of this fermion energy density $\rho_f$ in the presence of the general scalar-field configuration using the leading partial decay widths (using Heaviside functions where appropriate to make the evolution physically reasonable):
\begin{equation}\label{eq:decays}
\begin{split}
    \Gamma (\sigma_j \rightarrow N_1 N_1) &= \theta(1 - x_j) \frac{y_{N_1}^2}{16\pi}  m_j  \sqrt{ 1- x_j} \left(1 - \frac{x_j\langle\sigma_j\rangle^2}{\langle \sigma \rangle^2} \right) \\
    \Gamma (h \rightarrow t\overline{t}) &= \theta(1-x_{t} )\frac{3y_t^2}{16\pi} m_h \left( 1 - x_t \right)^{3/2} \\
    \Gamma (X \rightarrow b\overline{b}) &\simeq  \frac{3y_b^2}{16\pi} m_{X} \\
    \Gamma (X^+ \rightarrow t\overline{b}) &\simeq  \theta \left(1-x^+_t \right) \frac{3y_b^2}{32\pi} m_{X^+} \left(1 - x^+_t\right)^{3/2} \\
    \Gamma (h \rightarrow VV \rightarrow 4f) &\simeq \theta(1 - x_V) \frac{\delta_V g^2}{128\pi}\frac{m_h^3}{m_W^2}\sqrt{1-x_V}\left(1-x_V+\frac{3}{4}x_V^2\right)
\end{split}
\end{equation}
where $X \in \{H, A\}$,  $X^+ \in \{(H^+)_1, (H^+)_2\}$ and $VV \in \{ZZ, W^+W^-\}$. The mass ratios are:
\begin{equation}
     x_j = \frac{4m_{N_1}^2}{m_j^2},\quad x_t = \frac{4m_t^2}{m_{h}^2},\quad x^+_t = \frac{m_t^2}{m_{X^+}^2},\quad x_V = \frac{4m^2_V}{m_h^2},
\end{equation}
while the masses are given by:
\begin{equation}
    m_{N_1} = \frac{y_{N_1}}{\sqrt{2}}  \langle \sigma \rangle,\quad  m_t = \frac{y_t}{\sqrt{2}}  \langle h \rangle,\quad m^2_W =  \frac{g^2\langle h^2\rangle}{4},\quad m^2_Z = \frac{(g^2+g'^2)\langle h^2\rangle}{4}
\end{equation} 
and:
\begin{equation}
    m_{[\sigma_jh,X,X^+]} = \theta \left( \frac{\partial^2 V }{\partial[\sigma_j,h,X,X^+]^2} \right) \left|\frac{\partial^2 V }{\partial[\sigma_j,h,X,X^+]^2} \right|^{1/2}.
\end{equation}
where $y_t \equiv \left|(y_{u1})_{3}\right|$,  $y_b \equiv \left|\left(y_{d}\right)_{3,3}\right|$, $\delta_Z =1$ and $\delta_W=2$. Note that $m_b \sim 0$, and hence $x_b \sim 0$, follows from $\langle (\Phi_2)_i\rangle \sim 0$. As in Ref~\cite{Ballesteros:2021bee}, we cut off the apparent divergence in the final decay with $x_W > 10^{-3}$ to ensure numerical stability.\footnote{Again, our analysis of gauge boson production is strictly perturbative. In this approximation, the disappearance of the longitudinal polarisation during axes-crossings of $\langle h^2 \rangle$ simply invalidates the decay rate given and justifies the use of a simple cutoff, on which (in the regime considered) our results have little dependence.}

\section{Thermalisation details}
\label{sec:appB}

In this Appendix, we include additional expressions for the lattice-informed reheating analysis in Section~\ref{sec:reh}, to which we refer in the body of the paper.

\begin{figure*}[t]
\begin{center}
\includegraphics[width = 0.45\textwidth]{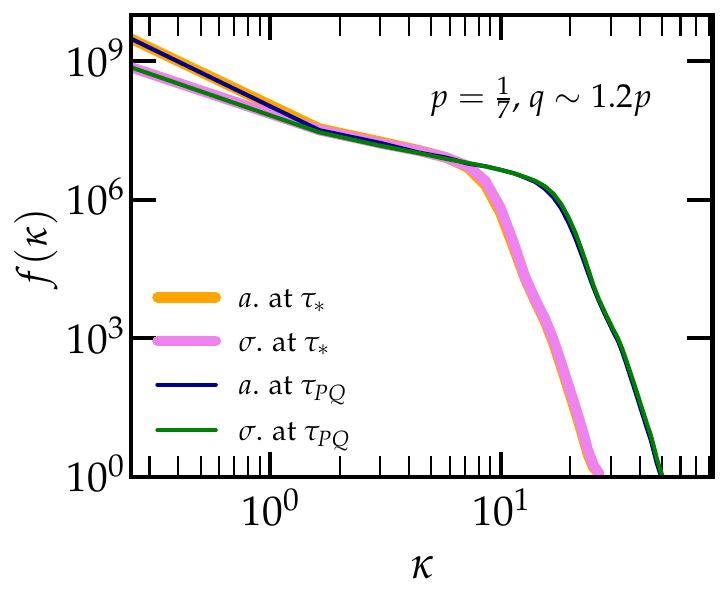}
\end{center}
\caption{\label{fig:nkextrap}In this plot we illustrate the good fit our assumed values of $p = \frac{1}{7}$ and $q \simeq 1.2p$ provide to the turbulent scaling of the axion and modulus distribution functions. In thick orange and violet, we provide the final lattice occupation number spectrum for the simulation (\ref{eq:sim2}). In blue and green, we present extrapolations using Eq. (\ref{eq:selfsimilar}) with the indicated values of $p=\frac{1}{7}$ and $q\sim 1.2p$.  }
\end{figure*}
\paragraph{Decay rates}

The leading contributions to the decay rates of the inflatons and $N_i$ are
\begin{equation}\label{eq:decayrates}
    \Gamma_{\sigma\rightarrow aa} = \frac{1}{32\pi}\frac{m_\sigma^3}{f_a^2}, \ \ \Gamma_{\sigma\rightarrow N_1N_1} \simeq \frac{y^2_{N_1}m_{N_1}}{16\pi}\left[1-\frac{4m_{N_1}^2}{m^2_{\sigma}} \right]^{\frac{3}{2}},\ \ \Gamma_{N_i \rightarrow\Phi_2 \ell} \simeq \frac{(y_\nu^\dagger y_\nu)_{ii}}{8\pi} m_{N_i}.
\end{equation}
Before the phase transition, the axion has a mass, Eq. (\ref{eq:effmass}), and hence a decay width which is $\Gamma_{a\rightarrow N_1N_1}\simeq \Gamma_{\sigma\rightarrow N_1N_1}$ up to a kinematic factor. In the parameter region of interest, this does not become efficient before vanishing, and is hence not included in Eq. (\ref{eq:boltz}).

\paragraph{Interaction rates}

In Eq. (\ref{eq:boltz}), we must consider the two kinds of 2-2 processes which can change the number density of axions and inflatons, which we designate as ``absorptions'' when  $\alpha\beta_1\rightarrow \beta_2\beta_3$ with $\beta_i$ a bath particle and $\alpha_i = a$ or $\sigma$; and (co)annihilation when it is $\alpha_1\alpha_2\rightarrow \beta_1\beta_2$. The axion absorption processes are dominated by the SM QCD contributions with $N_f = 6$ quark flavours~\cite{Graf:2010tv,Masso:2002np,Salvio:2013iaa,Mazumdar:2016nzr}
\begin{align}\label{eq:qcdabsorption}
    \sigma_{ag\rightarrow q\overline{q}} &\simeq \frac{\alpha_s^3}{\pi^2f_a^2}, \\ 
    \sigma_{ag\rightarrow gg} &\simeq \frac{\alpha^3}{2\pi^2f_a^2} \left[15\log\left(\frac{s}{m_D^2}\right) - \frac{55}{4} \right],
\end{align}
where a Debye mass $m_D = \sqrt{8\pi\alpha_s}T$ for the gluon has been used in the second case to render the $t$-channel divergence finite. This is because the additional axion and inflaton absorption processes mediated by the $N_i$; namely $aN_i\rightarrow \Phi_2  \ell$, $\sigma N_i\rightarrow \Phi_2  \ell$, and the other processes obtained by crossing $N_i, \Phi_2, \ell$, are all $\propto (y_\nu^\dagger y_\nu)_{ii} y^2_{N_{i}}$, which is highly suppressed in the parameter space of interest, Eq. (\ref{eq:naturalyukn}).

The (co)annihilation processes are then $aa \rightarrow N_i N_i$, $a\sigma \rightarrow N_i N_i$ and $\sigma\sigma\rightarrow N_iN_i$, for kinematically accessible $N_i$ final states. In all cases there are $t$- and $u$-channel processes mediated by $N_i$, $s$-channel contributions mediated by either the inflaton and axion, which, in the polar basis, must be summed with contact terms generated in the Yukawa sector.

This is made possible by the $\sigma (\partial a)^2$ and $\sigma^3$ vertices generated by the background value of $\sigma$, which is $f^{\mathrm{eff}}_a \gg f_a$ before the PT,
\begin{equation}
    \mathcal{L}_{PQ} \supset |\partial S|^2 - \frac{\lambda_S}{2}|S|^4 \supset \frac{\sigma}{f_a^{\mathrm{eff}}} (\partial a)^2 - \frac{\lambda_S}{2}f^{\mathrm{eff}}_a\sigma^3,
\end{equation}
where we approximate $\partial f^{\mathrm{eff}}_a \sim 0$ using an adiabatic approximation for the time-dependent spatial average for the inhomogeneous oscillations of $\sigma$. After the PT, $f^{\mathrm{eff}}_a \rightarrow f_a$, and the result is exact. This operator also realises the later decay of inflatons to axions.

In the limit $s\gg m^2_{\sigma}$ where the heavy $N_i$ are kinematically accessible as final states, the cross sections for the annihilation rates become
\begin{equation}\label{eq:sigmaann}
    \sigma_{aa/\sigma\sigma\rightarrow N_{i} N_{i}} \underset{s\gg m^2_\sigma}{\simeq} \frac{y^4_{N_i}}{32\pi s^2} \left\{(s+4m^2_{N_i})\log \left[\frac{1+ \sqrt{1-\frac{4m^2_{N_i}}{s}}}{1 - \sqrt{1-\frac{4m^2_{N_i}}{s}}} \right] - 8m^2_{N_i}\sqrt{1-\frac{4m^2_{N_i}}{s}}
    \right\}
\end{equation}
\begin{equation}\label{sigmacoann}
    \sigma_{a\sigma\rightarrow N_{i} N_{i}}  \underset{s\gg m^2_\sigma}{\simeq} \frac{y^4_{N_i}}{32\pi s^2} \left\{(s+4m^2_{N_i})\log \left[\frac{1+ \sqrt{1-\frac{4m^2_{N_i}}{s}}}{1 - \sqrt{1-\frac{4m^2_{N_i}}{s}}} \right] - 2s \sqrt{1-\frac{4m^2_{N_i}}{s}}\right\}
\end{equation}
In particular, the annihilation cross-sections are $\propto y^4_{N_i}$ and are enhanced for heavier $N_i$.\footnote{There is an exception, namely that the mediator can go on-shell for the $aa\rightarrow N_1N_1$ process for smaller $s$. This is included in Eq. (\ref{eq:boltz}) using a Breit-Wigner approximation for the associated $s$-channel resonance, but does not become efficient for our benchmark parameters.} However, they cannot be arbitrarily large due to kinematic blocking, as the UV tail of the non-thermal spectrum is exponentially suppressed. This results in a finite mass for $N_i$ where the non-thermally averaged rate peaks, which is around $y_{N_i} \sim \text{few} \times 10^{-4}$ for $f_a \sim 10^{11}$ GeV.

The cross-sections are then integrated over using the non-thermal distribution functions for the inflatons and axions, where in the collision term for the energy density we have 
\begin{equation}
    \Gamma_{12\rightarrow34} \equiv \frac{1}{\rho_1} \int \mathrm{d}\Pi_1 \int \mathrm{d}\Pi_2 \ E_1 f_1 f_2 \Lambda_{12} \sigma_{12\rightarrow34},
\end{equation}
where $\Lambda_{12} = 4g_1 g_2\sqrt{(p_1 \cdot p_2)^2- m_1^2 m_2^2}$ and $\mathrm{d}\Pi_i = \frac{\mathrm{d}^3\mathbf{p}_i}{(2\pi)^32E_i}$.

\paragraph{Effective masses}

The masses of the $N$'s and the PQ scalar components satisfy
\begin{equation}\label{eq:effmass}
    m^2_{\sigma} = \frac{\lambda_S}{2}\left[3(f_a^{\text{eff}})^2 - f_a^2 \right],\qquad m^2_{a} = \frac{\lambda_S}{2}\left[(f_a^{\text{eff}})^2 - f_a^2 \right],\qquad m_{N_i}=\frac{y_{N_i}}{\sqrt{2}}f_a^{\text{eff}},
\end{equation}
with $\tau_*$ the final lattice integration time, and
\begin{equation}
    f^{\text{eff}}_a = \begin{cases} \sqrt{\langle \sigma^2(\tau_*) \rangle }\left( \frac{\tau_*}{\tau} \right)^{8/7}&, \quad \tau < \tau_{PQ} \\ 
    \qquad \qquad  f_a&, \quad \tau \geq \tau_{PQ}
    \end{cases}
\end{equation}
where $\tau_{PQ}$ is defined so that the function is smooth. These expressions are used in the interaction rates and relevant estimates for $w$ and $\kappa$. 

\paragraph{Non-thermal equations of state}

Due to the red-shifting inflaton oscillation amplitude before the phase transition, each of the N's, inflatons and (massive) axions will red-shift as radiation regardless of their non-thermal distributions. One must then consider the equations of state post-PT, other than the radiation baths in SM and axions, on a case by case basis. 

\begin{figure*}[t]
\begin{center}
\includegraphics[width = 0.4\textwidth]{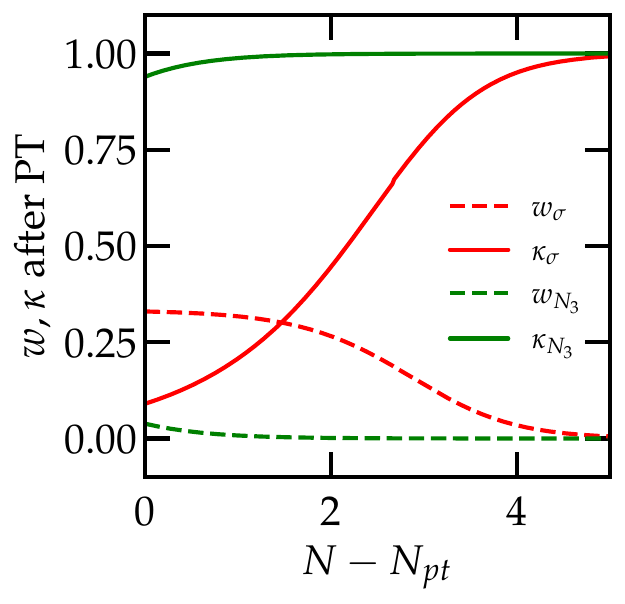} \quad\includegraphics[width = 0.4\textwidth]{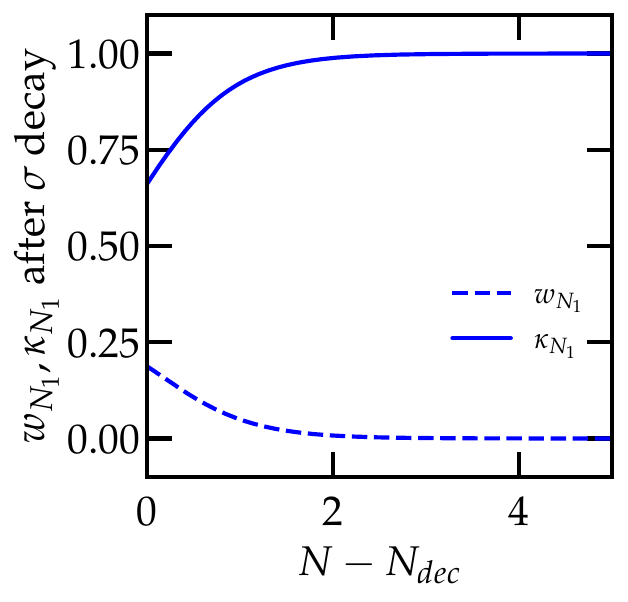}
\caption{\label{fig:timedil} Here we present the equations of state and time dilation factors resulting from numerical integrations of the extrapolated non-thermal distributions. In the left plot, the time-dependence of $w_\sigma$ and $\kappa_\sigma$ for each scenario, alongside $w_{N_3}$ and $\kappa_{N_3}$ resulting from freeze-in production with $y_{N_3} \sim 7.5 \times 10^{-4}$ are plotted after the phase transition at $N_{pt}$. In the right plot, $w_{N_1}$ and $\kappa_{N_1}$ are plotted after inflaton decay using an approximation where initially $E_{N_1} \sim \frac{m_\sigma}{2}$.  }
\end{center}
\end{figure*}

We directly use the extrapolated distribution function for the inflaton, i.e. with Eq. (\ref{eq:selfsimilar}), to obtain the time-dependent equation of state after the PT (see Figure~\ref{fig:timedil}), 
\begin{equation}
    w_{\sigma} =  \begin{cases}\qquad \qquad
        \frac{1}{3}, \quad &\text{if}\quad a<a_{PQ} \\
        \frac{\int \frac{d^3p}{(2\pi)^3}n_{k=ap}(t)\frac{p^2}{3\sqrt{p^2+m_\sigma^2}}}{\int \frac{d^3p}{(2\pi)^3}n_{k=ap}(t)\sqrt{p^2+m_\sigma^2}}, \qquad &\text{otherwise}.
    \end{cases}
\end{equation}
In principle, the $N_1$ decay products of the inflaton have a non-thermal spectrum which can be obtained by solving the Boltzmann equation at the phase-space level. We do not consider this level of detail, because the $N_1$ only play an important role after $\Gamma_{\sigma \rightarrow N_1 N_1}$, in which case we take them to have typical momenta $|p| \sim \frac{m_\sigma}{2}\frac{a_{dec}}{a}$ for $a > a_{\text{dec}}$ (the moment when the decay becomes efficient). The resulting equation of state is also plotted in Figure~\ref{fig:timedil}.

The equation of state for the neutrinos produced during preheating is well-approximated by
\begin{equation}
    w^{\mathrm{pre}}_{N_{2,3}} = \begin{cases}
        \frac{1}{3}, \quad &\text{if}\quad a<a_{PQ}  \\
        \ 0,\quad &\text{otherwise}
    \end{cases}
\end{equation}
For $N_3$ produced by freeze-in, we assume that the non-thermal spectrum satisfies a similarity relation with that of the initial states, i.e. with $f_{N} \sim \frac{n_N}{n_{a}}f_{\mathrm{ax}}$. In ratios of non-thermal averages, the ratio of the number densities drops out as a prefactor. The resulting equation of state after the PT is also plotted in Figure~\ref{fig:timedil}.

\paragraph{Non-thermal time dilation factors} In integrating the Boltzmann equation for the energy density, the decay contributions to the collision term (e.g. of the inflaton) may be brought into the form $\kappa_\sigma\Gamma_\sigma\rho_\sigma$, where $\Gamma_\sigma$ is the rest-frame decay rate, and
\begin{equation}\label{eq:timedilinf}
    \kappa_\sigma \equiv \frac{m_\sigma n_\sigma}{\rho_\sigma }.
\end{equation}
This is the analogue of the usual inverse Lorentz time dilation factor $\gamma^{-1}\sim \frac{m}{\langle E\rangle}$ which appears when the number density is considered. (The expressions may be seen to agree for a mono-energetic spectrum.) Using the extrapolated distribution function from the lattice, we solve for this function numerically (see Figure~\ref{fig:timedil}). It is then used in our solver so long as the inflatons remain non-thermal.

For $N_{2,3}$ produced by preheating, it is the case that $\kappa_{N_{2,3}}\sim1$. In the other cases, $N_1$ from inflaton decay and $N_{3}$ from freeze-in of annihilations, we use the same assumptions as for the equation of state (see Figure~\ref{fig:timedil}).

\paragraph{Bose enhancement}

As discussed in Section~\ref{sec:latticept}, we may crudely estimate the Bose enhancement of the number-changing non-thermal inflaton-axion processes by defining a rate 
\begin{equation}
    \Gamma^{\text{en}}_{\sigma\sigma \rightarrow aa} \equiv \Gamma_{\sigma\sigma\rightarrow aa}[1+f_{\text{ax}}(m_\sigma)]^2.
\end{equation}
This is $\gg H$ for a few e-folds after the PT, but then rapidly falls to $\ll H$  due to red-shifting and the exponential suppression of the UV tails on the non-thermal distributions.

\newpage
\bibliographystyle{jhep.bst}
\bibliography{vishnu2.bib}

\end{document}